\documentclass{jfm}
\usepackage{graphicx,pdflscape,rotating,psfrag}
\usepackage{epstopdf, epsfig, placeins,upgreek,url}
\usepackage{amsmath,mathtools,mathptmx,tikz}
\usepackage{amsfonts,oldgerm}
\usepackage{amssymb} 
\usepackage{bm}
\usepackage[mathlines]{lineno}
\usepackage{rotating}
\usepackage{authblk}
\usepackage{tabularx}
\usepackage{float,flafter}
\usepackage{booktabs}
\usepackage{multirow}
\usepackage{multicol}
\usepackage{makecell}
\usepackage{natbib}
\usepackage{pdfpages}
\usepackage{hyperref}
\hypersetup{
colorlinks=true,
linkcolor=black,
filecolor=black,
urlcolor=black,
citecolor=black,
}
\renewcommand{\arraystretch}{1.2}

\newcommand{\myquad}[1][1]{\hspace*{#1em}\ignorespaces}

\shorttitle{Polarity transitions induced
by symmetry-breaking outer boundary 
	heat flux}
\shortauthor{D. Majumder and B. Sreenivasan}
\title{Polarity transitions induced by
symmetry-breaking outer boundary heat 
	flux in rapidly rotating dynamos}
\author[]{Debarshi Majumder}
\author[]{Binod Sreenivasan
	\corresp{\email{bsreeni@iisc.ac.in}}}
\affil[]{Centre for Earth Sciences, 
	Indian Institute of Science, Bangalore 560012, India}

\begin{document}
\maketitle
\begin{abstract}

This study investigates, analytically and numerically,
  the role of a   non-axisymmetric  
  equatorially anti-symmetric 
  lateral variation in heat flux
   at the outer boundary in 
  polarity transitions in 
  rapidly rotating dynamos.
  In an unstably stratified fluid,
  the frequencies
  of vertical and horizontal (lateral)
   buoyancy complement each other such that
  a polarity transition is
   induced by the
  boundary anti-symmetry through the
  suppression of the slow magnetic-Archimedean-Coriolis (MAC)
   waves at relatively
  small vertical buoyant forcing. A dipole-dominated dynamo
 in the low-inertia limit
  transitions to polarity-reversing and multipolar states 
  in succession as the relative intensity of
  horizontal buoyancy is progressively
  increased for a fixed vertical buoyancy.
  In the same parameter space, an 
  equatorially symmetric heat flux variation does not induce a
polarity transition. 
  A composite boundary heterogeneity consisting of
comparable magnitudes of symmetric and anti-symmetric
variations induces the polarity transition at a 
horizontal buoyancy of the same order
as that for the transition induced by a 
purely anti-symmetric variation. This
makes the analysis relevant to
Earth's core, which convects
in response to a composite heat flux variation
in the lowermost mantle. 
A heterogeneity with a dominant 
equatorially symmetric component 
does not favour polarity transitions, 
likely producing long periods 
in Earth's history
without reversals.
The fact that
compositional buoyancy is much stronger than
thermal
buoyancy, together with the known
order of magnitude of the peak field
intensity
in the inertia-free limit,
indicates 
a
large lower-mantle heat flux heterogeneity
  of $O(10)$ times the mean superadiabatic
  heat flux at the core--mantle
   boundary.

\end{abstract}

\section{Introduction}
\label{intro}
It is well-known that the Earth's 
magnetic field is generated by thermochemical convection 
in the outer core. Compositional convection arises 
from the continual release of light elements that
accompanies the progressive growth of the
inner core. 
Thermal convection occurs primarily by
secular cooling of the core as the planet cools in time.  
Mantle convection produces a laterally 
heterogeneous heat flux at the core--mantle
boundary (CMB), which changes in geological time
 \citep{olson2010geodynamo}. Since the convective 
turnover time in the lower mantle is significantly 
longer than in the outer core 
\citep{zhang1992convection,holme2015large}, 
these lateral variations create a quasi-stationary
boundary condition that shapes outer core 
convection.

Based on several observations, it has been argued 
that the lower mantle may influence the geodynamo. 
The persistence of high-latitude flux patch 
concentrations in the paleomagnetic time average
\citep{gubbins1993persistent,
	carlut1998complex,johnson2003mapping}, the 
preferred paths of virtual geomagnetic poles 
during polarity transitions \citep{laj1991geomagnetic,
	love2000statistical}, 
and the low secular variation in the Pacific 
\citep{fisk1931isopors,doell1971} are examples. 
Paleomagnetic data also suggest that 
the frequency of geomagnetic reversals changes on the 
time scale of mantle convection 
\citep{jones1977thermal,mcfadden1984lower,
	larson1991mantle,lhuillier2013statistical}. 
In addition, recent laboratory experiments 
\citep{sahoo2020response} 
indicate that a large lower-mantle heterogeneity gives rise
to an east--west dichotomy in core convection, which in turn
explains the relative instability of the high-latitude magnetic
flux lobes in the Western hemisphere \citep{jackson2000}.

The first self-consistent dynamo models 
with inhomogeneous CMB heat flux 
\citep{glatzmaier1999role} showed that the frequency of 
geomagnetic reversals depends on the  
heat flux pattern. 
This CMB heat flux variation
was estimated from
the shear wave velocity anomalies obtained from seismic tomography, 
under the assumption that shear wave velocity is determined
by temperature and not by composition 
\citep{su1994degree,masters1996}.
	However, compositional heterogeneity can also 
	cause
significant seismic wave velocity variations in the lower
mantle	\citep{koelemeijer2012normal,gulcher2021coupled}. 
Several
numerical dynamo models \citep{
	olson2002time,
	christensen2003secular,aubert2007detecting,
	pepi11,
	olson2014magnetic,
	mound2023longitudinal} have used CMB heat
flux variations that linearly correlate with
seismic shear wave velocity anomalies. 
Additionally, various single 
harmonics have been used to understand their effect in isolation
on convection and the magnetic field
 \citep{kutzner2004simulated,takahashi2008effects,
 	coe2006symmetry,courtillot2007mantle,stanley2008mars,
 	amit2011influence}.
It has been shown that both
an enhanced equatorial heat flux 
\citep{glatzmaier1999role,olson2010geodynamo} 
and a North--South hemispherical
heat flux variation \citep{frasson2025geomagnetic} 
can produce reversals in numerical dynamos.
Mantle convection models indicate that the dominant 
spherical harmonic in the mantle flow switches between 
non-axisymmetric $l= 1$ and $l= 2$ 
\citep{zhong2007supercontinent,
	yoshida2008mantle,olson2010geodynamo}, 
where $l$ is spherical harmonic degree. 
During supercontinent aggregation, the non-axisymmetric $l=1$ 
mantle flow tends to dominate, whereas 
$l=2$ becomes predominant after supercontinent breakup. 
Therefore, it makes sense to focus on the effect of
non-axisymmetric degree 1 and 2 heat flux patterns at 
the outer boundary. 
Of these, equatorially anti-symmetric 
conditions are likely to trigger polarity 
transitions \citep[e.g.][]{takahashi2008effects}. 
The equatorially anti-symmetric part 
of the lower-mantle heat flux can become
significant for large lateral 
variations in boundary heat flux 
and thus influence the polarity of
the dynamo.
The present study analyses this problem in the
rapidly rotating, strongly driven regime of a planetary
dynamo. 

 A heterogeneous heat flux pattern at the 
outer boundary induces a steady mean flow, termed the
thermal wind.
 A study of the onset of convection 
in a plane layer in the 
	presence of a thermal wind \citep{teed2010rapidly} 
	suggests that dynamo action might be possible even 
	in subadiabatic conditions.  
	Near convective onset,
	an equatorially symmetric boundary heterogeneity
	in heat flux 
	likely favours dynamo action at large magnetic 
	Prandtl number whereas an equatorially anti-symmetric 
	heterogeneity can induce polarity
	reversals \citep{sahoo2016dynamos}. 
	In a thermally driven dynamo, 
	an equatorially anti-symmetric heat flux
	heterogeneity induces a 
	mean axial temperature gradient
	at the equator, which
	 modifies the buoyancy profile
of the homogeneous state, and hence the buoyancy
frequency. 
For a sufficiently large anti-symmetric heterogeneity, 
the resultant buoyancy frequency $\omega_A$ matches the Alfv\'en wave 
frequency $\omega_M$, causing the suppression of the slow 
Magnetic--Archimedean--Coriolis (MAC), or magnetostrophic, 
waves in the dynamo. In a recent study \citep{jfm24}, 
variations in the mean 
outer boundary heat flux were interpreted as 
changes in the Rayleigh number, and the condition  
$|\omega_A| \approx |\omega_M|$ was shown to be associated with 
the loss of kinetic helicity from these waves, in turn leading to 
the collapse of the axial dipole magnetic field. 

%
In an unstably stratified fluid, 
convection occurs when a superadiabatic gradient is present
in the basic state. 
Within this medium, isolated density perturbations give rise 
to fast and slow MAC waves whose frequencies are obtainable
 in the Boussinesq limit \citep{brag1967,
07bussechapter,jfm24}. In a rapidly rotating fluid
where the magnitude of the inertial wave frequency $|\omega_C|
\gg |\omega_M|, \, |\omega_A|$, the 
real part of the slow wave frequency is
approximated by \citep[e.g.][]{jfm24},
\begin{equation}
\omega_s \approx \dfrac{\omega_M^2}{\omega_C} \left(1+ \dfrac{\omega_A^2}{\omega_M^2} \right)^{1/2}.
\label{wsapprox}
\end{equation}
The spontaneous generation of slow MAC waves in a convection-driven
dynamo that evolves from
a small seed magnetic field 
(figure S1, Supplementary material) indicates that these
waves exist in an unstably stratified 
fluid where $\omega_A^2 <0$, $\omega_M^2 >0$
and $|\omega_M| \ge |\omega_A|$ in \eqref{wsapprox}. 
As the dynamo passes through
a chaotic multipolar state, 
the slow waves are excited by localized balances 
between the Lorentz, buoyancy and Coriolis forces, 
where the Lorentz force is made up
of the non-dipolar field (figure S2, Supplementary Material). 
The helicity of the slow
waves, which is at least as high as that of the fast waves, 
is essential for the
formation of the dipole field from this
 multipolar state \citep{aditya2022}. The hydrodynamic
dynamo, where the Lorentz force is absent, 
does not produce the dipole from a seed field
\citep{prf18} since the slow MAC waves 
are not generated in the first place. 
As the buoyant forcing is progressively increased
in the 
dipolar regime, $|\omega_M|$
attains its highest value, upon which a
state where $|\omega_A| \approx |\omega_M|$ ensues. 
Here, \eqref{wsapprox} predicts the disappearance
of the slow MAC waves,
which brings about the collapse of the axial
dipole field \citep{jfm24}.

%
%

The present study builds on previous work, 
with a focus on examining how heterogeneity in the 
outer boundary heat flux causes the dipole--multipole transition.
 Since polarity transitions occur when 
$|\omega_A| \approx |\omega_M|$, and the 
resultant buoyancy frequency $\omega_A$ 
is made up of both vertical (radial) and 
horizontal (lateral) parts, it is 
in principle possible for a weak 
vertical buoyancy to be compensated 
by a strong horizontal buoyancy 
in order to match $|\omega_M|$. 
This complementarity between vertical 
and horizontal buoyancy suggests 
that polarity transitions can arise 
even under weakly driven thermal convection, 
provided there is a sufficiently 
large heat flux heterogeneity.
For a fixed
vertical buoyancy, polarity reversals
may exist in a narrow range of horizontal buoyancy states
that lie between the dipolar and multipolar regimes.
As in dynamos with uniform boundary heat flux \citep{jfm24},
we anticipate that the
dipole--multipole transition would be
self-similar, in that the resultant Rayleigh
number based on the characteristic energy-containing
length scale bears the same linear relationship with
the square of the peak magnetic field at the transition,
regardless of the scale of energy injection.

	The formation and growth of the inner core make the primary source of buoyancy for
outer core convection compositional, which is thought to contribute up to 80\% of the total 
convective power \citep{lister1995strength}. 
The weaker source of buoyancy, derived from thermal convection, 
is influenced by heat flux heterogeneity at the CMB. 
 In two-component convection, 
	the vertical buoyancies of composition 
	and temperature, together with the horizontal 
	buoyancy from the lateral heat flux variation, 
	make up the resultant buoyant forcing.
Using this complementarity of buoyancies and the
condition for vanishing slow MAC waves, an upper
bound for the lower-mantle heterogeneity that 
would admit an axial dipole field may be obtained.

In \S\ref{linear}, we consider the evolution
of an isolated density disturbance 
in an unstably stratified fluid subject to a 
uniform magnetic field, background rotation and a lateral
temperature variation. Here, the gravity and magnetic
field axes coincide in Cartesian geometry, 
which is referred to as the \lq equatorial
radial configuration' by \cite{loper2003buoyancy}.
This analysis quantifies the resultant buoyancy
based on the vertical and horizontal buoyancies
and shows their complementary role in the
evolution of fast and slow MAC waves.
The Cartesian 
magnetoconvection model with non-zero mean
axial temperature gradient
serves as the basis for 
the study on the effect of equatorially anti-symmetric,
 or symmetry-breaking, 
 outer boundary heat flux on wave motions in 
rapidly rotating spherical dynamos, described in 
\S\ref{nonlinear}. In a dynamo
where the nonlinear inertial forces are small,
an equatorially anti-symmetric
condition causes polarity transitions
 not by breaking the symmetry of the
columnar vortices but rather by selectively
suppressing the slow MAC waves that already exist
in the dipolar dynamo. This study also considers
composite boundary heat 
flux patterns obtained by adding equatorially
symmetric and anti-symmetric patterns in a known 
proportion in order to extend the
analysis to lower-mantle heat flux anomalies.
 \S\ref{twocomp} analyses two-component 
 linear magnetoconvection and suggests an upper bound
 for the horizontal buoyancy that would admit a
 dipolar dynamo for a given thermal power ratio.
In \S\ref{concl}, we discuss the 
implications of our results for Earth 
and propose future work.
\section{Evolution of a density disturbance under rapid rotation,
lateral temperature variation and a magnetic field}
\label{linear}
	
\subsection{Problem set-up and governing equations }
\label{setup}
\begin{figure}
\centering
\hspace{1in} 	\includegraphics[width=0.5\linewidth]
{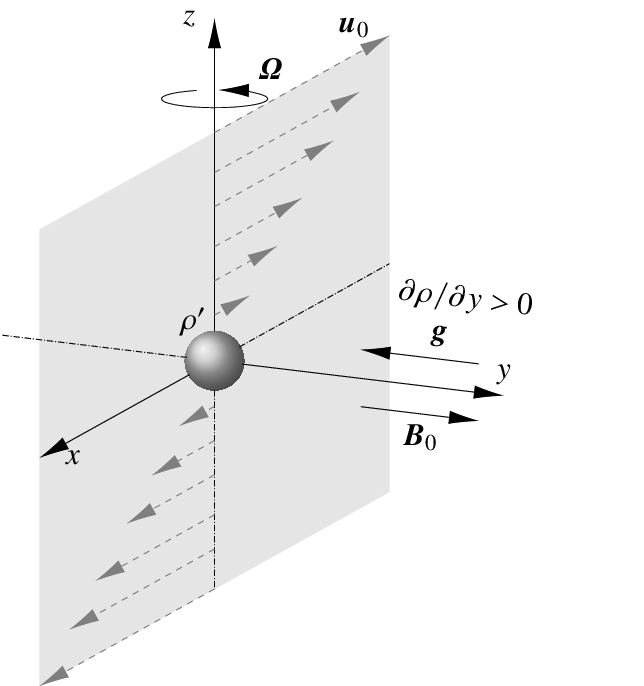}
\caption{The initial state of a density perturbation
	$\rho^{\prime}$ that evolves in an unstably 
	stratified
	fluid subject to a uniform magnetic
	field $\bm{B}_0 = B_0 \hat{\bm{e}}_y$, background rotation 
	$\bm{\varOmega} = \varOmega \hat{\bm{e}}_z$, and gravity 
	$\bm{g} = -g \hat{\bm{e}}_y$ in
	Cartesian coordinates $(x,y,z)$. The lateral
	variation in temperature produces a mean flow 
	$\bm{u}_0=u_c z \,\hat{\bm{e}}_x$.}
\label{coordinate}
\end{figure}
	
A localized density perturbation $\rho^\prime$ is 
situated in an unstably stratified fluid subject
to a uniform magnetic field and rapid rotation.
Since $\rho^\prime$ is related to a temperature perturbation
$\Theta$ by $\rho^\prime = -\rho \alpha \Theta$, 
where $\rho$ is the ambient density and $\alpha$ is 
the coefficient of thermal
expansion, an initial temperature perturbation is 
chosen in the form
\begin{equation}
\Theta_0=C \, \exp \left[-(x^2+y^2+z^2)/\delta^2 \right],
	\label{pert}
\end{equation}
where $C$ is a constant and $\delta$ is the length scale 
of the perturbation.
Figure \ref{coordinate} shows the initial perturbation, 
which evolves under the influence of 
gravity $\bm{g} = -g \hat{\bm{e}}_y$, 
background rotation $\bm{\varOmega} = \varOmega \hat{\bm{e}}_z$, 
and a uniform magnetic field $\bm{B}_0 = B_0 \hat{\bm{e}}_y$ 
in Cartesian coordinates $(x,y,z)$. 
The basic state temperature $T_0$ is assumed to have
constant variations in  
both the vertical ($y$) and a horizontal ($z$) direction
 \citep{teed2010rapidly}, which results in 
a steady mean flow $\bm{u}_0$ satisfying the 
thermal wind equation
\begin{equation}\label{twind}
2\varOmega\frac{\partial \bm{u}_0}{\partial z}=
-g \alpha \hat{\bm{e}}_y \times \nabla T_0.
\end{equation}
Here, the mean flow corresponds to a constant shear,
\begin{equation}\label{u0}
\bm{u}_0=u_c z\, \hat{\bm{e}}_x,
\end{equation}
which on substitution in \eqref{twind} in scalar form,
\begin{equation}\label{twind2}
\begin{aligned}
2 \varOmega \,u_c =
- g \alpha \dfrac{\partial T_0}{\partial z},
\end{aligned}
\end{equation}
gives the basic state temperature,
\begin{equation}\label{t0}
T_0 =\beta_y y+\beta_z z,
\end{equation}
where $\beta_y = \partial T_0/\partial y <0$ is the mean vertical
temperature gradient and  $\beta_z = - 2\varOmega u_c/g \alpha >0$ is
the mean horizontal (lateral) temperature gradient.

The initial temperature 
perturbation \eqref{pert} gives rise to a velocity field $\bm{u}$, 
which interacts with the ambient magnetic field
$\bm{B}_0$ to produce the induced magnetic 
field $\bm{b}$. The initial velocity perturbation 
and induced field are both zero. Since all the variables
are decomposed into their mean and perturbation parts, e.g.
\begin{subequations}\label{vars}
\begin{gather}
T= T_0 + \Theta, \quad \bm{U}= \bm{u}_0 + \bm{u}, 
\quad \bm{B}= \bm{B}_0 + \bm{b},
\tag{\theequation a--c}
\end{gather}
\end{subequations}
the following linearized equations 
give the evolution of $\bm{u}$, $\bm{b}$ and $\Theta$:
\begin{eqnarray}
&& \frac{\partial \bm{u}}{\partial t}
+u_c\, z \frac{\partial \bm{u}} {\partial x}+u_c u_z \hat{\bm{e}}_x
+2\varOmega\hat{\bm{e}}_z\times\bm{u}=
-\frac{1}{\rho}\nabla p^\star+ \frac{1}{\rho}
\bm{j} \times \bm{B}_0
+g\alpha \Theta \hat{\bm{e}}_y + \nu\nabla^2 \bm{u},
\label{mom1} \\
&& \frac{\partial \bm{b}}{\partial t} + 
u_c \,z \frac{\partial \bm{b}}{\partial x} = 
u_c b_z\hat{\bm{e}}_x
+(\bm{B}_0\cdot\nabla)\bm{u}+\eta \nabla^2 \bm{b},
\label{ind1} \\
&& \frac{\partial \Theta}{\partial t}
+u_c z \frac{\partial\Theta}{\partial x}
+u_y\beta_y+u_z\beta_z=\kappa \nabla^2 \Theta,
\label{temp1} \\
&& \nabla \cdot \bm{u} =0,
\label{divcond1}\\
&& \nabla \cdot \bm{b}=0,
\label{divcond2}
\end{eqnarray}
where $\bm{j}$ is the electric current density, 
$\nu$ is the kinematic viscosity,
$\kappa$ is the thermal diffusivity, 
$\eta$ is the magnetic diffusivity
and $p^*=p -(\rho/2) |{\bm{\Omega}} \times
{\bm{x}}|^2$ is a modified pressure.
\subsection{Solutions for the perturbation velocity field}
\label{pert1}
Taking the curl of equation \eqref{mom1} and
\eqref{ind1} and eliminating $\bm{j}$, we obtain
\begin{equation}\label{a3}
\begin{aligned}
P_\nu&
P_\eta\bm{\zeta}
=2(\bm{\Omega}\cdot\nabla) P_\eta \bm{u}\\
&-u_c P_\eta\bigg[-\frac{\partial u_y}{\partial x}
\hat{\bm{e}}_x+\bigg(\frac{\partial u_x}{\partial x}
+\frac{\partial u_z}{\partial z}\bigg)\hat{\bm{e}}_y
-\frac{\partial u_z}{\partial y}\hat{\bm{e}}_z\bigg]
+g \alpha P_\eta\bigg(-\frac{\partial \Theta}{\partial z}\hat{\bm{e}}_x
+\frac{\partial \Theta}{\partial x}\hat{\bm{e}}_z\bigg)\\
&+u_c\frac{(\bm{B}_0 \cdot\nabla)}{\rho \mu} 
\bigg[\frac{\partial b_y}{\partial x}\hat{\bm{e}}_x
+\bigg(-\frac{\partial b_x}{\partial x}
+\frac{\partial b_z}{\partial z}\bigg)\hat{\bm{e}}_y
-\frac{\partial b_z}{\partial y}\hat{\bm{e}}_z\bigg]
+\frac{(\bm{B}_0 \cdot \nabla)^2}{\rho \mu}\bm{\zeta}.
\end{aligned}
\end{equation}
where $P_\nu=\frac{\partial}{\partial t}
+u_c z \frac{\partial}{\partial x}-\nu \nabla^2$,
$P_\eta=\frac{\partial}{\partial t}
+u_c z \frac{\partial}{\partial x}-\eta \nabla^2$,
$\bm{\zeta}=\nabla\times\bm{u}$ is the vorticity
and $\mu$ is the magnetic permeability. 
Successive elimination of $\bm{\zeta}$ and $\Theta$ through
the $z$ component of the curl of \eqref{a3}, the
$z$ component of \eqref{ind1} and \eqref{temp1} 
gives
	\begin{equation}\label{a7}
		\begin{aligned}
			&P^2 P_\kappa (-\nabla^2 u_z)=
			P\bigg[-g \alpha P_\eta
			\frac{\partial^2}{\partial y\partial z}(\beta_y u_y
			+\beta_z u_z)+u_c P_\kappa\frac{B_0}{\rho \mu}
			\frac{\partial}{\partial y}
			\bigg(-\frac{\partial^2 b_x}{\partial x^2}
			+\frac{\partial^2 b_z}{\partial x \partial z}
			-\frac{\partial^2 b_y}{\partial x \partial y}\bigg)\bigg]\\
			&\myquad[2]+2\varOmega \frac{\partial}{\partial z} 
			P_\kappa \bigg[2 \varOmega \frac{\partial}{\partial z}
			P_\eta^2u_z+u_c P_\eta^2\frac{\partial u_z}{\partial y}
			-u_c\frac{B_0^2}{\rho \mu}
			\frac{\partial^3 u_z}{\partial y^3}\bigg]
			-2\varOmega g \alpha\frac{\partial^2}{\partial x\partial z} 
			P_\eta^2(\beta_y u_y+\beta_z u_z),
		\end{aligned}
	\end{equation}
where $P= P_\nu P_\eta-B_0^2/\rho \mu$ and
$P_\kappa=\frac{\partial}{\partial t}
+u_c z \frac{\partial}{\partial x}-\kappa \nabla^2$.
For zero horizontal variation in temperature, $\beta_z, \, u_c \to 0$,
and \eqref{a7} reduces to the classical MAC wave
equation \citep[pp. 165--168]{brag1967,07bussechapter}.
On the other hand, since $x$ is not a preferred
horizontal direction, setting $\partial(\,\,)/\partial x = 0$
in \eqref{a7} \citep[e.g.][]{hathaway1979convective} gives
\begin{equation}\label{a8}
	\begin{aligned}
		&\bigg[\bigg(\frac{\partial}{\partial t}
		-\nu \nabla_*^2\bigg)\bigg(\frac{\partial}{\partial t}
		-\eta \nabla_*^2\bigg)
		-\frac{B_0^2}{\rho\mu} \, \frac{\partial^2}{\partial y^2} \bigg]^2
		\bigg(\frac{\partial}{\partial t}-\kappa \nabla_*^2\bigg)
		(-\nabla_*^2  \hat{u}_z)\\
		&= -g \alpha \bigg(\frac{\partial}{\partial t}
		-\eta \nabla_*^2\bigg) \bigg[\bigg(\frac{\partial}{\partial t}
		-\nu \nabla_*^2\bigg)\bigg(\frac{\partial}{\partial t}
		-\eta \nabla_*^2\bigg)
		-\frac{B_0^2}{\rho\mu}\, \frac{\partial^2}{\partial y^2} \bigg]
		\, \frac{\partial^2}{\partial y \partial z} 
		(\beta_y \hat{u}_y+\beta_z \hat{u}_z)\\
		&+2 \varOmega \, \frac{\partial}{\partial z}
		\bigg(\frac{\partial}{\partial t}-\kappa \nabla_*^2\bigg)
		\bigg[2\varOmega \frac{\partial}{\partial z}\bigg(\frac{\partial}{\partial t}
		-\eta \nabla_*^2\bigg)^2+u_c \frac{\partial}{\partial y}
		\bigg(\frac{\partial}{\partial t}
		-\eta \nabla_*^2\bigg)^2 -u_c 
		\frac{B_0^2}{\rho\mu} \frac{\partial^3}{\partial y^3}\bigg]
		\hat{u}_z,
	\end{aligned}
\end{equation}
where $\nabla_*^2=\partial^2/\partial y^2+\partial^2/\partial z^2$.

Now, applying the two dimensional Fourier transform defined by
\begin{equation}
	\mathcal{F}\big(\mathbf{A}\big) = \hat{\mathbf{A}}=  
	\int_{-\infty}^{\infty}\int_{-\infty}^{\infty}\mathbf{A}
	\mbox{e}^{- \mathrm{i} \, (k_y \, y+k_z \, z)}\,\mathrm{d}y\,
	\mathrm{d}z,
	\label{a9}
\end{equation}
to \eqref{a8}, the following homogeneous equation is obtained:
\begin{equation}\label{a11}
\begin{aligned}
\bigg[\bigg(\frac{\partial}{\partial t}&
+\nu k^2\bigg)\bigg(\frac{\partial}{\partial t}
+\eta k^2\bigg)+\frac{B_0^2\, k_y^2}{\rho \mu}\bigg]^2
\bigg(\frac{\partial}{\partial t}+\kappa k^2\bigg)k^2
\, \hat{u}_z\\
=& - g \alpha k_z^2 \bigg(\frac{\partial}{\partial t}
+\eta k^2\bigg) \bigg[\bigg(\frac{\partial}{\partial t}
+\nu k^2\bigg)\bigg(\frac{\partial}{\partial t}+\eta k^2\bigg)
+\frac{B_0^2\, k_y^2}{\rho \mu}\bigg] \left(\beta_y 
 - \beta_z \frac{k_y}{k_z} \right)\, \hat{u}_z\\
&+2 \varOmega \, \mathrm{i}\, k_z \bigg(\frac{\partial}{\partial t}
+\kappa k^2\bigg)\bigg[2 \varOmega \, \mathrm{i}\, k_z
\bigg(\frac{\partial}{\partial t}+\eta k^2\bigg)^2+u_c\mathrm{i} 
k_y\bigg(\frac{\partial}{\partial t}+\eta k^2\bigg)^2 
+u_c\mathrm{i} k_y\frac{B_0^2\, k_y^2}{\rho \mu} \bigg]\, \hat{u}_z,
\end{aligned}
\end{equation}
where $k^2=k_y^2+k_z^2$ and the following substitution
is made from the continuity equation \eqref{divcond1}:

\begin{equation}
	\hat{u}_y=-\frac{k_z}{k_y}\hat{u}_z.
	\label{uy}
\end{equation}
Seeking a plane wave solution  of the form
$\hat{u}_z \sim \mbox{e}^{\mathrm{i} \lambda t}$, we obtain,
\begin{equation}\label{a13}
	\begin{aligned}
		\big[\big(\mathrm{i}\lambda
		+\omega_\nu\big)&\big(\mathrm{i}\lambda
		+\omega_{\eta}\big)+ \omega_{M}^2\big]^2
		\big(\mathrm{i}\lambda+\omega_\kappa\big)=-\omega_{A, V}^2\big(\mathrm{i}\lambda
		+\omega_{\eta}\big)\big[\big(\mathrm{i}\lambda
		+\omega_\nu\big)\big(\mathrm{i}\lambda
		+\omega_{\eta}\big)+ \omega_{M}^2\big]\\
		&+\omega_{A, H}^2\big(\mathrm{i}\lambda
		+\omega_{\eta}\big)\big[\big(\mathrm{i}\lambda
		+\omega_\nu\big)\big(\mathrm{i}\lambda
		+\omega_{\eta}\big)+\omega_{M}^2\big]-\omega_{C}^2
		\big(\mathrm{i}\lambda+\omega_\kappa\big)\big(\mathrm{i}\lambda
		+\omega_{\eta}\big)^2\\
		&+\omega_{A, H}^2\big(\mathrm{i}\lambda
		+\omega_{\kappa}\big)\big[\big(\mathrm{i}\lambda
		+\omega_\eta\big)^2+ \omega_{M}^2\big],
	\end{aligned}
\end{equation}
where
\begin{subequations}
\begin{gather}\label{freq}
\begin{aligned}
	&\omega_C^2=4 \varOmega^2 k_z^2/k^2, \quad
	\omega_{A,V}^2=g\alpha\beta_y k_z^2/k^2, 
	\quad \omega_{A,H}^2=g\alpha\beta_z k_y k_z/k^2, \\
	&\omega_M^2= B_0^2\, k_y^2/ \rho \mu = V_M^2 k_y^2,
	\quad \omega_{\eta}^2=\eta^2 k^4, \quad \omega_\kappa^2 =\kappa^2 k^4, 
	\quad \omega_\nu^2=\nu^2 k^4,
\end{aligned}
\tag{\theequation a--g}
\end{gather}
\end{subequations}
represent the squares of the frequencies of linear inertial waves, 
vertical and horizontal parts of the buoyancy, 
Alfv\'en waves, 
magnetic diffusion, thermal diffusion and 
viscous diffusion, respectively. Here,
$V_M= B_0/\sqrt{\rho \mu}$ is the Alfv\'en wave
velocity. 
In line with a recent study \citep{jfm21}
where both viscous and thermal diffusion are much 
smaller than magnetic diffusion,
the characteristic equation has the following
form in
the limit of $\omega_\kappa,\omega_\nu \to 0$:
\begin{equation}\label{characteristic}
\begin{aligned}
\lambda^5- 2 \mathrm{i}\omega_\eta \lambda^4&
- (\omega_C^2+\omega_\eta^2+2\omega_M^2
+\omega_{A,V}^2-2\omega_{A,H}^2) \lambda^3
+2 \mathrm{i} \omega_\eta(\omega_C^2+\omega_M^2
+\omega_{A,V}^2-2\omega_{A,H}^2) \lambda^2\\&
+(\omega_C^2\omega_\eta^2+\omega_M^4+(\omega_M^2
+\omega_\eta^2)(\omega_{A,V}^2-2\omega_{A,H}^2)) 
\lambda-\mathrm{i}\omega_\eta\omega_M^2(\omega_{A,V}^2
-\omega_{A,H}^2)=0.
\end{aligned}
\end{equation}	
The general solution for $\hat{u}_z$ is
given by
\begin{equation}
\hat{u}_z = \sum_{m=1}^{5}D_{m} \mbox{e}^{\mathrm{i} \lambda_{m}t},
\label{solutionyhat}
\end{equation}
where the coefficients $D_m$ are evaluated
from the initial 
conditions of $\hat{u}_z$ and its
 derivatives ($\S$\ref{init} below). 
Of the five terms in the expansion on the right-hand 
side of equation \eqref{solutionyhat}, two terms 
represent oppositely travelling fast MAC waves, 
two other terms represent oppositely travelling slow MAC waves, 
and the fifth term represents the overall 
growth of the velocity perturbation. 
\subsection{Evaluation of spectral coefficients}
\label{init}
From \eqref{solutionyhat}, the initial conditions 
for $\hat{u}_z$ and its time derivatives are given by
\begin{equation}
\begin{aligned}
\mathrm{i}^n \sum_{m=1}^{5}D_m \lambda_{m}^n
=\bigg(\frac{\partial^n\hat{u}_z}{\partial t^n}\bigg)_{t=0}
=a_{n+1}, \ n=0,1,2,3,4.
\end{aligned}
\label{axzn}
\end{equation}

Algebraic simplifications give the right-hand sides of 
\eqref{axzn} as follows:

\begin{equation}
\begin{aligned}
a_1&=\hat{u}_z|_{t=0}=0,\\
a_2&=\frac{\partial{\hat{u}_z}}{\partial{t}}|_{t=0}=
-g \alpha \frac{k_y k_z}{k^2}\hat{\Theta}_0,\\
a_3&=\frac{\partial^2 \hat{u}_z}{\partial{t^2}}|_{t=0}=0,\\
a_4&=\frac{\partial^3{\hat{u}_z}}{\partial{t^3}}|_{t=0}=
-(\omega_M^2+\omega_C^2+\omega_{A,V}^2-2\omega_{A,H}^2)~a_2,\\
a_5&=\frac{\partial^4{\hat{u}_z}}{\partial{t^4}}|_{t=0}=
\omega_M^2\omega_\eta ~a_2.
\label{initial_an}
\end{aligned}
\end{equation}

The coefficients $D_m$ may now be 
obtained using the roots of equation 
\eqref{characteristic}.
For example, we obtain,
\begin{eqnarray}
D_{1}&=\dfrac{a_{5}- \mathrm{i} \, a_{4}(\lambda_2
+\lambda_3+\lambda_4+\lambda_5)
+\mathrm{i} \, a_{2}(\lambda_2\lambda_4\lambda_5
+\lambda_3\lambda_4\lambda_5
+\lambda_2\lambda_3\lambda_4
+\lambda_2\lambda_3\lambda_5)}{(\lambda_1-\lambda_2)
(\lambda_1-\lambda_3)(\lambda_1-\lambda_4)(\lambda_1-\lambda_5)},
\label{d1coeff}\\
D_{3}&=\dfrac{a_{5}- \mathrm{i} \, a_{4}
(\lambda_1+\lambda_2+\lambda_4+\lambda_5)
+ \mathrm{i} \,a_{2}(\lambda_1\lambda_4\lambda_5
+\lambda_2\lambda_4\lambda_5
+\lambda_1\lambda_2\lambda_4
+\lambda_1\lambda_2\lambda_5)}{(\lambda_3-\lambda_1)
(\lambda_3-\lambda_2)(\lambda_3
-\lambda_4)(\lambda_3-\lambda_5)},
\label{d3coeff}
\end{eqnarray}
for the forward-travelling fast and slow wave 
solutions respectively.
We separate the fast and slow MAC wave parts of the 
general solution, which is a linear superposition of
the two wave solutions \citep{jfm21}. 
For example,
\begin{equation}
\begin{aligned}
\hat{u}_{z,f}&=D_{1} \mathrm{e}^{ \mathrm{i} {\lambda}_1 t}
+D_{2} \mathrm{e}^{\mathrm{i}{\lambda}_2 t},\\
\hat{u}_{z,s}&=D_{3} \mathrm{e}^{\mathrm{i}{\lambda}_3 t}
+D_{4}\mathrm{e}^{\mathrm{i} {\lambda}_4 t},
\label{separate}
\end{aligned}
\end{equation}
where the subscripts $f$ and $s$ in the left-hand sides
of \eqref{separate} denote the fast and slow
wave parts of the solution.
\subsection{Complementarity of vertical and horizontal buoyancies}
\label{compl1}
For the rapidly rotating
regime given by $|\omega_C| \gg |\omega_M|
\gg |\omega_{A,V}|,|\omega_{A,H}| \gg |\omega_\eta|$, 
the roots of
the characteristic equation \eqref{characteristic} are approximated by
\begin{eqnarray}
\lambda_{1,2} &\approx& \pm \bigg(\omega_C
+ \frac{\omega_M^2}{\omega_C} \bigg) 
+\mathrm{i} \, \frac{\omega_M^2\omega_\eta}{\omega_C^2}, 
\label{l12approx}\\ 
\lambda_{3,4} &\approx& \pm \bigg(\frac{\omega_M^2}{\omega_C}+
\frac{\omega_{A,V}^2-2\omega_{A,H}^2}{2\omega_C}\bigg) 
+ \mathrm{i} \, \omega_\eta \,
\bigg(1-\frac{\omega_{A,V}^2-\omega_{A,H}^2}{2\omega_M^2}\bigg),
\label{l34approx} \\
\lambda_{5} &\approx& \mathrm{i}\, 
\frac{\omega_\eta(\omega_{A,V}^2-\omega_{A,H}^2)}{\omega_M^2},
\label{l5approx}
\end{eqnarray}
following the procedure in \cite{jfm21}.  Here, the slow
MAC waves of frequency $\lambda_{3,4}$ are damped on the
time scale $(\omega_\eta)_0^{-1} = 
(\eta k_0^2)^{-1} \sim \delta^2/\eta$, where the subscript \lq 0'
represents the initial state of the buoyancy disturbance
\citep[see also][]{jfm21}.

From \eqref{l34approx}, the resultant buoyancy frequency is  given by
\begin{equation}\label{wastar}
\begin{aligned}
{\omega_A}^2=&\omega_{A,V}^2-2\omega_{A,H}^2
=\omega_{A,V}^2\bigg(1 - 2\,\beta^\star \frac{k_y}{k_z}\bigg),
\end{aligned}
\end{equation}
where $\beta^\star=\beta_z/\beta_y$. If $\beta^\star=0$, 
the approximate classical MAC wave roots are recovered. 
In the limit of $|\omega_C| \gg |\omega_M|$, 
$|\omega_A|$, the real parts 
of $\lambda_{3,4}$ may be expressed as \citep{brag1967},
\begin{equation}\label{eq:slow_root}
\begin{aligned}
\mbox{Re}(\lambda_{3,4}) \approx \pm 
\frac{\omega_M^2}{\omega_C}\,
\bigg(1+ \frac{{\omega_A}^2}{\omega_M^2} \bigg)^{1/2},
\end{aligned}
\end{equation}
which implies that slow MAC waves would vanish
 in an unstably stratified fluid with 
${\omega_A}^2<0$   as $|\omega_A|$ nears $|\omega_M|$. 
Since $\beta_y < 0$ 
and $\beta_z > 0$ in equation~\eqref{t0},
$\omega_{A,V}^2 < 0$ and $\omega_{A,H}^2 > 0$ 
in equation~\eqref{wastar}. These two buoyancy 
frequencies contribute to the 
resultant buoyancy frequency. 
An increase in either $|\omega_{A,H}|$ or 
$|\omega_{A,V}|$ results in an increase in
$|\omega_A|$, which may 
match $|\omega_M|$ and thereby 
suppress the slow MAC waves.
While a large vertical buoyancy
requires a small horizontal buoyancy to suppress these waves,
a small vertical buoyancy together with a large horizontal
buoyancy would produce the same result. 
As we shall see later
in \S \ref{twocomp}, the complementarity of the
buoyancy frequencies places
a bound on the magnitude of the horizontal buoyancy in
two-component convection.

\subsection{Fast and slow MAC waves under 
unstable stratification and a thermal wind}
\label{linmac}
The evolution of velocity and the induced magnetic field 
is obtained from the solution of the initial value problem. 
The solution to the problem is obtained for times much shorter 
than the time scale for the exponential 
increase of the perturbations. 
The Lehnert number $Le$ and the 
magnetic Ekman number $E_\eta$
based on the length scale of the perturbation 
are used to describe the parameter regime,
\begin{subequations}
\begin{gather}\label{ekmanle}
Le= \dfrac{V_M}{2 \varOmega \delta} \approx
\bigg(\frac{\omega_{M}}{\omega_{C}}\bigg)_{\!0}, \quad 
E_\eta= \dfrac{\eta}{2 \varOmega\delta^2}
\approx\bigg(\frac{\omega_{\eta}}{\omega_{C}}\bigg)_{\!0},
\tag{\theequation a,b}
\end{gather}
\end{subequations}
The relative intensity of
horizontal buoyancy is measured by the ratio $Ra_{\ell,H}/Ra_\ell$,
where $Ra_\ell$ is the resultant local Rayleigh number given by
the sum of the local vertical and horizontal Rayleigh numbers,
\begin{subequations}
\label{radelta}
\begin{gather}
Ra_{\ell,V} = 
\dfrac{g \alpha |\beta_y| \delta^2}{2 \varOmega \eta}, 
\quad Ra_{\ell,H} = 
2 \dfrac{g \alpha |\beta_z| \delta^2}{2 \varOmega \eta}
\frac{k_y}{k_z},
\tag{\theequation a,b}
\end{gather}
\end{subequations}
so that
\begin{equation}
Ra_{\ell} = 
\dfrac{g \alpha |\beta| \delta^2}{2 \varOmega \eta},
\label{raell}
\end{equation}
based on the resultant temperature gradient,
\begin{equation}
\beta = \beta_y - 2 \beta_z k_y/k_z.
\label{rbeta}
\end{equation}
\begin{figure}
\centering
\hspace{-1.1 in}	(a)  \hspace{1.1 in} (b) \hspace{1.1 in} (c) \\
\includegraphics[width=0.3\linewidth]{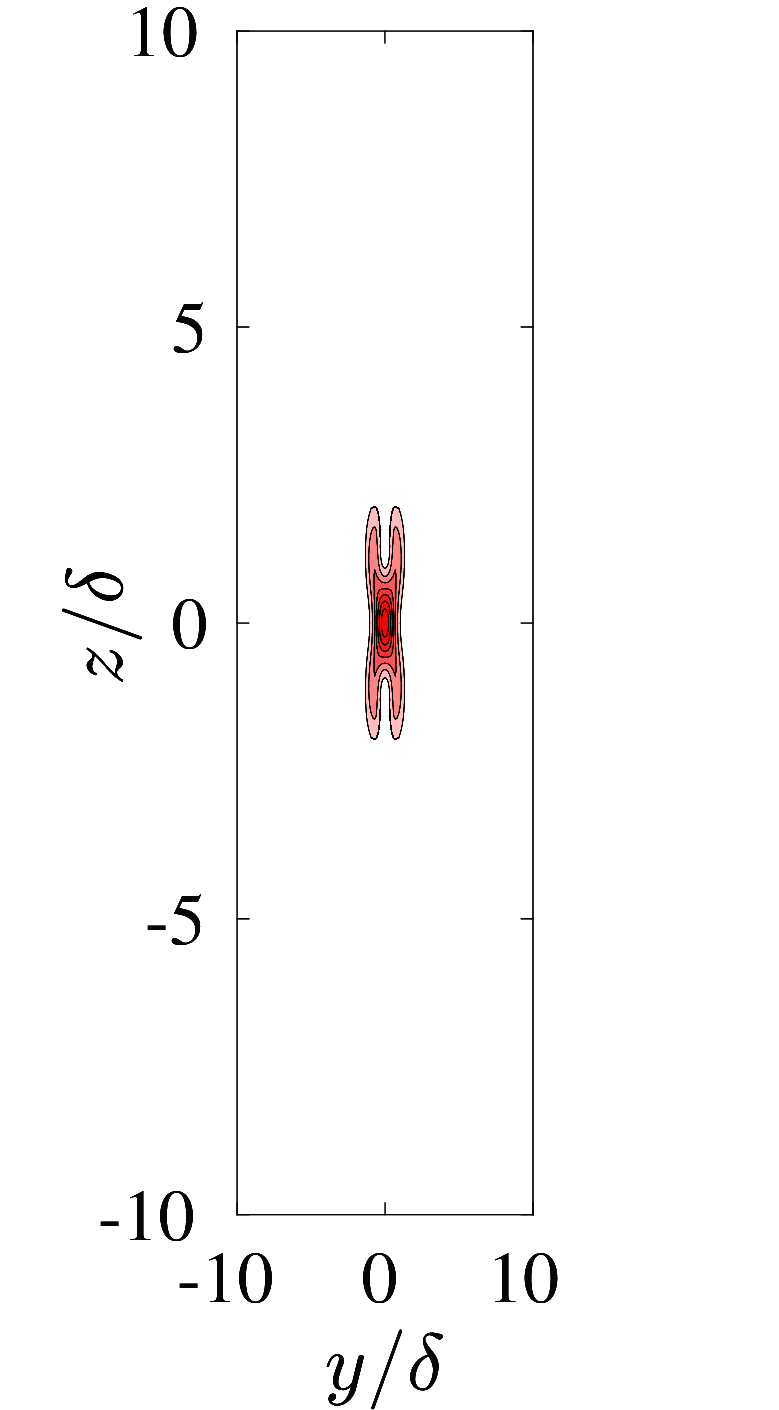}
\hspace{-1.1cm}
\includegraphics[width=0.3\linewidth]{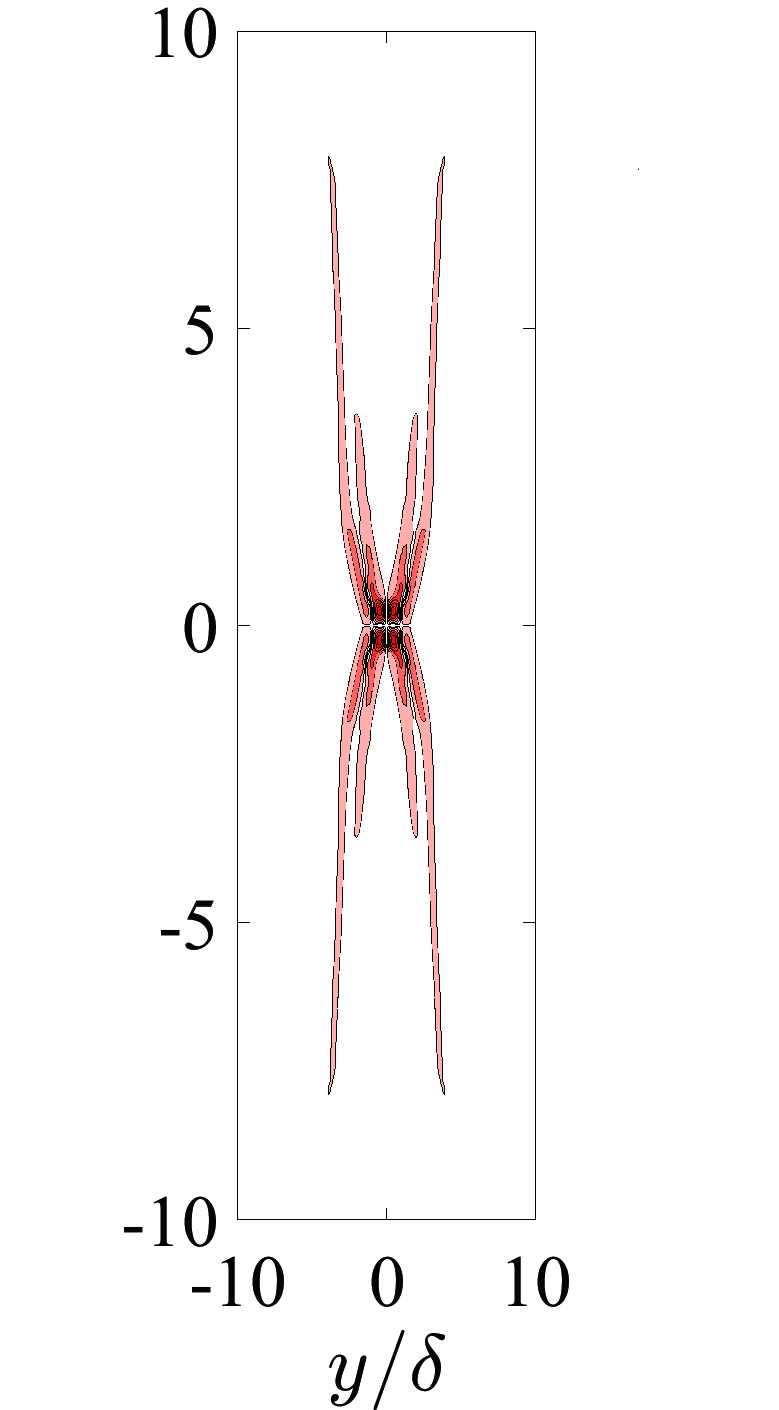}
\hspace{-1.1cm}
\includegraphics[width=0.3\linewidth]{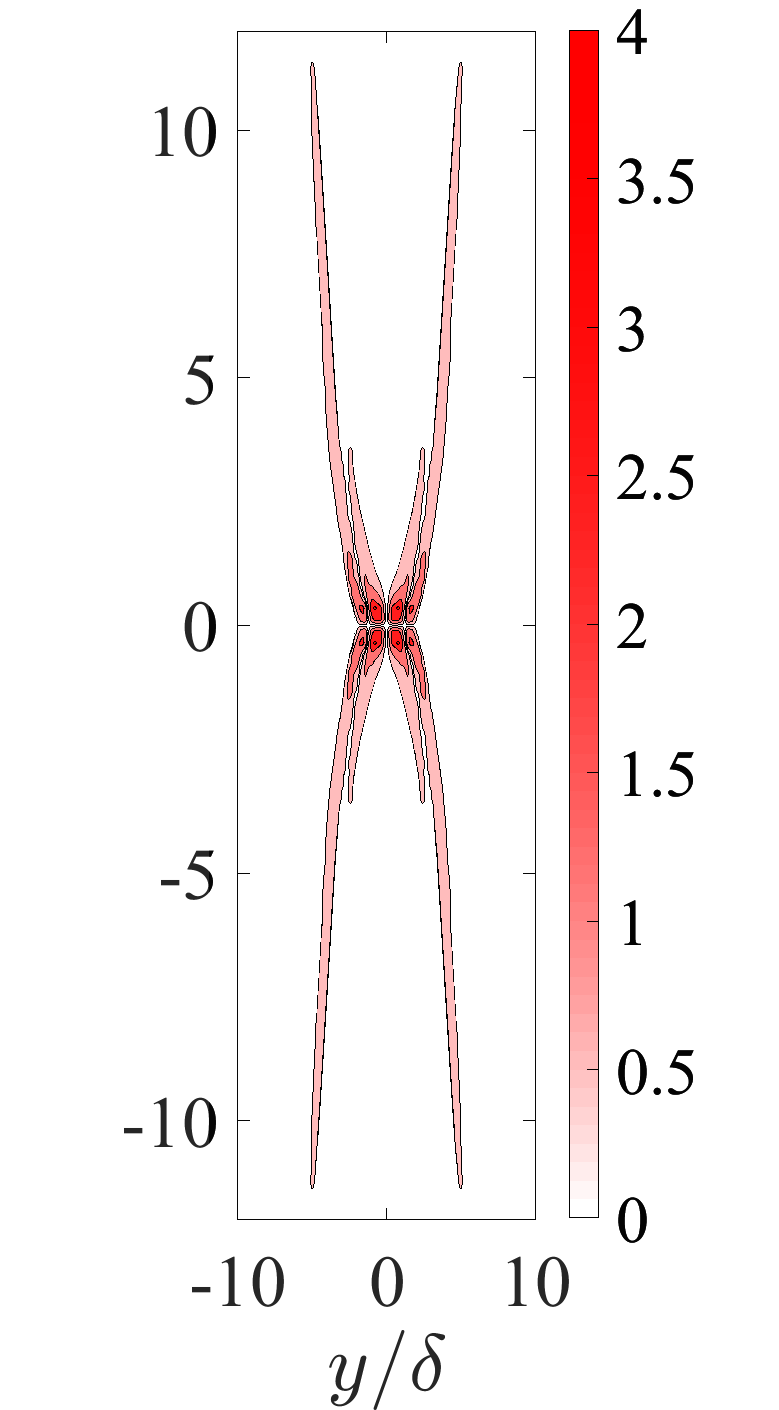}\\
\caption{Evolution of $u_z^2$ on the $y$-$z$ plane 
	at $x=0$ with time (measured in units of the 
	magnetic diffusion time $t_\eta$)
	for $Le=0.03$ and $E_\eta=2\times 10^{-5}$. 
	The snapshots are at 
	(a) $t/t_\eta =1\times10^{-4}$, 
	(b) $t/t_\eta=1\times10^{-3}$
	and (c) $t/t_\eta = 1\times10^{-2}$ 
	for $Ra_{\ell, {H}}/Ra_\ell=0.4$ 
	and $|\omega_{A}^2/\omega_M^2|=0.1$.}
\label{evolutionuz}
\end{figure}
Figure \ref{evolutionuz} shows the evolution of 
the axial kinetic energy density
 of the perturbation. Here,
the real-space $z$ velocity is obtained from
the inverse Fourier transform of \eqref{solutionyhat},
\begin{equation}
u_z =\frac{1}{(2\pi)^2}
\int_{-\infty}^{\infty}\int_{-\infty}^{\infty}\hat{u}_z
\mbox{e}^{ \mathrm{i} \, (k_y \, y+k_z \, z)}\,\mathrm{d}k_y\,
\mathrm{d}k_z,
\label{invfourier}
\end{equation}
where a truncation value of $\pm 5/\delta$ is used for 
the two wavenumbers in the computed integrals since the initial 
wavenumber $k_0 =\sqrt{3}/\delta$ \citep{jfm24}.
The evolution of blobs into columnar structures 
through the propagation of damped waves 
is evident in  the  $y-z$ plane at $x=0$.
\begin{figure}
	\centering  
	\includegraphics[width=1\linewidth]{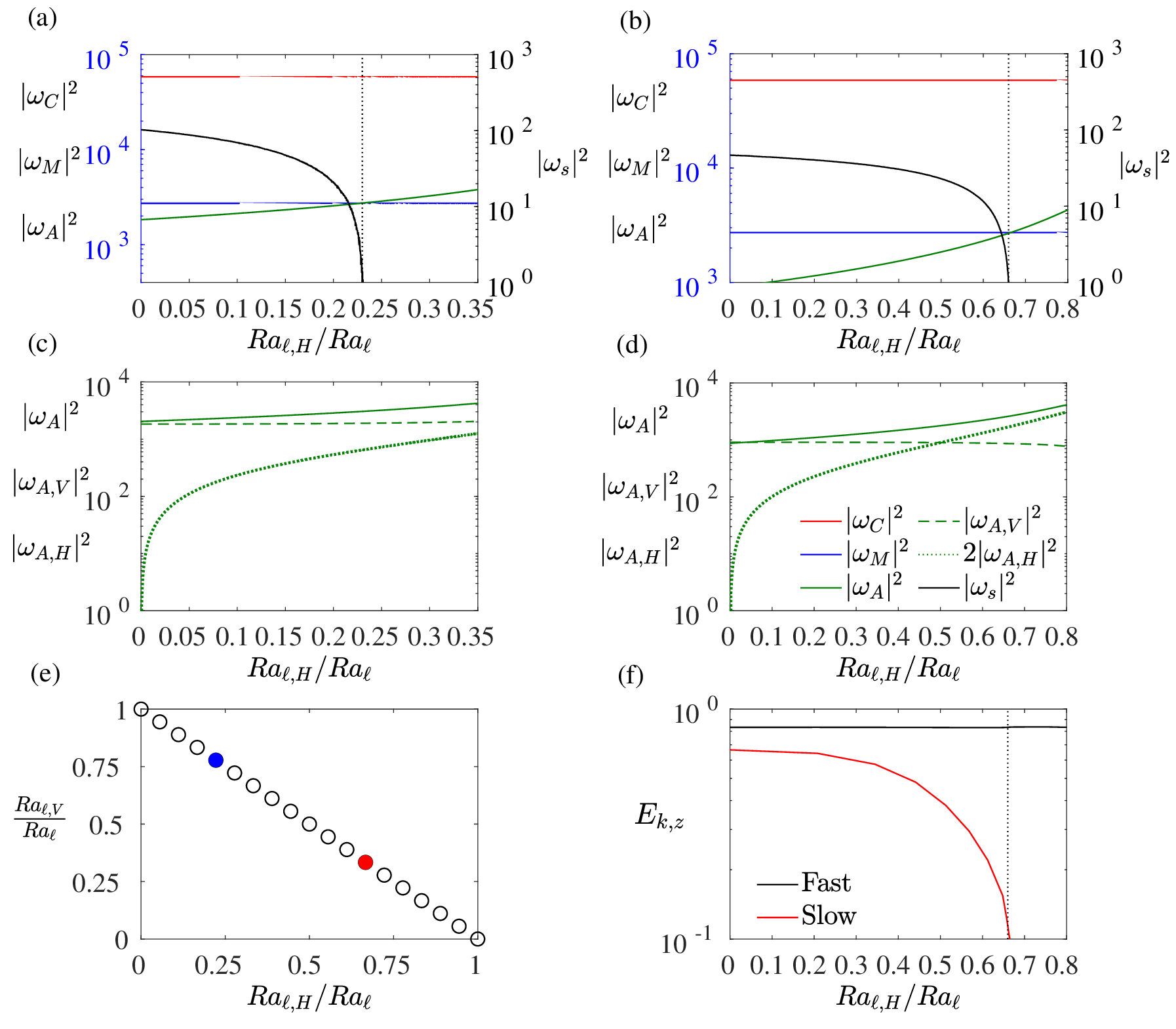}\\
	\caption{Variation of the squares of the fundamental 
		frequencies with $ Ra_{\ell,H}/Ra_\ell $ under 
		constant vertical buoyancy conditions. Panels (a,c) and (b,d) 
		correspond to cases with 
		$ -\omega_{A,V}^2/\omega_{M}^2 = 0.8 $ 
		and $ -\omega_{A,V}^2/\omega_{M}^2 = 0.3 $, respectively. 
		In panels (a) and (b), $\omega_{C}^2$, $\omega_{M}^2$, 
		and $\omega_{A}^2$ are plotted while panels (c) and (d) 
		decompose $\omega_{A}^2$ into its vertical ($\omega_{A,V}^2$) 
		and horizontal ($\omega_{A,H}^2$) parts. 
		The dotted vertical lines indicate the values of 
		$ Ra_{\ell,H}/Ra_\ell $ where the slow MAC wave 
		is suppressed as $ |\omega_M| \approx |\omega_A| $. 
		Panel (e) shows the suppression points of 
		the slow wave for various strengths of 
		$ -\omega_{A,V}^2/\omega_{M}^2 $. 
		The blue and red points in panel (e) correspond 
		to the slow wave 
		suppression points in panels (a) and (b), respectively. 
		(f) The axial kinetic energy $E_{k,z}$ of the fast 
		and slow MAC waves versus $ Ra_{\ell,H}/Ra_\ell $ 
		where $ -\omega_{A,V}^2/\omega_{M}^2 = 0.3 $. 
		The parameters used are $ E_\eta = 2 \times 10^{-5} $, 
		$ Le = 0.03$ ($\omega_{M}/\omega_{C}=0.18$), and $ t/t_\eta = 1 \times 10^{-2} $.}
	\label{freqlinear}
\end{figure}

In figure \ref{freqlinear}(a)--(d), the fundamental 
frequencies are plotted against the relative intensity
of horizontal buoyancy. Both $Ra_{\ell,V}$ and $Le$
are kept constant. 
The frequencies and the local Rayleigh numbers are based on the 
mean wavenumbers obtained from ratios of $L^2$ norms; e.g.
\begin{subequations}\label{l2norm}
\begin{gather}
\bar{k}_y = \frac{|\!|k_y \, \hat{u}_z|\!| }{|\!| \hat{u}_z|\!|},
\quad \bar{k}_z = \frac{|\!|k_z \, \hat{u}_z|\!| }{|\!| \hat{u}_z|\!|},
\quad \bar{k} = \frac{|\!| k \, \hat{u}_z|\!|}{|\!| \hat{u}_z|\!|}.
\tag{\theequation a--c}
\end{gather}
\end{subequations}
By progressively increasing $Ra_{\ell,H}$ for a fixed $Ra_{\ell,V}$
and $Le$, 
$\omega_{A,H}$ increases, and  
the resultant buoyancy frequency 
$\omega_{A}$ crosses the Alfv\'en frequency $\omega_{M}$ at  certain 
$Ra_{\ell,H}/Ra_{\ell}$, where the slow 
MAC waves of frequency $\omega_s$ disappear 
($\omega_s = \omega_{3,4}$; 
	see \eqref{l34approx}). 
The same exercise performed with a different value of 
$Ra_{\ell,V}$ reveals the complementarity 
between $Ra_{\ell,V}/Ra_{\ell}$ and $Ra_{\ell,H}/Ra_{\ell}$ 
(figure \ref{freqlinear}(e)). 
The red and blue points in figure \ref{freqlinear}(e) indicate the 
value of $Ra_{\ell,H}/Ra_{\ell}$ from figure \ref{freqlinear}(a)
and (b), 
where $|\omega_{A}| \approx |\omega_{M}|$ so that $\omega_{s}$ 
approaches zero. In \S \ref{nonlinear}, 
it is shown that the inequality $|\omega_M| > |\omega_A|$ 
represents the dipole-dominated regime while 
$|\omega_M| \approx|\omega_A|$ represents the transition
to a multipolar state, with polarity reversals occurring
in a narrow range of $Ra_{\ell,H}/Ra_{\ell}$ that lies between
the dipolar and multipolar states.

The axial kinetic energy of the MAC waves is given by
\begin{equation}\label{parseval}
E_{k,z}=\frac{1}{(2\pi)^2}\int_{-\infty}^{\infty}
\int_{-\infty}^{\infty}|\hat{u}_z(k_y,k_z)|^2 \mbox{d}k_y \mbox{d}k_z,
\end{equation}
where the limits of the integration are set 
to $\pm 5/\delta$. While the energy of the
fast MAC waves is practically unaffected by increasing horizontal
buoyancy, the slow MAC wave energy goes to zero when 
$|\omega_A|$ approximately matches $|\omega_M|$
(figure \ref{freqlinear}(f)). 
\begin{figure}
\centering
\includegraphics[width=0.45\linewidth]{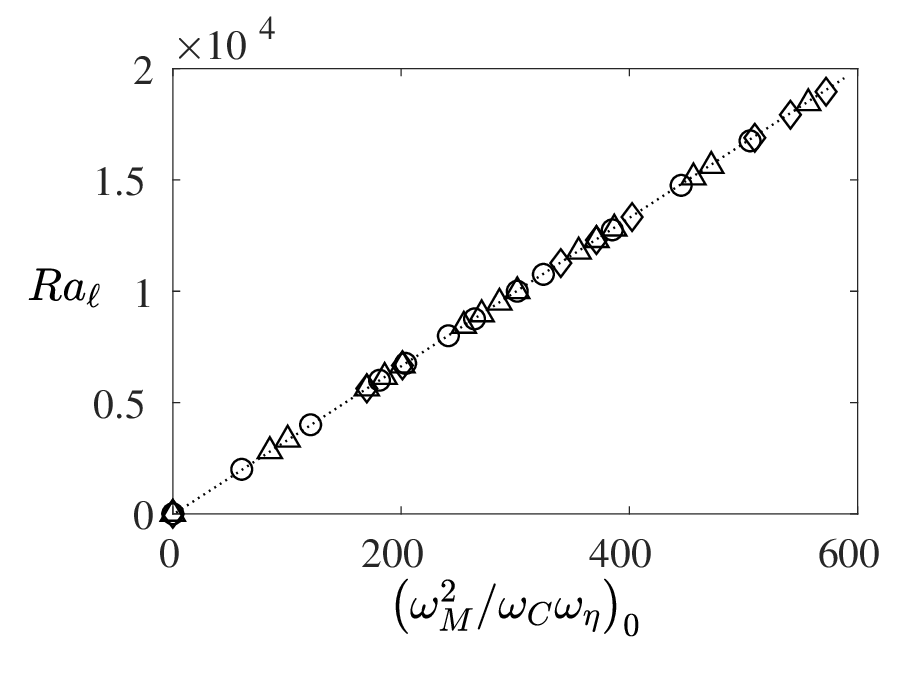}
\caption{Variation of $Ra_{\ell}$ with Elsasser number
	$\varLambda=
	\left(\omega_M^2/\omega_C\omega_\eta\right)_0$ for the 
	state of vanishing slow wave axial kinetic energy. 
	Three values of $E_\eta$ are considered: diamonds 
	represent $E_\eta=1\times 10^{-6}$, circles represent 
	$E_\eta=2\times 10^{-5}$, and triangles represent 
	$E_\eta=5\times 10^{-6}$.
}
\label{selfsimilar}
\end{figure}
The relation between the resultant local
Rayleigh number $Ra_\ell$ and 
the ratio of frequencies given by
\begin{equation}\label{els1}
\bigg(\dfrac{\omega_M^2}{\omega_C \omega_\eta}\bigg)_{\!0}
\sim \dfrac{V_M^2}{2 \varOmega \eta}
\end{equation}
is shown in figure \ref{selfsimilar} 
for the condition of vanishing slow wave axial 
kinetic energy for three 
different values of $E_\eta$. Each point is obtained by 
increasing the forcing through $Ra_\ell$ 
and noting the transition from the regime
$|\omega_{M}|>|\omega_{A}|$ to the regime 
$|\omega_{A}|\gtrsim|\omega_{M}|$.
The resultant forcing needed to
suppress the slow MAC waves increases with $(\omega_M^2/\omega_C \omega_\eta)_0$,
which is a measure of the square of the peak dimensionless magnetic field
in the dynamo (see \S \ref{complement} below).
The linear relation obtained in figure \ref{selfsimilar}
 reflects the parity between
$|\omega_{M}|$ and $|\omega_{A}|$ at the state
of vanishing slow MAC waves. Since the relative
orders of magnitude of the basic frequencies 
do not
depend on the 
orientation of the gravity, rotation and magnetic field axes,
the Cartesian model serves as the basis for the study
of the role of equatorially anti-symmetric
boundary heat flux in the nonlinear dynamo in
\S \ref{nonlinear}, where the
state of vanishing slow waves is shown to be
a proxy for 
polarity transitions. We anticipate a self-similar 
behaviour for the dipole--multipole transition 
in the dynamo, where both $Ra_\ell$ and 
the square of the peak magnetic field are 
measurable quantities.
\section{Nonlinear dynamo simulations}
	\label{nonlinear}
We consider an electrically conducting fluid 
confined between two concentric, co-rotating 
spherical surfaces. 
The ratio of the inner radius to the outer radius is 0.35. 
Fluid motion is driven by thermal 
buoyancy-driven convection. 
Lengths are scaled by the depth of the spherical shell $L$ 
and time is scaled by magnetic diffusion time 
$L^2/\eta$.
The velocity $\bm{u}$ and magnetic field 
$\bm{B}$ are scaled by $\eta/L$ and $(2\Omega\rho\mu\eta)^{1/2}$
respectively. 
The temperature is scaled by $\beta_s L$, where $\beta_s$ is the 
mean equatorial radial
temperature gradient within the shell. In the 
Boussinesq approximation, the nondimensional 
magnetohydrodynamic (MHD) equations for velocity, magnetic field, 
and temperature are as follows:
	
\begin{align}
E Pm^{-1}  \Bigl(\frac{\partial {\bm u}}{\partial t} + 
(\nabla \times {\bm u}) \times {\bm u}
\Bigr)+  \hat{\bm{e}}_z \times {\bm u} = - \nabla p^\star +
Ra_V \, Pm Pr^{-1} \, T \, {\bm r} \,  \nonumber\\ 
+  (\nabla \times {\bm B})
\times {\bm B} + E\nabla^2 {\bm u}, \label{momentum} \\
\frac{\partial {\bm B}}{\partial t} = 
\nabla \times ({\bm u} \times {\bm B}) 
+ \nabla^2 {\bm B},  \label{induction}\\
\frac{\partial T}{\partial t} +({\bm u} \cdot \nabla) T 
=  Pm Pr^{-1} \,
\nabla^2 T,  \label{heat1}\\
\nabla \cdot {\bm u}  =  \nabla \cdot {\bm B} = 0,  \label{div}
\end{align}
The modified pressure $p^*$  in equation \eqref{momentum}
is given by $p+ \frac1{2} E \, Pm^{-1} \, |\bm{u}|^2$.
The dimensionless parameters in the above equations 
are the Ekman number $E=\nu/2\varOmega L^2$,
the Prandtl number, $Pr=\nu/\kappa$, 
the magnetic Prandtl number, $Pm=\nu/\eta$ and
the modified vertical Rayleigh number 
$Ra_V=g \alpha \beta_s L^2/2 \Omega \kappa$. 
Here, $g$ is the gravitational acceleration, 
$\nu$ is the kinematic viscosity, 
$\kappa$ is the thermal diffusivity and $\alpha$ 
is the coefficient of thermal expansion.
The basic state conductive temperature 
profile is one of basal heating, given by  
$T_0(r) = r_i r_o/r$, where $r_i$ and $r_o$ are the
inner and outer radii of the spherical shell. 
The velocity and magnetic fields satisfy the no-slip and 
electrically insulating conditions respectively at the 
two boundaries. 
The inner boundary is isothermal while a 
fixed
heat flux at the outer boundary is given as the 
sum of the uniform mean and a lateral variation.
The heterogeneity at the outer boundary
is measured by the ratio
\begin{equation}
q^*=\frac{\text{Maximum heat flux variation}}
{\text{Mean heat flux}}.
\label{qstar}
\end{equation}
 The mean basic state heat flux in the 
denominator of \eqref{qstar}
 represents the 
superadiabatic heat flux from the 
core \citep[e.g.][p.~11]{olson2015treatise}.
The maximum heat flux 
variation at the outer boundary
represents the magnitude of the peak-to-peak variation
at the CMB relative to the mean superadiabatic value.
%
The calculations are performed by a pseudospectral 
code that uses
spherical harmonic expansions in 
the angular coordinates $(\theta,\phi)$
and finite differences in radius $r$ 
\citep{07willis}.
	
The parameter space in focus is that of
 low
inertia, wherein the Rossby number 
based on the characteristic length scale of convection,
$Ro_\ell$ \citep{chraub2006} is small.
Two Ekman numbers are considered, with the value of
$Pm = Pr$ set
to keep $Ro_\ell \ll 0.1$, ensuring 
that the
simulations remain in the low-inertia regime. 
We explore the effect 
of inhomogeneous 
heat flux boundary conditions at the outer boundary
on the dipole--multipole transition. 
To this end, we progressively
increase the magnitude of the
heat flux 
heterogeneity at the outer 
boundary for a given $Ra_V$. 
For a given $q^*$, the critical vertical Rayleigh
number for the onset of nonmagnetic
convection, $Ra_{V,\, c}$, is obtained for
neutral stability of the perturbations subject 
to a steady mean flow and temperature field produced
by the heterogeneity \citep{sahoo2017}. The ratio
$Ra_V/Ra_{V,\, c}$ (table \ref{tableruns}) is a measure
of the supercriticality of convection.


 The mean spherical harmonic degrees for 
	convection and energy injection, 
denoted by $l_{c}$ \citep{chraub2006} and $l_{E}$ 
\citep{sreeni2014,aditya2022,jfm24} respectively, are given by,

\begin{equation}
l_{c}= \dfrac{\Sigma \hspace{1pt} l \hspace{1pt}
	E_{k}(l)}  {\Sigma \hspace{1pt} E_{k}(l)}; \hspace{5pt}
l_{E}= \dfrac{\Sigma \hspace{1pt} l \hspace{1pt}
	E_{T}(l)}{\Sigma \hspace{1pt} E_{T}(l)},
\label{elldef}
\end{equation}
where $E_{k}(l)$ is the kinetic energy spectrum and
$E_{T}(l)$ is the spectrum obtained from the product of 
the transform of $u_{r}T$
and its conjugate, showing the spectral 
	distribution of scales at which energy is 
	injected by buoyancy. 
The total kinetic and magnetic energies in the saturated
dynamo are given by the volume integrals
	
\begin{equation}
E_k= \dfrac{1}{2} \int \bm{u}^2 \mbox{d}V; 
\quad E_m= \dfrac{Pm}{2E} \int \bm{B}^2 \mbox{d}V.
\label{energies}
\end{equation}

The relative dipole field strength $f_{dip}$, 
which is the ratio of the mean dipole field
strength to the field strength in harmonic degrees 
$l =$ 1--12 at the outer boundary
\citep{chraub2006}, takes values $>0.35$ in 
all the dipole-dominated runs (see table \ref{tableruns}).

The square of the peak magnetic field, denoted by $B^2_{peak}$
(tables 1 \& 2), 
is obtained from the time-averaged value of the magnetic field at 
the peak-field location in the saturated state of each run.
%
\subsection{The effect of equatorially symmetric 
	and anti-symmetric heat flux patterns on polarity transitions}
\begin{figure}
	\centering
	\hspace{-1.75 in}	(a)  \hspace{1.75 in} (b) \\
	\includegraphics[width=0.35\linewidth]{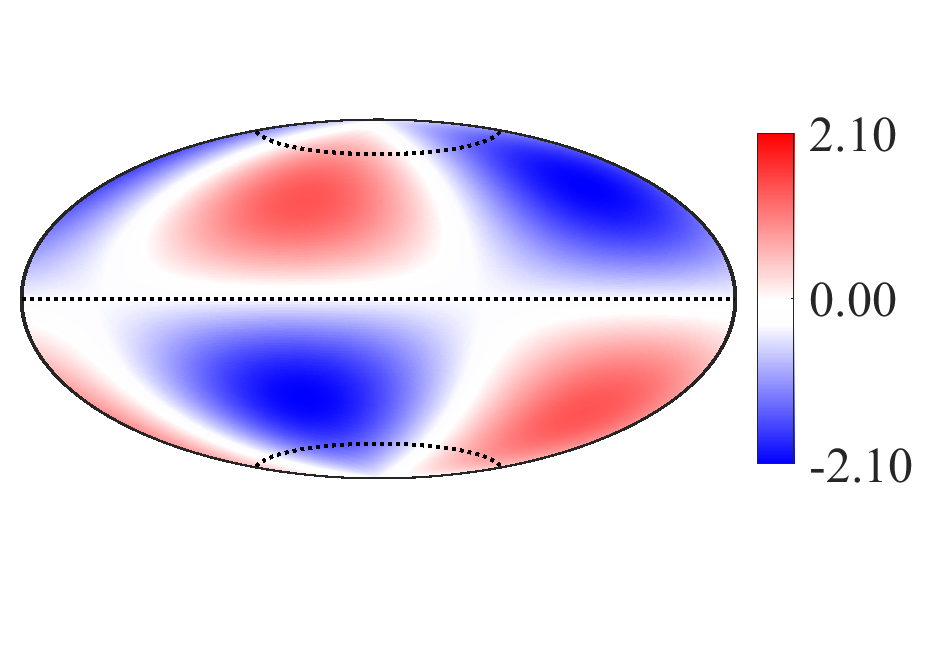}	
	\includegraphics[width=0.35\linewidth]{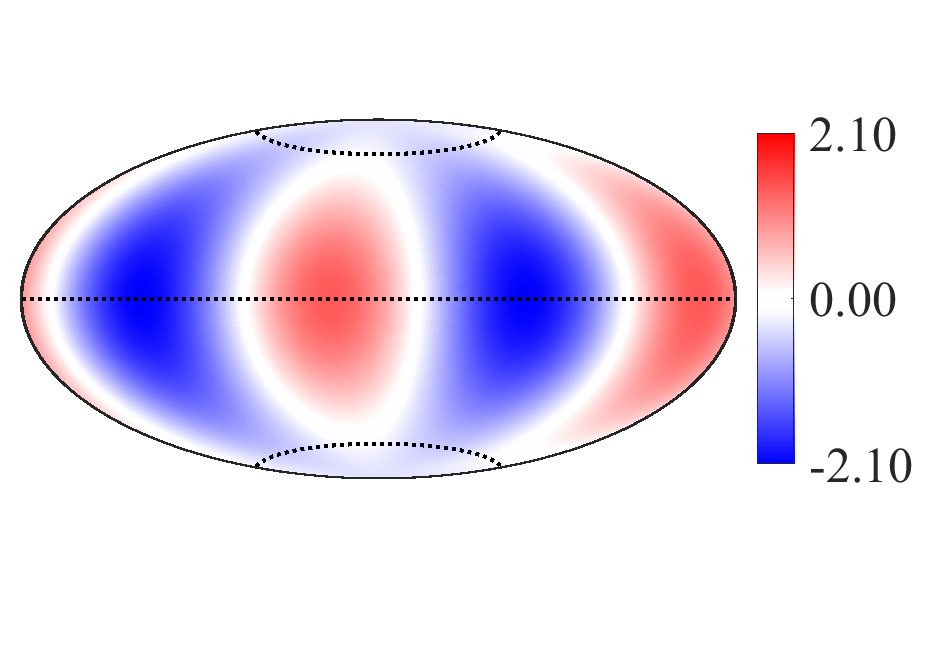}	\\
	\hspace{-1.75 in}	(c)  \hspace{1.75 in} (d) \\
	\includegraphics[width=0.35\linewidth]{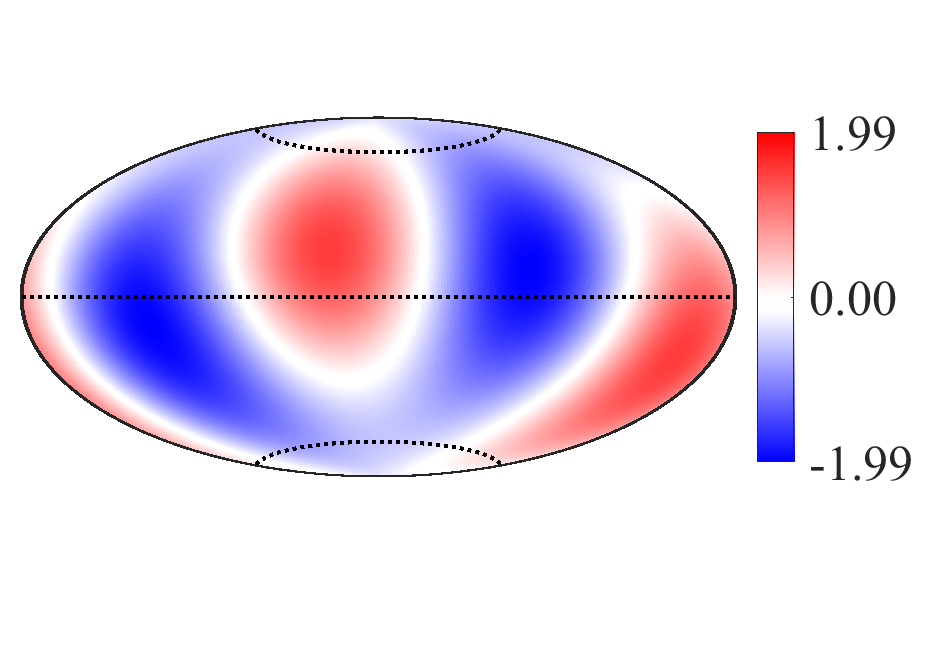}	
	\includegraphics[width=0.35\linewidth]{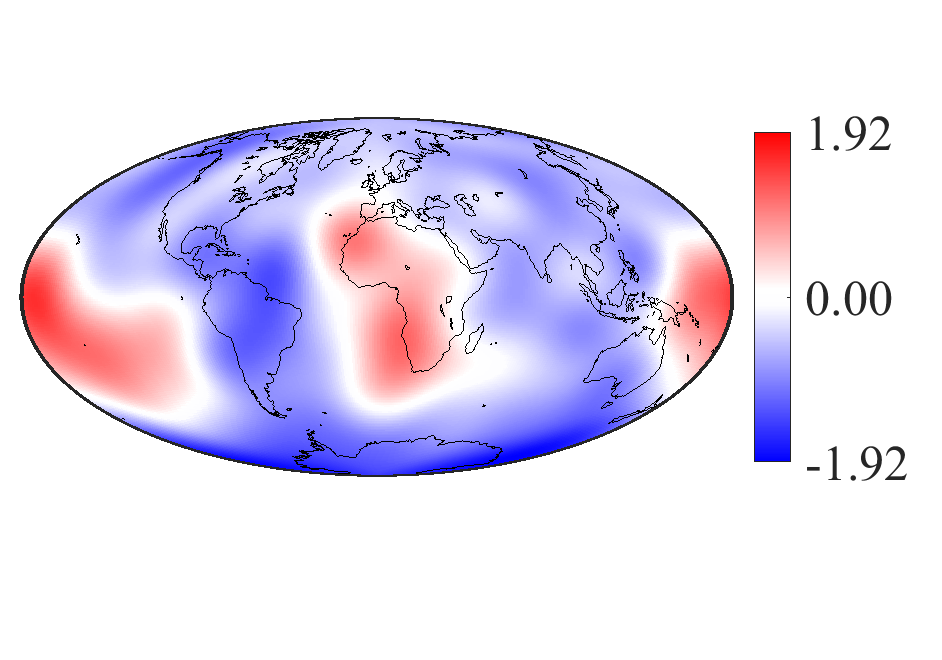}	
	\caption{The distribution of heterogeneous
	radial temperature gradient $\partial T/\partial r$
	at the outer boundary for different conditions. 
		(a) $Y_2^1$, (b) $Y^2_2$, 
		(c)  $Y_2^2:Y_2^1=2:1$ and 
		(d) tomographic condition derived from
		the seismic shear wave velocity variation
		in the Earth's lower mantle \citep{masters1996}.}
	\label{dtdr}
\end{figure}
\begin{figure}
\centering
\hspace{-2.1 in}	(a)  \hspace{2.1 in} (b) \\
\includegraphics[width=0.45\linewidth]{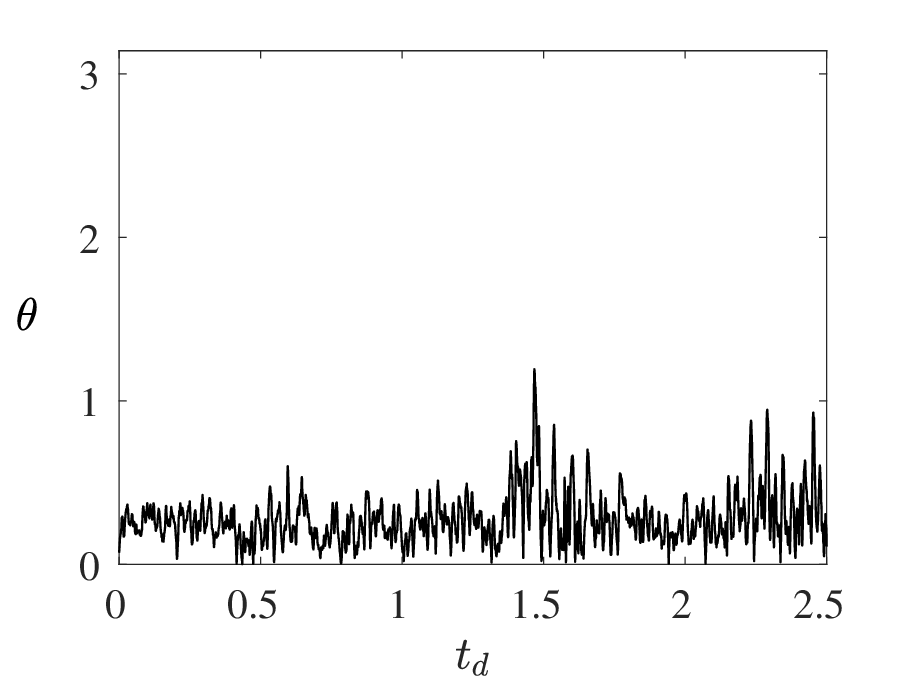}
\includegraphics[width=0.45\linewidth]{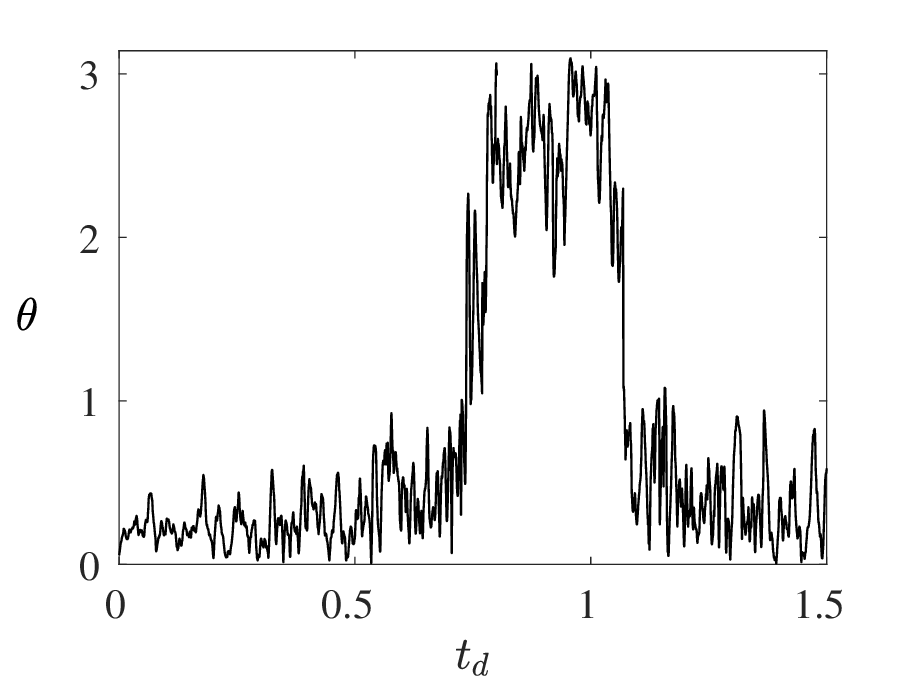}\\
\hspace{-2.1 in}	(c)  \hspace{2.1 in} (d) \\
\includegraphics[width=0.45\linewidth]{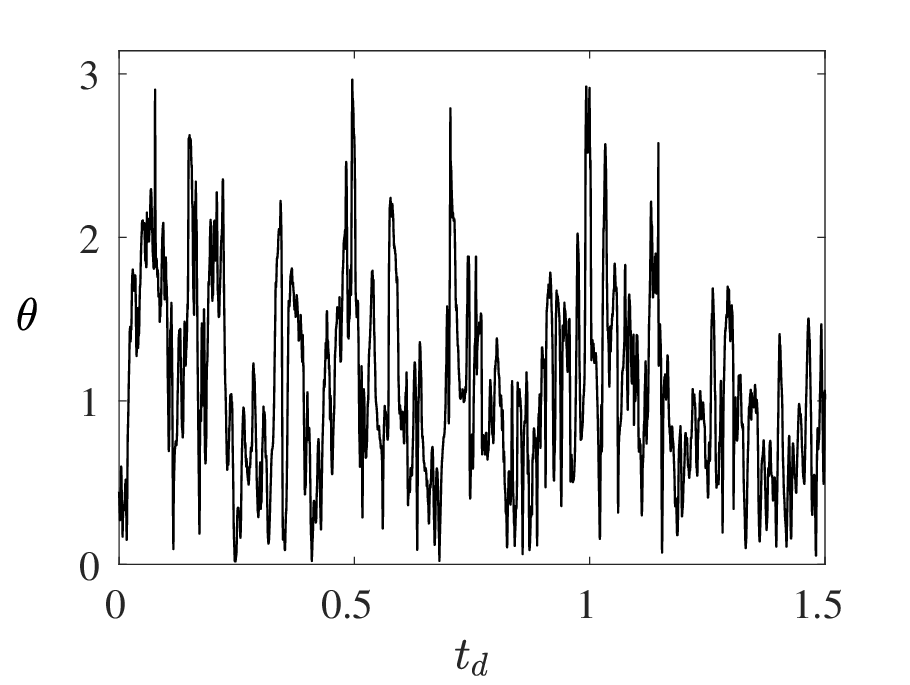}
\includegraphics[width=0.45\linewidth]{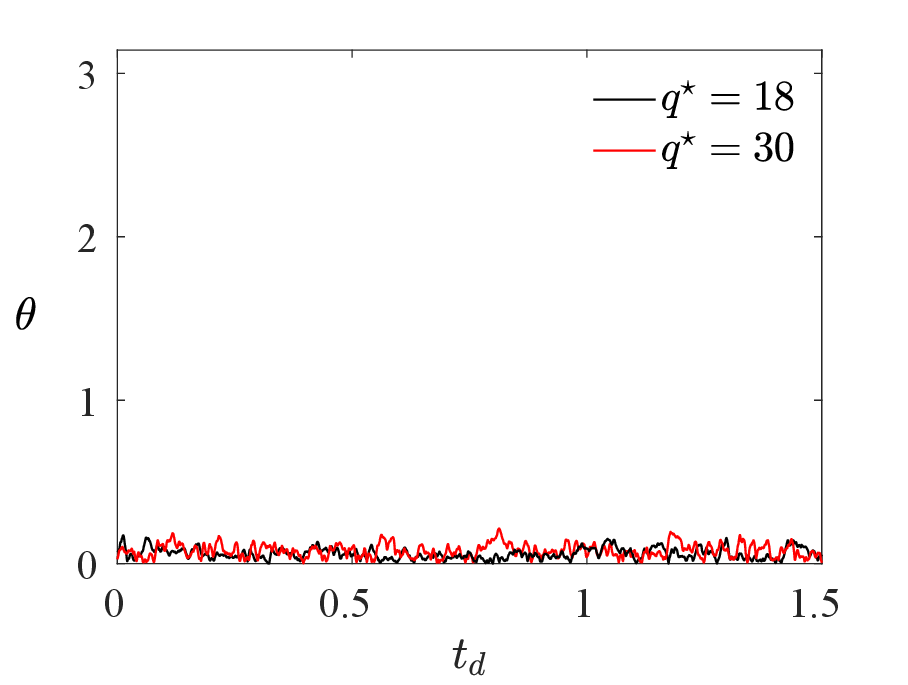}\\
\caption{Evolution of dipole colatitude 
	with magnetic diffusion time for 
	(a) $q^*=17$, $Y_2^1$ (stable dipolar), 
	(b) $q^*=18$, $Y_2^1$ (reversing), 
	(c) $q^*=20$, $Y_2^1$ (multipolar) 
	and (d) $q^*=18$ and $30$, $Y_2^2$ 
	(stable dipolar). The other dynamo 
	parameters are $Ra_V = 2500$, 
	$E = 1.2 \times 10^{-5}$, $Pm = Pr = 1$.}
\label{ra2500tilt}
\end{figure}
\begin{figure}
	\centering
	\hspace{-1.5 in}	(a)  \hspace{1.5 in} (b)\hspace{1.5 in} (c) \\
	\includegraphics[width=0.32\linewidth]{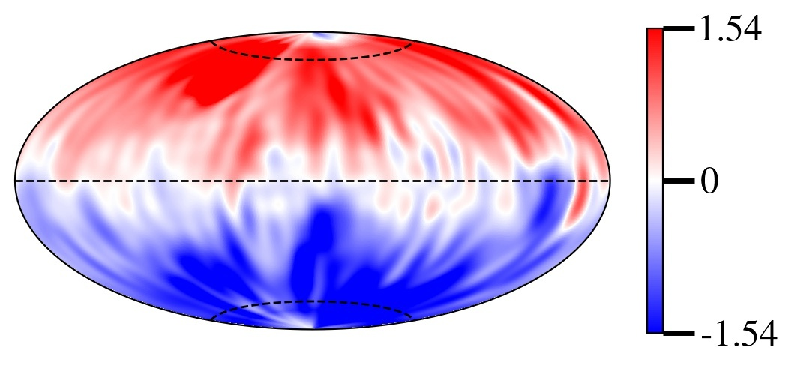}
	\includegraphics[width=0.32\linewidth]{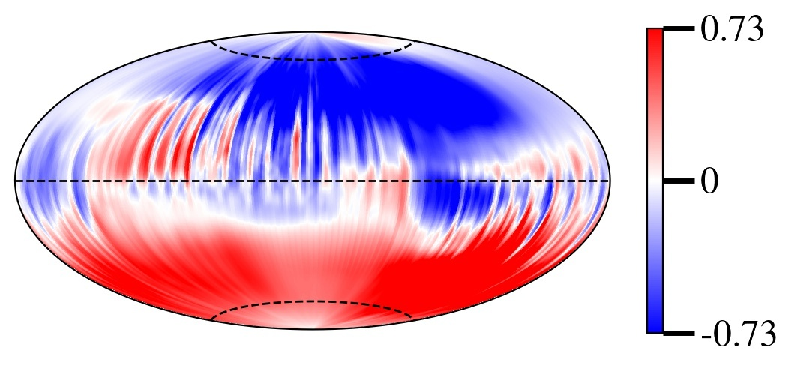}
	\includegraphics[width=0.32\linewidth]{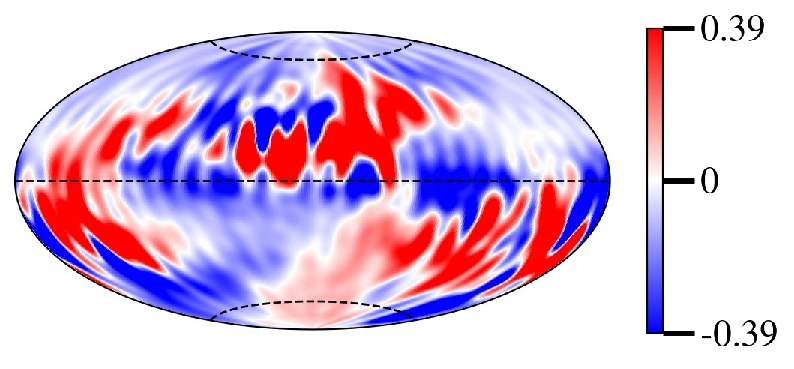}\\
	\hspace{-1.5 in}	(d)  \hspace{1.5 in} (e)\hspace{1.5 in} (f) \\
	\includegraphics[width=0.32\linewidth]{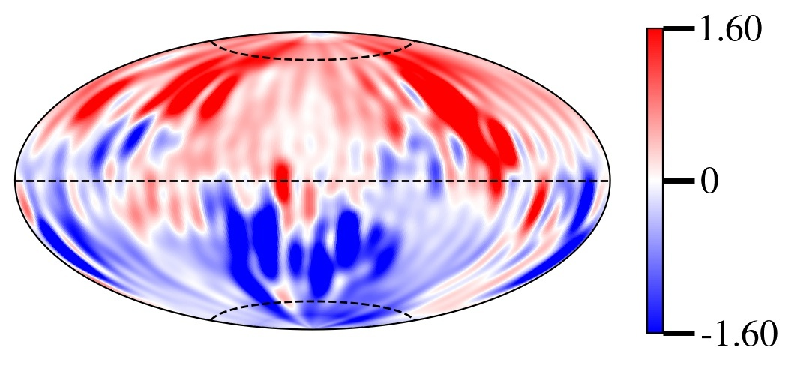}
	\includegraphics[width=0.32\linewidth]{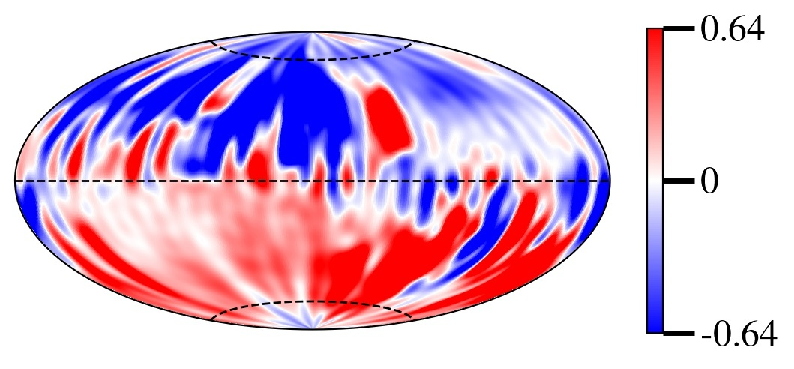}
	\includegraphics[width=0.32\linewidth]{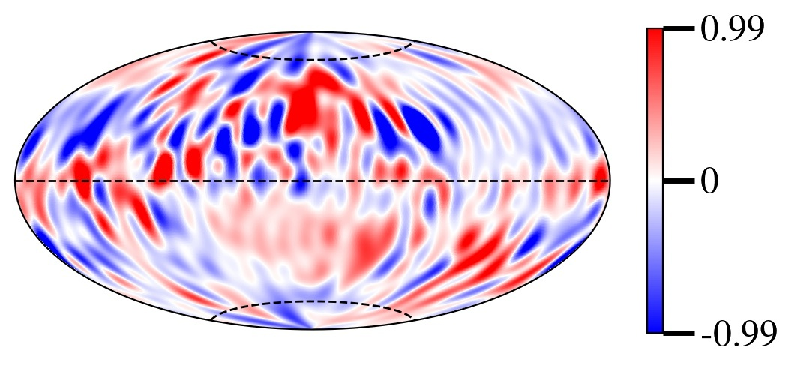}\\
	\caption{Contours of the radial magnetic field at the outer boundary 
		for $Ra_V = 2500$ at $q^* = 17$ (a), 18 (b), 20 (c); and for 
		$Ra_V =25000$ at $q^* = 5$ (d), 6 (e), 7 (f). 
		The other dynamo parameters are 
		$E = 1.2 \times 10^{-5}$, $Pm = Pr = 1$.
		A $Y_2^1$ heat flux heterogeneity is applied at
		the outer boundary.}
	\label{br}
\end{figure}	
We begin our simulations with a strong field dynamo 
($B^2_{rms} > 1$) and increase 
the heterogeneity of the heat flux at the outer 
boundary in steps while keeping all other parameters fixed. 
Figure \ref{dtdr} (a)--(d) 
gives examples of the equatorially
anti-symmetric, symmetric and composite boundary conditions
whose effects on polarity transitions are
analyzed in this study.
The simulations are performed for at least one 
magnetic diffusion time in the saturated state. 
	
In figure \ref{ra2500tilt}, the  
colatitude of the axial dipole field, $\theta$, 
at the outer boundary is obtained from the Gauss 
coefficients of the spherical harmonics, as follows:
\begin{equation}
\cos\theta = g_1^0/|\bm{m}|, \quad 
\bm{m} = (g_1^0, g_1^1, h_1^1),
\label{diplat}
\end{equation}
where $g_1^0$, $g_1^1$, and $h_1^1$ are 
derived from the Schmidt-normalized expansion 
for the scalar potential of the field 
\cite[][pp. 142--143]{glatz2013}. 
	
For $E = 1.2 \times 10^{-5}$, $Pm = Pr = 1$, 
and $Ra_V = 2500$, a strong dipole-dominated 
dynamo undergoes well-defined polarity reversals as
shown in figure \ref{ra2500tilt}(b)
by progressively increasing $q^*$ for the 
equatorially anti-symmetric $Y_2^1$ heat flux pattern. 
With a further increase in $q^*$, 
the dynamo exhibits a multipolar solution 
(figure \ref{ra2500tilt}(c)). However, the 
equatorially symmetric $Y_2^2$ heat flux pattern  does 
not produce polarity transitions even 
with a strong heterogeneity 
(see figure \ref{ra2500tilt}(d)). 
For the same
parameters, the Earth-like composite
heat flux pattern in figure \ref{dtdr}(d),
which consists of both symmetric and anti-symmetric
parts, produces polarity transitions. From the 
evolution of the dipole
axis tilt 
(figure \ref{tomogtilt1}, Appendix \ref{tomogrev}),
well-defined polarity reversals are noted for $q^*=13$.
For the anti-symmetric $Y_2^1$
boundary heat flux variation,
the radial magnetic fields in the dipolar,
reversing and multipolar states are 
shown in figure \ref{br} for 
two series with different 
vertical thermal forcing. 
In table \ref{tableruns}, 
the runs for two values of $Ra_V$, $2500$ and $20000$, are 
presented with different heat flux patterns. 
In the reversing runs, the ratio of 
	magnetic to kinetic energy $E_m/E_k$ falls below unity, 
	 as noted in earlier studies 
	\citep{kutzner2002,tassin2021} although
	a recent study \citep{frasson2025geomagnetic}
	suggests that
	multipolar solutions 
	can exist for $E_m/E_k>1$. 
	Reversing dynamo models in general have not 
	realized $E_m/E_k >1$ probably because their
	magnetic Ekman number $E_\eta$
	is much higher than that in the Earth's core 
	in the relatively small length scales. 
	In contrast, the criterion of
	vanishing slow MAC wave helicity considered in this study
	applies to the energy-containing
	scales where density perturbations are continually
	produced, and is independent of the
	choice of
	$E_\eta$ in the low-inertia limit that rapidly rotating
	planets operate in.

\begin{figure}
\centering
\includegraphics[width=0.5\linewidth]{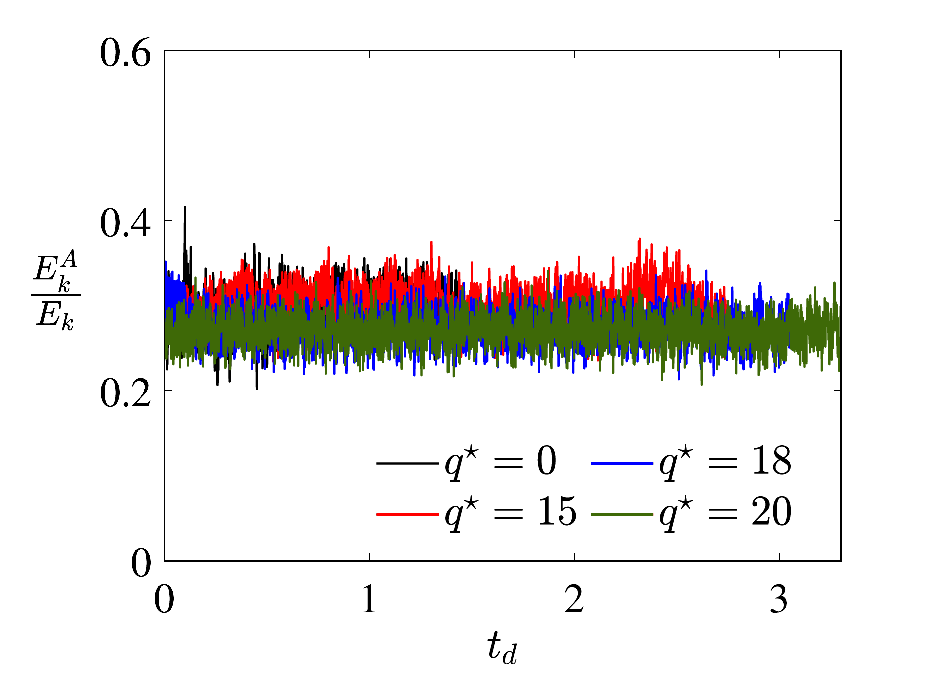}
\caption{Ratio of
anti-symmetric to total 
kinetic energy for $q^*=0$ (black), 
	$q^*=15$ (red), $q^*=18$ (blue) and 
	$q^*=20$ (green) with the $Y_2^1$
	heat flux heterogeneity at the outer boundary.
	 The other dynamo parameters 
	are $Ra=2500,E = 1.2 \times 10^{-5}, Pm=Pr=1$.}
\label{supantisymenergy}
\end{figure}

The main output parameters of the polarity-reversing 
dynamo runs with the anti-symmetric $Y_2^1$ heat flux 
boundary condition 
 are reported in table \ref{tablereversal}. 
\begingroup
\setlength{\tabcolsep}{1.4pt} 
\renewcommand{\arraystretch}{0.8} 
\begin{table}
	\centering
	\begin{tabular}{ ccccccccccccccccccc } 
		$q^*$&$\dfrac{Ra_V}{Ra_{V,\, c}}$ &$N_r$ & $l_{max}$ & $Rm$ & $Ro_\ell$ 
		& $l_C$ & $l_E$  & $\overline{m}$ &$\bar{k}_s$&$\bar{k}_z$
		& $E_k$&$E_m$&$f_{dip}$&$B^2_{peak}$&$B^2_{rms}$&
		$\bigg|\dfrac{\omega_{A}}{\omega_{M}}\bigg|$
		&$\dfrac{Ra_{\ell,H}}{Ra_\ell}$& Type \\
		&&&&&&&&&&&$\times10^5$&$\times10^5$&&&&&&D/R/M\\ 
		
		\multicolumn{19}{c}{ $Ra_V=2500$, $Y_2^1$ }\\
		0 &75.75&168&160&174&0.011&19&26&7&4.04&2.82
		&2.22&27.25&0.81&95&3.59&0.49&0.00&D\\
		5 &80.64&176&176&181&0.013&19&28&7&4.46&4.43
		&2.43&26.30&0.77&81&3.51&0.60&0.36&D\\    
		10&80.64&176&176&232&0.018&19&28&7&4.41&3.94
		&4.06&16.80&0.71&78&2.26&0.76&0.51&D\\    
		15&80.64&176&176&250&0.020&19&28&7&4.01&3.92
		&4.81&14.16&0.67&72&1.66&0.86&0.56&D\\    
		17&80.64&176&176&261&0.021&19&28&7&4.39&3.97
		&5.15&13.11&0.65&67&1.52&0.92&0.60&D\\    
		18&80.64&176&176&341&0.025&19&28&7&4.55&3.92
		&8.58&2.41 &0.25&41&0.40&1.00&0.62&R\\    
		20&80.64&176&176&349&0.027&19&27&7&3.93&3.83
		&8.93&2.43 &0.13&43&0.38&1.19&0.64&M\\    
		\multicolumn{19}{c}{ $Ra_V=2500$, $Y_2^2$ }\\
		18&125.00&176&176&239&0.018&20&34&7&4.45
		&4.90&4.15&15.68&0.82&75&2.56&0.42&--&D\\    
		25&131.57&176&176&251&0.017&18&35&6&4.74
		&4.99&4.60&17.68&0.81&88&2.82&0.41&--&D\\    
		30&138.88&176&176&263&0.018&18&35&5&4.59
		&5.02&5.03&18.68&0.78&90&3.01&0.41&--&D\\  
		\multicolumn{19}{c}{$Ra_V=2500$, $Y_2^2:Y_2^1=2:1$ }\\
		5&78.12&176&176&199&0.016&21&29&8&4.67
		&3.31&2.93&17.81&0.78&72&2.92&0.53&0.31&D\\          
		7&78.12&132&144&240&0.023&22&31&8&4.86
		&3.44&4.20&7.86&0.79&58&1.45&0.60&0.38&D\\   
		9&78.12&176&176&280&0.026&22&32&8&4.98
		&3.53&6.54&1.32&0.40&46&0.36&0.76&0.43&D\\    
		10&78.12&176&176&301&0.027&22&33&8&5.22
		&3.43&6.62&1.25&0.33&39&0.20&1.01&0.50&R\\    
		12&78.12&176&176&328&0.029&22&33&8&5.17
		&3.41&7.93&0.68&0.31&27&0.11&1.23&0.51&R\\   
		\multicolumn{19}{c}{$Ra_V=2500$, $Y_2^2:Y_2^1=5:1$ }\\
		10&80.64&132&144&222&0.017&21&34&7&3.43
		&3.83&3.47&18.83&0.69&86&2.65&0.30&0.24&D\\            
		30&80.64&176&176&264&0.018&22&34&7&3.82
		&4.11&5.07&15.01&0.79&116&2.51&0.41&0.44&D\\    
		\multicolumn{19}{c}{$Ra=2500$, 
			variation derived from seismic tomography }\\  
		5&113.63&176&176&182&0.012&17&27&6&5.27
		&4.36&2.43&24.13&0.64&64&3.86&0.54&0.19&D\\ 
		10&113.63&132&144&199&0.012&16&30&6&5.12
		&4.25&2.88&21.30&0.52&70&3.46&0.62&0.33&D\\   
		12&113.63&132&144&217&0.013&16&31&6&5.3
		&3.41&3.49&17.50&0.44&58&2.92&0.82& 0.43&D\\   
		13&113.63&132&144&270&0.025&25&41&6&5.5
		&3.51&6.62&0.12&0.32&15&0.23&1.03&0.50&R\\  
		\multicolumn{19}{c}{$Ra_V=20000$, $Y_2^1$ }\\
		0&606.06&192&180&573&0.052&25&35&10&4.43
		&3.12&23.84&33.57&0.78&253&5.46&0.64&--& D\\
		5&645.16&176&176&592&0.052&21&30&8&4.25
		&4.03&27.62&32.81&0.65&243&5.43&0.85&0.38&D\\    
		10&645.16&176&176&622&0.051&21&30&8&4.23
		&4.08&28.46&32.58&0.57&235&5.35&0.94&0.49&D\\    
		11&645.16&176&176&687&0.054&19&28&7&4.31
		&4.01&34.51&25.53&0.46&231&3.94&0.98&0.52&D\\    
		12&645.16&176&176&785&0.054&17&27&6&5.25
		&4.33&42.48&20.32&0.16&199&3.49&1.10&0.55&R\\   
		13&645.16&176&176&773&0.054&18&29&9&4.56
		&4.00&43.61&20.04&0.08&207&2.99&1.18&0.57&M\\   
		\multicolumn{19}{c}{$Ra_V=20000$, $Y_3^2$ }\\
		4&645.16&176&176&607&0.053&23&31&9&4.14
		&4.09&27.21&30.32&0.77&265&4.88&0.92&0.22&D\\  
		5&645.16&176&176&685&0.058&22&31&9&4.00&4.03
		&34.53&16.98&0.18&208&2.79&1.09&0.25&R\\  
		\multicolumn{19}{c}{$Ra_V=20000$, $Y_1^1$ }\\
		30&869.56&176&176&810&0.061&19&44&5&5.85&4.07
		&47.80&18.75&0.52&154&2.93&0.85&--&D\\    
		\multicolumn{19}{c}{$Ra_V=20000$, $Y_2^2$ }\\
		12&952.38&176&176&696&0.060&25&39&8&4.62&4.41
		&34.61&12.95&0.82&142&2.42&0.83&--&D\\  
		30&1111.11&176&176&776&0.073&24&40&7&4.61&4.76
		&44.39&7.80&0.57&104&1.26&0.92&--&D\\  
		\multicolumn{19}{c}{$Ra_V=20000$, $Y_3^3$ }\\
		30&952.38&176&176&782&0.072&25&37&8&4.23&4.45&37.62&8.77&0.52&94&0.98&0.95&--&D\\  
		\multicolumn{19}{c}{$Ra_V=20000$, $Y_3^1$ }\\
		30&645.16&176&176&782&0.07&25&38&8&4.18&4.52&31.44&35.61&0.54&156&3.18& 0.77&--&D\\  
	\end{tabular}
	\caption{Summary of the main input and 
		output parameters of 
		a few dynamo simulations. Here, $Ra_V$ is 
		the modified vertical Rayleigh number, 
		$Ra_{V,\, c}$ is the modified vertical critical 
		Rayleigh number for onset of 
		nonmagnetic convection, $q^*$ 
		is a dimensionless measure 
		of the boundary heterogeneity defined in \eqref{qstar}, 
		$N_r$ is the number of radial grid points, $l_{\max}$ is 
		the maximum spherical harmonic degree, $Rm$ is the 
		magnetic Reynolds number, $Ro_\ell$ is the local 
		Rossby number, $l_C$ and $l_E$ are the mean 
		spherical harmonic degrees of convection and 
		energy injection respectively, $\overline{m}$ is the 
		mean spherical harmonic order in the range 
		$l \leq l_E$, $\bar{k}_s$ and $\bar{k}_z$ are 
		the mean $s$ and $z$ wavenumbers in the range $l \leq l_E$, 
		$E_k$ and $E_m$ are the time-averaged total kinetic 
		and magnetic energies, $f_{dip}$ is the relative 
		dipole field strength, $Ra_{\ell,H}/Ra_\ell$ is 
		the ratio of the local horizontal to the resultant Rayleigh numbers
		(obtained from \eqref{ravrah2} and \eqref{raresultant}), 
		and $B^2_{rms}$ is the measured mean square 
		value of the field in the spherical shell. 
		Types `D', `R', and `M' denote dipolar, 
		reversing and multipolar
		dynamo states, respectively.
		The other dynamo parameters are 
		$E = 1.2 \times 10^{-5}$, $Pm = Pr = 1$.
	}
	\label{tableruns}
\end{table}
\endgroup
\begingroup
\setlength{\tabcolsep}{1.4pt} 
\renewcommand{\arraystretch}{1.25} 
\begin{table}
	\centering
	\begin{tabular}{ ccccccccccccccccccc} 
		$Ra_V$& $\dfrac{Ra_V}{Ra_{V,\, c}}$ & $q^*$&$N_r$ 
		& $l_{max}$ & $Rm$ & $Ro_\ell$ & $l_C$ 
		& $l_E$ &$\overline{m}$&$\bar{k}_s$&$\bar{k}_z$
		&$E_k$&$E_m$&$f_{dip}$&$B^2_{rms}$ 
		& $Ra_\ell$ &$B^2_{peak}$ &$\dfrac{Ra_{\ell,H}}{Ra_\ell}$ \\
		&&&&&&&&&&&&$\times10^5$&$\times10^5$&&&$\times10^4$&&\\
		\multicolumn{19}{c}{$E=6\times10^{-5},Pm=Pr=5$}\\
		2000&54.05&25 &176&176&235&0.012&14&21&5.0
		&3.74&3.81&4.03&1.05&0.11&0.12&0.71&85&0.66 \\
		3000&81.08&25 &176&176&235&0.012&14&21&5.2
		&3.65&3.93&6.39&3.51&0.12&0.24&0.97&130&0.65\\
		4500&121.62&22 &176&176&333&0.018&14&21&5.3
		&3.6&3.61&8.04&2.07&0.08&0.35&1.25&160&0.64\\
		6000&162.16&20 &176&176&378&0.020&14&21&5.4
		&3.87&3.83&10.50&2.18&0.08&0.27&1.63&170&0.63\\
		8000&216.21&18 &176&176&427&0.023&14&21&5.6
		&3.89&3.98&13.17&2.35&0.07&0.38&2.01&210&0.61\\
		10000& 270.27&18 &176&176&479&0.026&14&21&5.6
		&3.94&4.36&15.55&4.58&0.08&0.44&2.48&225&0.60\\
		12000&324.32&17 &176&176&509&0.028&15&22&5.8
		&4.25&4.58&18.83&4.83&0.12&0.59&2.84&250&0.58 \\
		15000&405.40&17 &176&176&562&0.030&14&21&5.4
		&4.2&4.75&23.02&5.54&0.11&0.90&3.55&274& 0.57\\
		18000&486.48&14&176&176&603&0.033&14&21&5.6
		&4.27&4.35&26.38&6.20&0.11&1.04&3.85&295&0.54\\
		20000&540.54&5&176&176&629&0.034&14&21&6.0
		&4.32&3.94&26.64&8.26&0.25&1.29&3.22&268&0.41\\
		21000&552.63&0  &160&180&549&0.039&19&25&8.1
		&4.22&3.38&20.54&11.47&0.30&0.38&2.25&227&0.00\\
		\multicolumn{19}{c}{$E = 1.2 \times 10^{-5},Pm=Pr=1$}\\
		1500 &48.38&18&132 &128&275&0.020&19&27&5.5
		&4.73&3.95&5.49&1.38&0.23&0.25&0.42&43&0.64\\
		2500&80.64&18&176&176&341&0.025&19&28&7.0
		&4.55&3.92&8.58&2.41&0.25&0.40&0.69&65&0.63\\    
		4000&129.03&17&176&176&423&0.030&19&28&7.1
		&4.49&3.98&12.82&3.94&0.22&0.67&1.04&91&0.61\\  
		6000&193.54&17&176&176&485&0.033&19&28&7.1
		&4.55&4.21&16.90&6.82&0.15&1.16&1.52&120&0.60\\   
		8000&258.06&17&176&176&546&0.037&18&28&6.7
		&4.66&4.58&21.77&8.51&0.14&1.21&2.00&137& 0.58\\   
		10000&322.58&17&176&176&605&0.040&18&28&6.5
		&4.39&4.49&26.52&10.70&0.10&1.74&2.42&169& 0.57\\  
		12000&387.09&16&176&176&639&0.044&18&28&6.5
		&4.53&4.12&29.82&13.17&0.09&2.01&2.77&186&0.57\\  
		15000&483.87&15&176&176&703&0.045&17&27&6.1
		&4.83&4.28&34.77&16.65&0.11&2.68&3.23&202&0.56\\   
		20000&645.16&12&176&176&785&0.054&17&27&6.2
		&5.25&4.33&42.48&20.32&0.16&3.49&4.06&221&0.53\\  
		25000&806.45&6&176&176&795&0.060&18&28&7.2
		&5.18&4.14&44.83&21.34&0.11&1.85&3.45&217&0.42\\ 
		27000&870.96&1&176&176&842&0.062&19&29&7.8
		&5.10&3.94&39.66&21.51&0.08&3.11&3.11&200&0.20\\
		28000&848.48&0&192&180&775&0.073&25&36&
		10.0&4.81&2.91&39.74&19.67&0.25&0.67&2.80&197&0.00\\
	\end{tabular}
	\caption{Summary of the main input and output parameters in the 
		reversing dynamo simulations considered in this study for 
		$Y_2^1$ heat flux condition. 
			These simulations have  
		$|\omega_{A}/\omega_{M}|\geq1$ throughout. 
		Here, $Ra_V$ is the modified 
		Rayleigh number, $Ra_{V,\, c}$ is the modified 
		vertical critical Rayleigh 
		number for onset of nonmagnetic convection, 
		$q^*$ is a dimensionless measure of the boundary 
		heterogeneity defined in \eqref{qstar}, $N_r$ is the 
		number of radial grid points, $l_{max}$ is the maximum 
		spherical harmonic degree, $Rm$ is the magnetic 
		Reynolds number, $Ro_\ell$ is the local Rossby number, 
		$l_C$ and $l_E$ are the mean spherical harmonic degrees 
		of convection and energy injection respectively 
		(defined in \eqref{elldef}), $\overline{m}$ is the  
		mean spherical harmonic order in the range $l \leq l_E$, 
		$\bar{k}_s$ and $\bar{k}_z$ are the mean $s$ and $z$ 
		wavenumbers in the range $l \leq l_E$, $E_k$ and $E_m$ 
		are the time-averaged total kinetic and magnetic energies 
		defined in \eqref{energies}, $f_{dip}$ is the relative 
		dipole field strength, $Ra_\ell$ is the local Rayleigh 
		number defined in \eqref{ravrah2} and 
		$B^2_{peak}$ is the square of the measured peak
		field when slow MAC 
		waves cease to exist when $|\omega_{A}|\approx
		|\omega_{M}|$.}
	\label{tablereversal}
\end{table}
\endgroup
 The ratio of anti-symmetric to total kinetic energy 
	remains nearly constant even at high $q^*$ 
	(figure \ref{supantisymenergy}), 
	so equatorially anti-symmetric boundary conditions 
	do not induce polarity transitions by breaking 
	the equatorial symmetry of the convection columns.
Before analysing the role of 
wave motions in polarity transitions,
we examine the mean
temperature and velocity fields
produced by the heterogeneous boundary heat flux, which
give useful comparisons with the linear convection
model.

\subsection{ Mean resultant temperature gradient and velocity field}
\label{meanfields}
\begin{figure}
	\centering
	\hspace{-2 in}	(a)  \hspace{2 in} (b) \\
	\includegraphics[width=0.4\linewidth,
	height=0.29\linewidth]{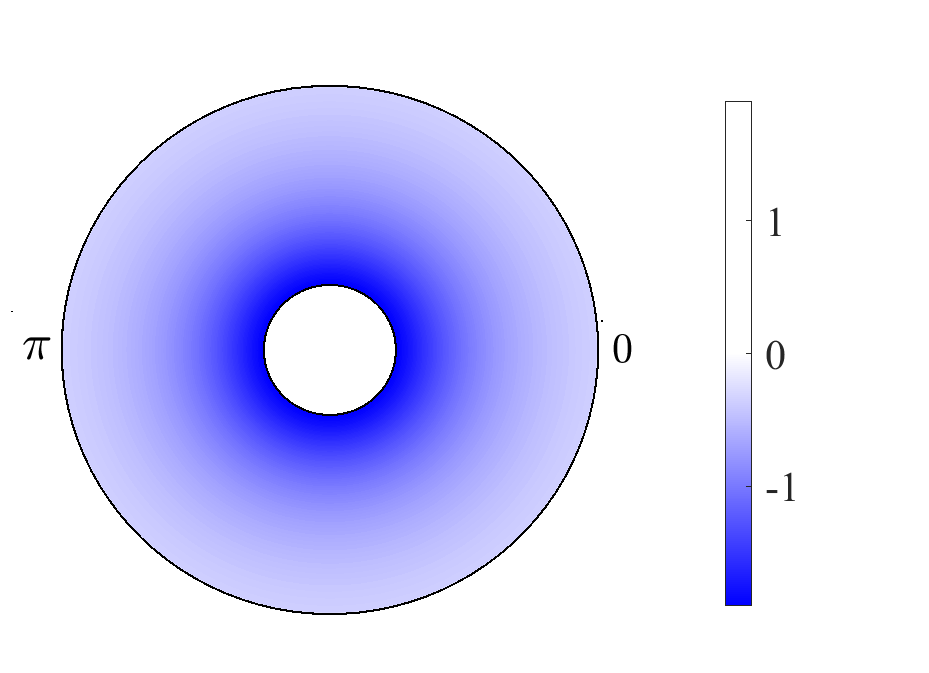}
	\includegraphics[width=0.4\linewidth,
	height=0.29\linewidth]{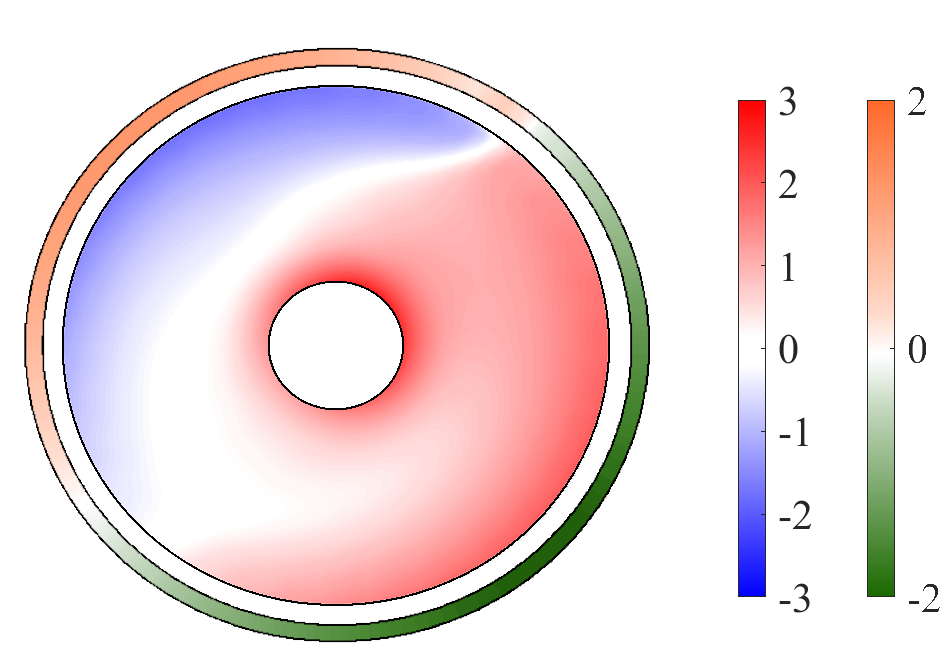}\\
	\hspace{-2 in}	(c)  \hspace{2 in} (d) \\
	\includegraphics[width=0.4\linewidth,
	height=0.29\linewidth]{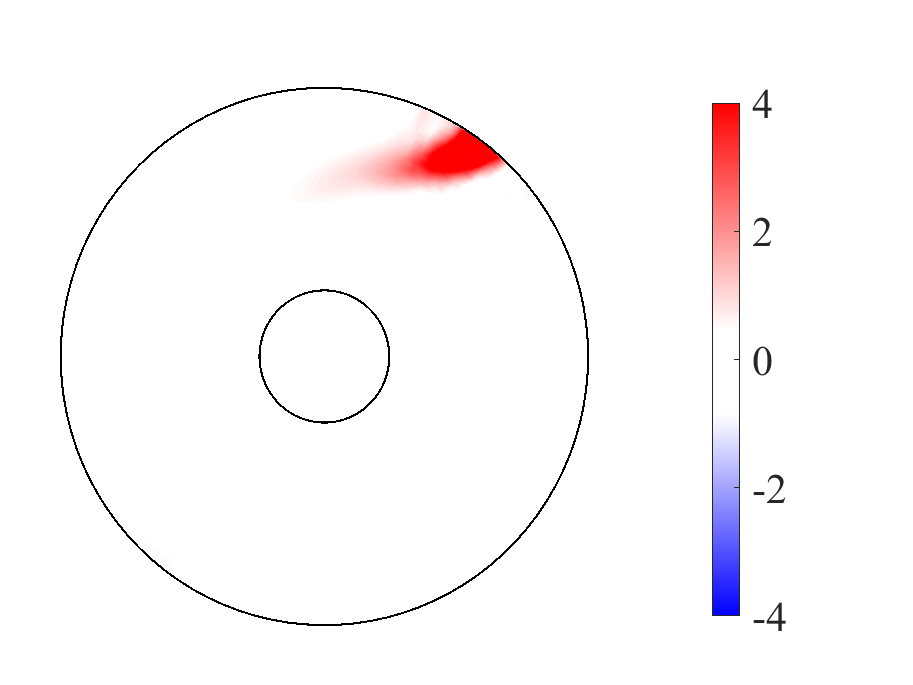}
	\includegraphics[width=0.4\linewidth,
	height=0.29\linewidth]{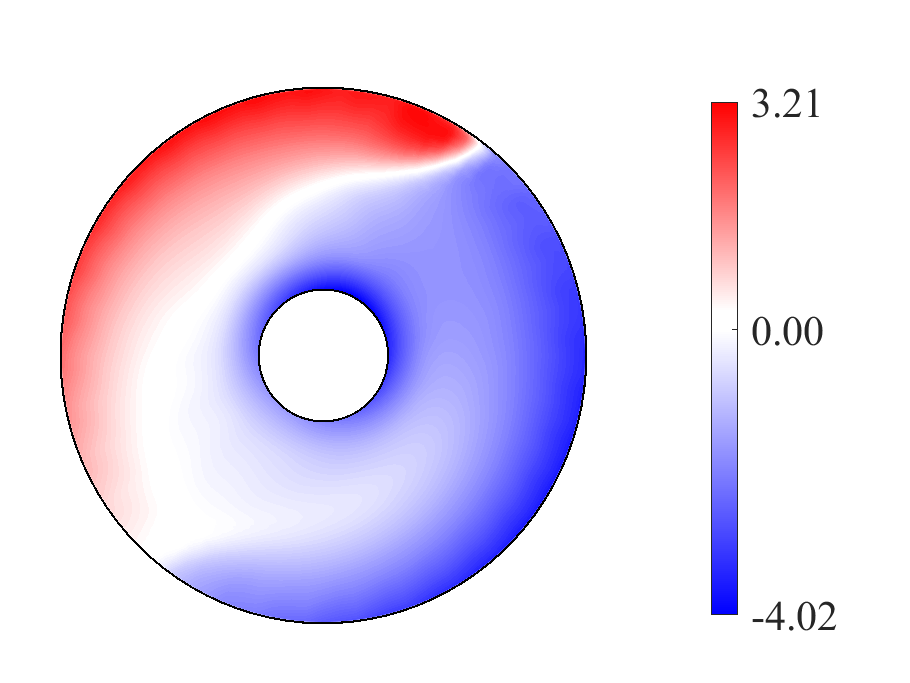}\\
	\hspace{-2 in}	(e)  \hspace{2 in} (f) \\
	\includegraphics[width=0.4\linewidth,
	height=0.29\linewidth]{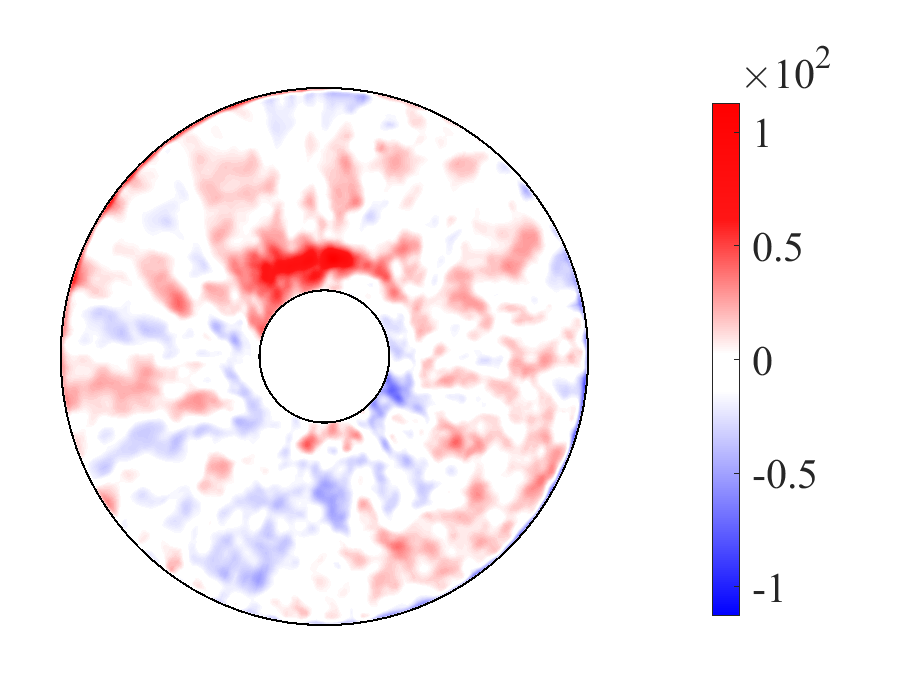} 
	\includegraphics[width=0.4\linewidth,
	height=0.29\linewidth]{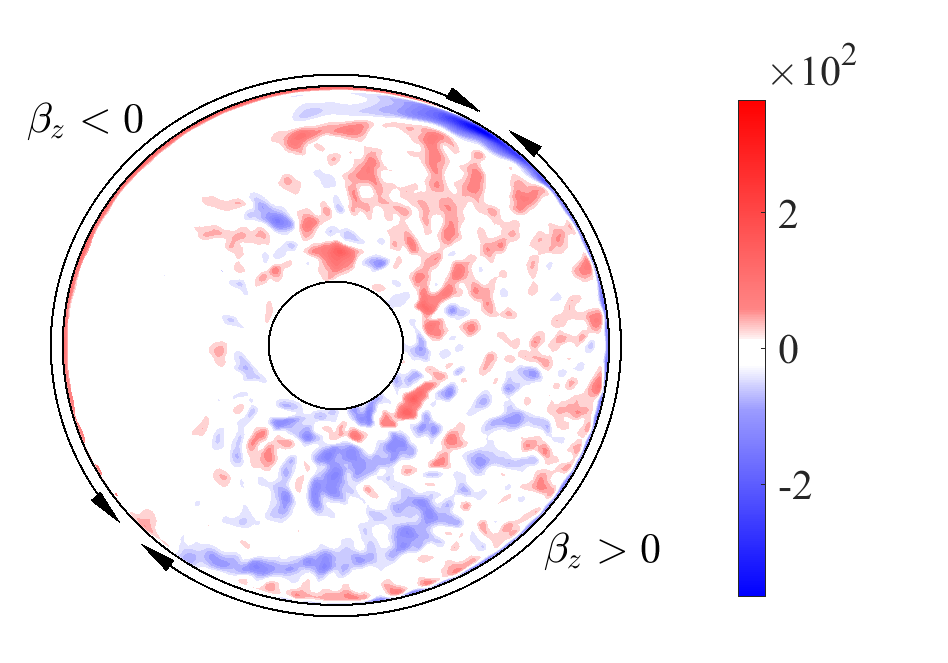}
	\caption{Horizontal section plots  at $z = 0.4$ 
	below the equator showing 
	(a) $\beta_s$ for homogeneous boundary heating, 
	(b) $\beta_z$ for $q^* = 18$, 
	(c) $\beta_\phi$ for $q^* = 18$, and 
	(d) resultant gradient $\beta$ for $q^* = 18$. 
	The temperature gradients are 
		calculated at vertical Rayleigh
		number $Ra_V = 30$, which is a state
		near the onset of nonmagnetic convection.
	A $Y_2^1$
	heterogeneity in outer boundary heat flux is applied. 
	The orange-green coloured strip at the periphery of 
	plot (b) represents
	 $\partial T_0 / \partial r$ at the
	 outer boundary.  
	Panels (e) and (f) show snapshots of
	the axial velocity $u_z$ at $q^* = 0$ 
	and $q^* = 18$ respectively for dynamo 
	simulations at $Ra_V = 2500$. The other parameters 
	are $ E = 1.2 \times 10^{-5} $, and $ Pm = Pr = 1 $.}
	\label{beta}
\end{figure}

In line with that obtained in equation \eqref{rbeta} in
the Cartesian model,
the mean (superadiabatic) 
resultant temperature gradient is given by
\begin{equation}
\beta = \beta_s - 2 \beta_z k_s/k_z,
\label{betadyn}
\end{equation}
where $\beta_s = \partial T_0/\partial s < 0$ 
in an unstably stratified layer with homogeneous 
boundary heat flux (see figure~\ref{beta}(a)).
In figure~\ref{beta}(b), $\beta_z = \partial T_0/\partial z$ 
is plotted for an equatorially 
antisymmetric $Y_2^1$ heat-flux heterogeneity at the outer boundary.
The temperature gradients are 
	calculated from a steady state just below nonmagnetic
	convective onset  
	($Ra_V = 30$, whereas onset occurs at $Ra_V = 31$
	 for $q^* = 18$) so that the
	 perturbations do not affect the calculation of the 
	mean gradients. 
An averaged $\beta_z$, obtained by
taking the average of the positively signed
$\partial T_0/\partial z$ 
within the spherical shell, is proportional 
to the intensity of the anti-symmetric heterogeneity.
Likewise, the averaged $\beta_s$ is obtained by taking the average of  
$\partial T_0/\partial s$  within the shell. 
The averaged resultant gradient $\beta$ is 
then evaluated from \eqref{betadyn} 
using the average $s$ and $z$ wavenumbers defined 
in \eqref{kzmean} (see figure \ref{beta}(d)).
The resultant $\beta$ calculated at $Ra_V = 30$ 
and the resultant $\beta$ averaged over 
$\approx$ 0.5 diffusion time from the strongly 
supercritical state at $Ra_V=310$ exhibit similar structures 
as well as nearly equal maximum and minimum values 
(see figure S4 in the Supplementary Material).
Convection is suppressed in stably stratified 
regions where $\partial T/\partial r > 0$. 
In contrast, regions beneath 
$\partial T/\partial r < 0$, where $\beta_z > 0$
and $\beta < 0$,
support convection.
	This unstably stratified region is in focus in the analysis
	in \S \ref{macwaves} below --  
	at moderately
	positive values of $\beta_z$, both fast and slow 
	MAC waves are present, giving the dipolar
	dynamo regime; 
	at large positive values of $\beta_z$, the 
	slow MAC waves are selectively suppressed whereas 
	the fast MAC waves persist, giving the reversing and multipolar 
	dynamo regimes.
	
Equation \eqref{betadyn} is derived 
from the theoretical model described in \S\ref{linear}, 
which predicts the presence or absence of 
slow MAC waves subject to the frequency inequality 
$|\omega_{C}| > |\omega_{M}|,|\omega_{A}| > |\omega_{\eta}|$, 
with slow MAC waves present when $|\omega_{M}| > |\omega_{A}|$ 
and absent when $|\omega_{M}| \le |\omega_{A}|$. 
Since the above frequency inequality does not predict the
localized suppression of convection observed in figure \ref{beta}(f), 
equation \eqref{betadyn} can only be applied to regions 
where $\beta_{z}>0$.
Interestingly,
$\beta_\phi= \partial T_0/\partial \phi$ (figure \ref{beta}(c))
has practically
no influence on $\beta$, which justifies our original assumption
of ignoring variations with respect to the non-preferred
horizontal direction (see \eqref{a8}, \S \ref{pert1}).
The convective region produced by $\beta_s <0$ and $\beta_z>0$ 
with the equatorially anti-symmetric heat flux at the
boundary corresponds to the unstably stratified system
modelled by the equatorial radial configuration
of the Cartesian linear model in \S \ref{setup}.

\begin{figure}
	\centering
	\hspace{-1.7 in} (a)  \hspace{1.5 in} (b) 
	\hspace{.8 in} (c) \\
	\includegraphics[width=0.35\linewidth]{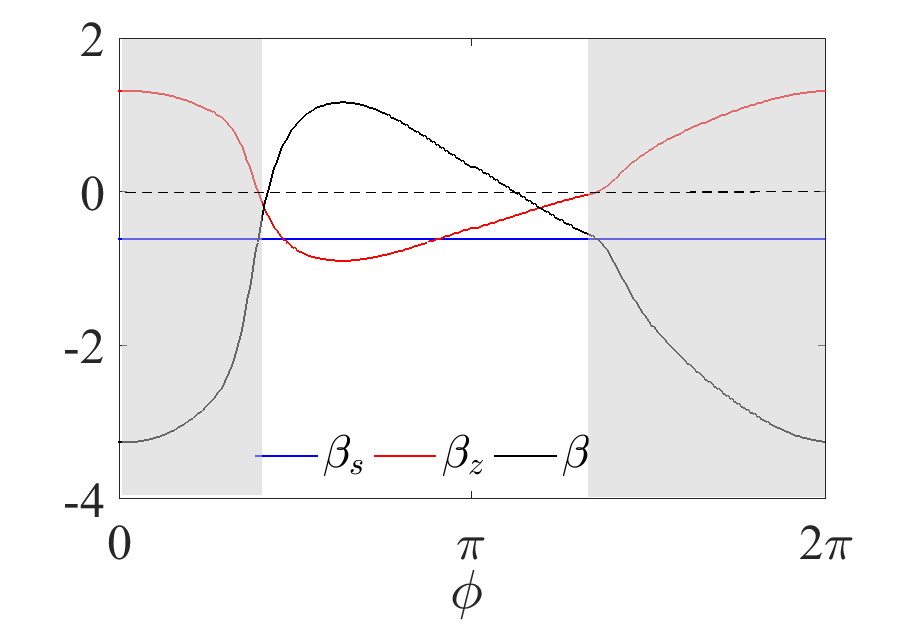}
	\includegraphics[width=0.16\linewidth]{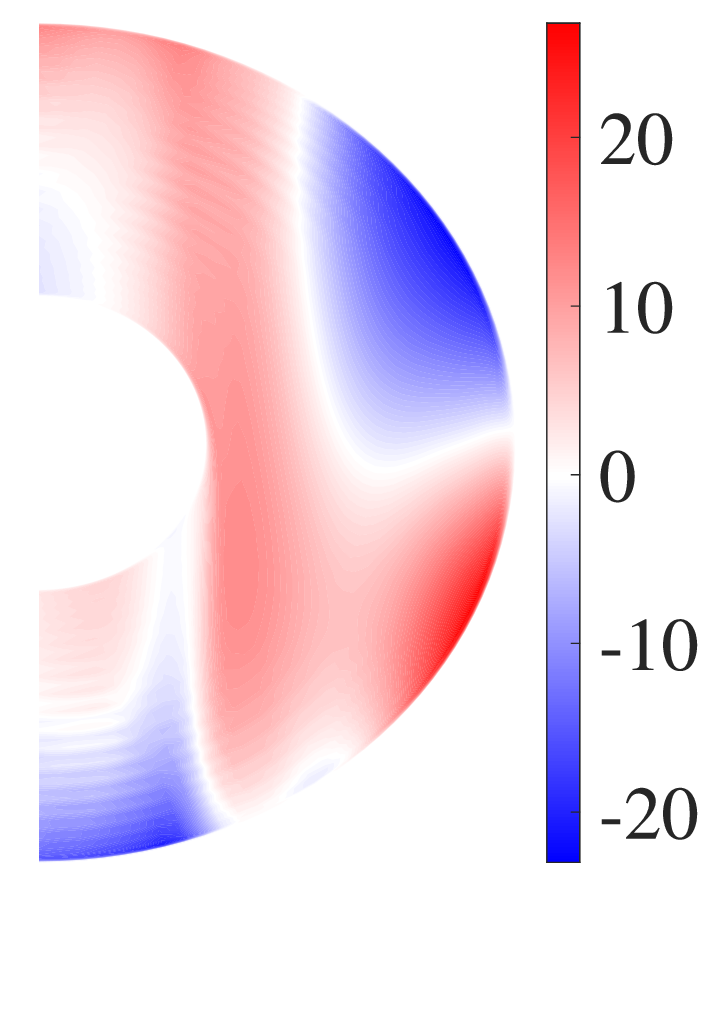}
	\includegraphics[width=0.35\linewidth]{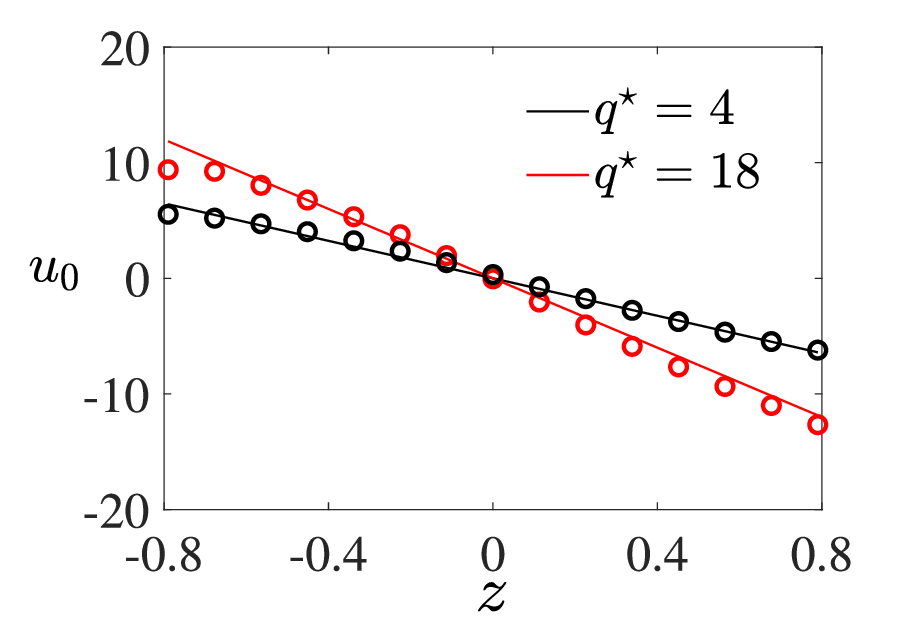}\\
	\caption{(a) Values of $\beta_s$ (blue), 
	$\beta_z$ (red) and the resultant 
	$\beta$ (black) at cylindrical radius
	$s = 1$ at the section $z = 0.4$ below the equator 
	for $q^* = 18$. (b)
	Meridional sections plot of 
	the $\phi$-component of the velocity, $u_0$ 
	for $q^* = 18$. 
	The plot shows $\phi$-averaged 
	values of $u_0$ in the unstably
	stratified region where $\beta < 0$ 
	($\phi= 1.31 \, \pi$ to $0.31 \, \pi$ through
	$\phi=0$) in the grey-shaded region of figure (a). 
	(c) The values of $u_0$ along a vertical 
	line passing through cylindrical radius 
	$s = 1$ (circles) are compared with the 
	theoretical dimensionless 
	thermal wind in equation \eqref{u02}. 
	The parameters are $Ra_V = 30$  (below onset),
	 $E = 1.2 \times 10^{-5}$, 
	and $Pm = Pr = 1$. 	A $Y_2^1$ heat flux 
	heterogeneity is applied at
	the outer boundary.}
	\label{betalineplot}
\end{figure}

Figure~\ref{betalineplot}(a) shows the azimuthal variation of the
mean temperature gradients 
at cylindrical radius 
$s = 1$ and at the section $z = 0.4$ below the equator
for the $Y_2^1$ heterogeneity with $q^* = 18$. 
 The grey-shaded region indicates $\beta_z > 0$, 
	where the resultant $\beta$ is 
	calculated from \eqref{betadyn}.
Figure~\ref{betalineplot}(b) gives the meridional plot
of the steady mean 
flow $u_0$ ($\phi$-component of the velocity) averaged over the unstably
stratified longitudes in figure
\ref{betalineplot}(a) ($ \phi= 1.31 \, \pi$ to $0.31 \, \pi$ through
$\phi=0$). In figure \ref{betalineplot}(c),
the magnitude of this equatorially anti-symmetric
flow measured
 at $s=1$ shows a fair agreement with the theoretical
value of the thermal wind
calculated from equation \eqref{u0} in dimensionless units, 
\begin{equation}\label{u02}
	u_0 = -\frac{g \alpha \beta_z L}{2\varOmega \eta} \,z
	=-Ra_V  \, \beta^* z, 
\end{equation}
where $\beta^* =\beta_z/\beta_s$ is obtained from the
respective averaged temperature gradients.
\subsection{The role of slow MAC waves 
	in the dipole--multipole transition}
	\label{macwaves}
In \S \ref{linear}, 
we examined the evolution of an isolated 
density perturbation in an 
unstably stratified fluid subject to background 
rotation, a uniform magnetic field and a horizontal
temperature gradient. 
Perturbations of this kind
excite MAC waves in the dynamo, 
the frequencies of which depend on the 
fundamental frequencies given below:
\begin{equation} 
\omega_{C}^2 = \frac{4(\bm{\varOmega} 
	\cdot \bm{k})^2}{ k^2},\quad
\omega_{M}^2 =\frac{(\bm{B} \cdot \bm{k})^2}{\mu \rho},
\quad	
-\omega_{A,V}^2= g\alpha\beta_s \, \frac{ k_{h}^2}{k^2},
\quad
\omega_{A,H}^2= g\alpha\beta_z
\frac{ k_{h}^2}{k^2}\frac{k_s}{k_z}
\label{om1}
\end{equation}
and scaling the frequencies by $\eta/L^2$,
we obtain in dimensionless units,
\begin{equation}
\omega_C^2 = \frac{Pm^2}{E^2} \frac{k_z^2}{k^2},\quad
\omega_M^2 = \frac{Pm}{E} (\bm{B} \cdot {\bm k})^2, \quad 
-\omega_{A,V}^2 = \frac{Pm Ra_V}{E} 
\,\frac{{k_h}^2}{k^2}, \quad
\omega_{A,H}^2 = \frac{Pm Ra_H}{2 \,E} 
\,\frac{{k_h}^2}{k^2}, \quad
\label{om2}
\end{equation}	
where $k_s$, $k_{\phi}$ and $k_z$ are the radial, 
azimuthal and
axial wavenumbers in cylindrical
coordinates $(s,\phi,z)$, $k_\phi=m/s$, 
where $m$ is the
spherical harmonic order,
$k^2=k_s^2+k_\phi^2+k_z^2$, and $k_h$ is the 
horizontal wavenumber 
in the equatorial region defined by 
$k_h^2=k_\phi^2+k_z^2$.
The $s$, $\phi$ and $z$ wavenumbers are calculated
in the saturated state of the dynamo runs in the 
energy-containing scales ($l\le l_E$).
For example, real space integration over $(s,\phi)$
gives the kinetic energy as a function of $z$, the Fourier
transform of which gives the one-dimensional spectrum
$\hat{u}^2 (k_z)$. Subsequently, we obtain,
\begin{equation}
	\bar{k}_z = \frac{\Sigma k_z \, \hat{u}^2 (k_z)}
	{\Sigma \hat{u}^2 (k_z)}.
	\label{kzmean}
\end{equation}
A similar approach yields $\bar{k}_s$ and $\overline{m}$. 
%
The magnetic (Alfv\'en) wave frequency 
$\omega_M$ is based on the three components of the
 measured magnetic
field at the peak-field location. 
The wavenumber $k_\phi$ is evaluated at $s=1$, 
approximately mid-radius of the spherical shell.
The horizontal Rayleigh number is given by
\begin{equation}
Ra_H=2 \, 
\frac{g \alpha \beta_z L^2}{2 \Omega \eta}\frac{k_s}{k_z}.
\label{ravrah}
\end{equation}
The local vertical and horizontal 
Rayleigh numbers in the dynamo based on the flow length scale
are given by,
\begin{equation}
Ra_{\ell,V}=\frac{g \alpha |\beta_s|}{2 \Omega \eta} 
\left(\frac{\pi}{k_s}\right)^2,\,  
Ra_{\ell,H}=2 \, \frac{g \alpha |\beta_z|}{2 \Omega \eta}
\frac{k_s}{k_z} 
\left(\frac{\pi}{k_s}\right)^2,
\label{ravrah2}
\end{equation}
and the resultant local Rayleigh number is then 
\begin{equation}
Ra_{\ell}=Ra_{\ell,V} +Ra_{\ell,H}.
\label{raresultant}
\end{equation}
\begin{figure}
\centering
\includegraphics[width=1\linewidth]{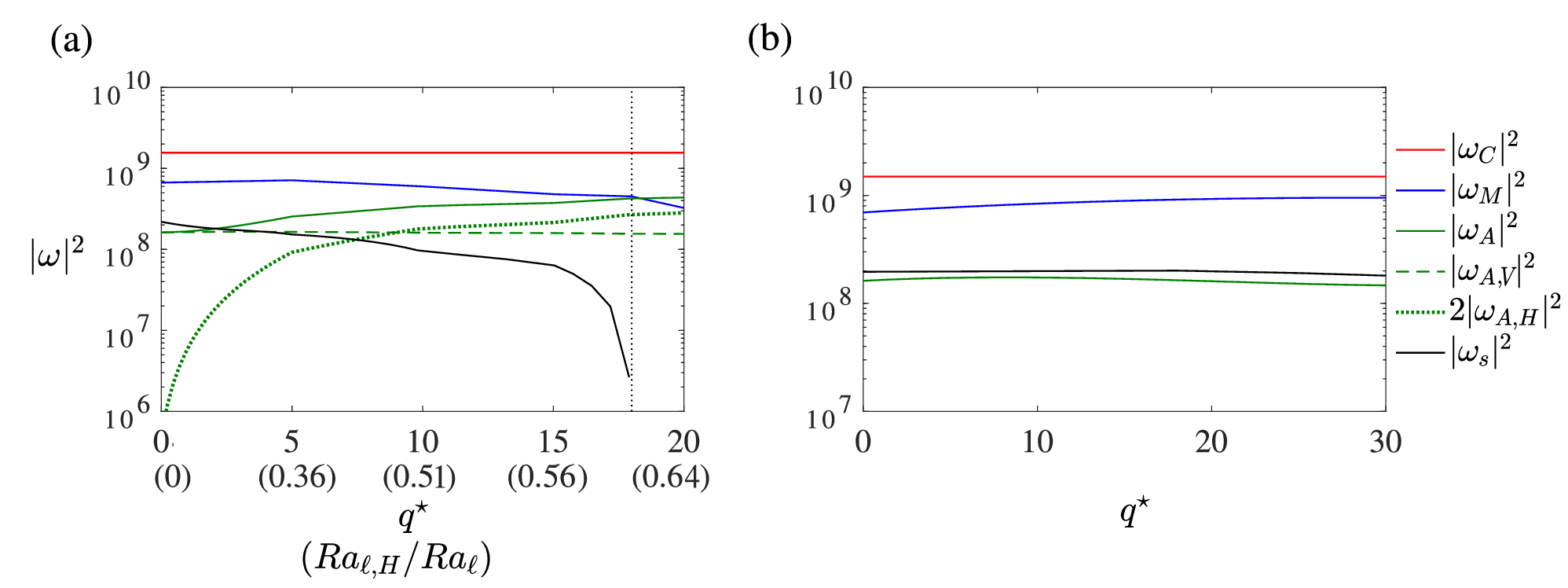}
\caption{Variation of the squares of the fundamental 
	frequencies with $q^*$ and $Ra_{\ell,H}/Ra_\ell$
	(within brackets) 
	for (a) $Ra_V=2500$, $Y_2^1$ heat flux heterogeneity,
	and (b) $Ra_V=2500$, $Y_2^2$ heat flux heterogeneity.
	The dotted vertical line marks the polarity-reversing
	state that
	lies between  the dipolar and multipolar regimes. 
	The other dynamo parameters are 
	$E = 1.2 \times 10^{-5}$, $Pm=Pr=1$. }
\label{ra2500freq}
\end{figure}
\begin{figure}
	\centering
\includegraphics[width=0.8\linewidth]{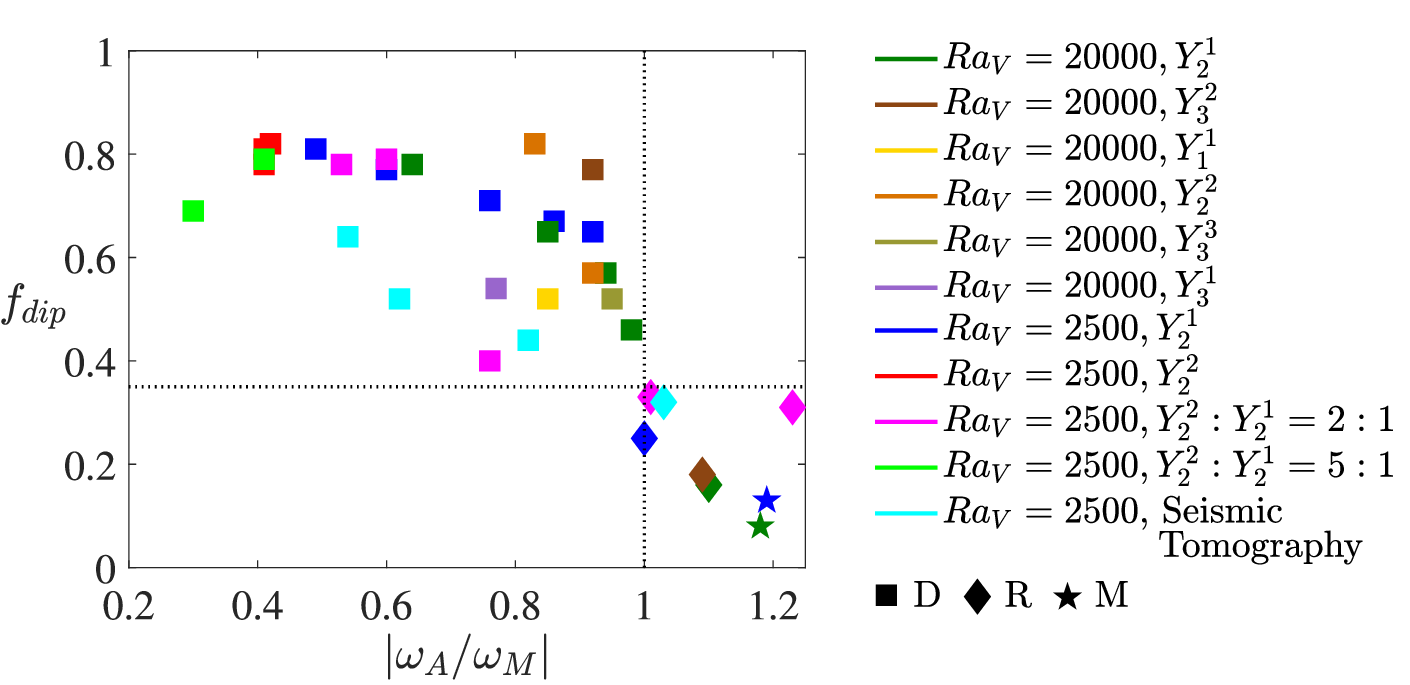}
\caption{Variation of $f_{dip}$ with 
	$|\omega_{A}/\omega_{M}|$ for different heat 
	flux boundary conditions at two values of $Ra_V$. 
	The vertical dotted line indicates 
	$|\omega_{A}/\omega_{M}| = 1$, while 
	the horizontal dotted line marks 
	$f_{dip} = 0.35$, which has been 
	proposed as the lower bound for the 
	existence of dipole-dominated numerical 
	dynamos \citep{chraub2006}. 
	Types `D', `R', and `M' denote dipolar, 
	reversing and multipolar
	dynamo states, respectively.
	The other dynamo parameters are 
	$E = 1.2 \times 10^{-5}$ and $Pm = Pr = 1$.}
\label{fdipwawm}
\end{figure}

In figure \ref{ra2500freq}, the magnitudes of 
fundamental frequencies are shown as a function 
of $q^*$ and $Ra_{\ell, H}/Ra_\ell$ for 
heterogeneous heat flux boundary conditions.
The frequencies and local Rayleigh numbers
 are computed from \eqref{om2},
  \eqref{ravrah2} and \eqref{raresultant} using
  the mean values of the wavenumbers as shown in equation \eqref{kzmean}.
In Figure \ref{ra2500freq}(a), we begin with a 
strong field dynamo and increase the $Y_2^1$ 
heat flux heterogeneity, which is reflected
in the  horizontal 
buoyancy frequency $\omega_{A,H}$, while keeping the vertical 
buoyancy constant at $Ra_V=2500$. The slow MAC waves are 
present at relatively low $q^*$ when 
$|\omega_{M}| > |\omega_A|$. 
In the polarity-reversing state at $q^* = 18$ 
($Ra_{\ell, {H}}/Ra_\ell \approx 0.62$), the magnitude
of the resultant buoyancy frequency $|\omega_A|$ 
approximately matches that of the Alfv\'en frequency 
$|\omega_{M}|$, causing the slow wave frequency
$\omega_{s}$ to go to zero.
While the 
generation of slow MAC waves ceases,
the existing slow MAC waves are damped on the 
time scale $\delta^2 / \eta$ (see \S \ref{compl1}),
which is much shorter 
than the magnetic diffusion time $L^2 / \eta$ 
because the ratio of the core depth to the length scale 
of buoyancy disturbances
would be $L / \delta \sim 10^2$ \citep{jfm21}.
Whether the rapid damping of
the slow MAC waves causes the
rapid decay of the axial dipole field
during a polarity reversal (see figure~\ref{ra2500tilt} b) 
is at present an open question that requires
 further investigation.
The equatorially symmetric $Y_2^2$ heat flux boundary 
condition has zero horizontal buoyancy at the equator, and even at 
large $q^*$, the condition 
$|\omega_A| > |\omega_{M}|$ is never met 
(figure \ref{ra2500freq}(b)).  

The 
variation of the parameter $f_{dip}$ against
$|\omega_A/\omega_M|$ in figure \ref{fdipwawm} shows
that reversing (R) and multipolar (M) states exist at values
of $f_{dip} \le 0.35$, the proposed lower bound for
the existence of dipole-dominated numerical dynamos
in the literature.
\begin{figure}
\centering
\includegraphics[width=1\linewidth]{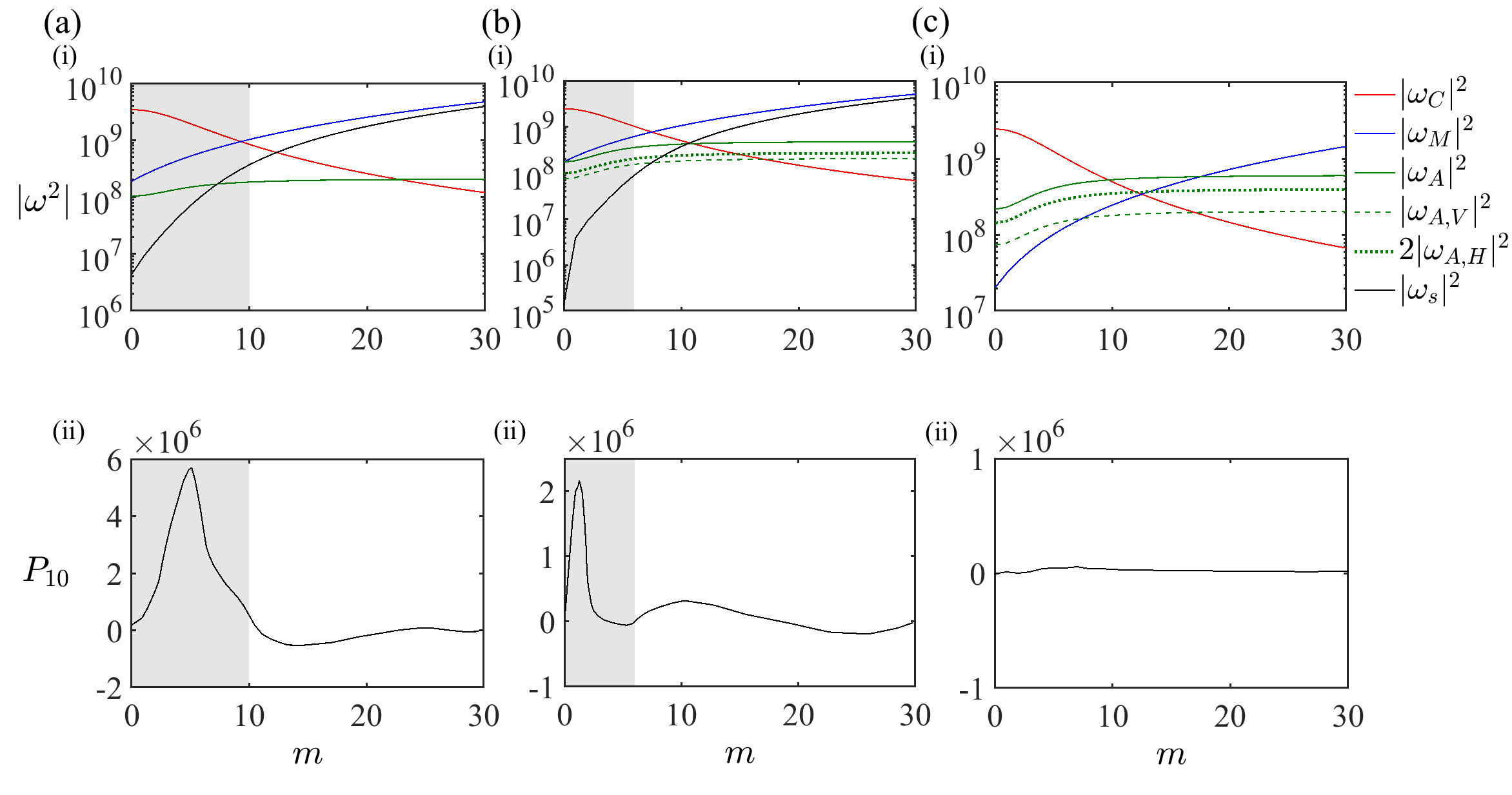}
\caption{Panels  (a)--(c) (i) show the absolute 
values of wave frequencies plotted for 
the saturated state 
of the dynamo run. (a) $q^*= 0$, (b) 
$q^*= 10$ and (c)  $q^*= 18$  for the $Y_2^1$ heat flux 
heterogeneity at the outer boundary.  The shaded grey area shows 
the range of scales where the helicity of 
the dynamo run is greater than that of the 
equivalent non-magnetic run.  Panels (a)--(c) (ii)
 show the spectral 
	distribution of the power supplied to 
	the axial dipole, defined in \eqref{ps10}.
The other dynamo parameters are $Ra_V=2500$,
$E = 1.2 \times 10^{-5}$, $Pm=Pr=1$. }
\label{ra2500spec}
\end{figure}

The top panel of figure \ref{ra2500spec} displays the 
square of the fundamental frequencies for a range of 
spherical harmonic order $m$ for (a) $q^*=0$, 
(b) $q^*=10$, and (c) $q^*=18$ in the saturated state
for the equatorially anti-symmetric $Y_2^1$ heat flux pattern.
Dynamo simulations with $q^* = 0$ 
and $q^* = 10$ exhibit stable axial dipoles 
while the simulation with $q^* = 18$ gives a reversing dynamo.
 The shaded area indicates the range of $m$ where the 
kinetic helicity in the dynamo 
exceeds that of the equivalent 
non-magnetic run.  \cite[For the governing
equations of equivalent 
non-magnetic run, see][]{jfm24}. 
The black line 
represents the square of the slow MAC wave
frequency under the condition 
$|\omega_C| > |\omega_M| > |\omega_A|$. As $q^*$ 
increases, the range of $m$ satisfying this condition narrows, 
and for $q^*=18$, this condition is not 
satisfied for any $m$. 
The bottom panel of figure \ref{ra2500spec} gives the 
spectral distribution of the power supplied to the poloidal 
component of the axial dipole field $B_{10}^P$ 
\citep[e.g.][]{buffett2002energetics} for the three
$q^*$ values,
\begin{equation}
	P_{\mathrm{10}}= \int_{V} \bm{B}^{P}_{\mathrm{10}} 
	\cdot [\mathbf{\nabla} \times (\bm{u} \times \bm{B})_{m}] \, dV,
	\label{ps10}
\end{equation}
where $\bm{u}$ and $\bm{B}$ have the same $m$ \citep{bullard1954}. 
The dipole power peaks at $q^* = 0$ and $q^* = 10$ 
within the range of $m$ where slow MAC waves are 
generated and the helicity of the dynamo  
is greater than that of the equivalent non-magnetic simulation. 
In contrast, the run with
$q^* = 18$ shows no such peak since there
is no helicity generation at any $m$. 
The $f_{dip}$ values for $Ra = 2500$ decrease with 
increasing heterogeneity, with a sharp drop observed at the 
polarity transition for $q^* = 18$ 
(table \ref{tableruns}). These results indicate 
the crucial role of the
slow MAC waves in helicity generation in the energy-containing
scales, and in turn, dipole formation.
\begin{figure}
\centering
\hspace{-2.1 in}	(a)  \hspace{2.1 in} (b) \\
\includegraphics[width=0.45\linewidth]{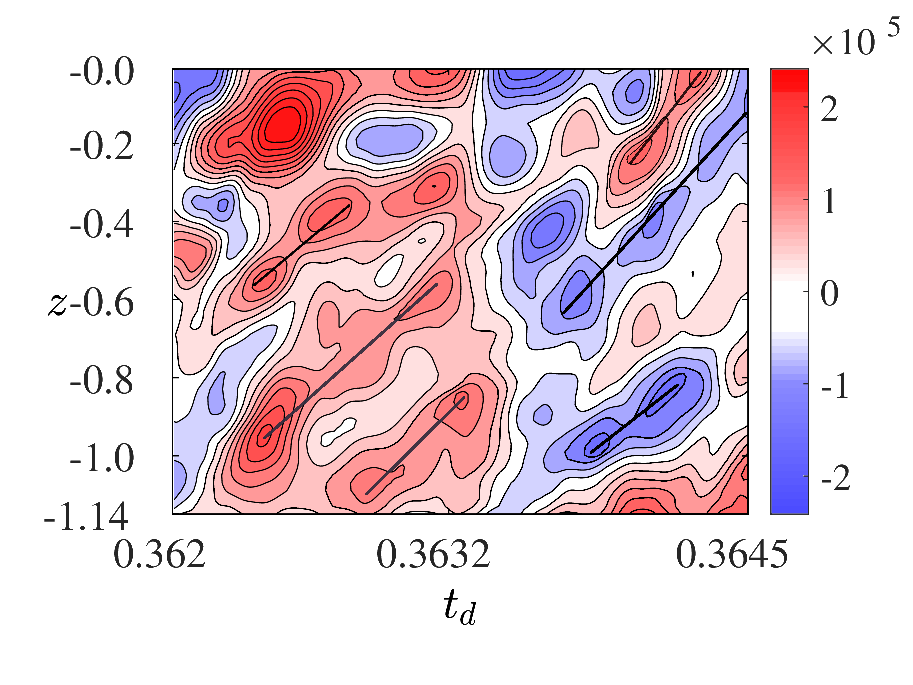}
\includegraphics[width=0.45\linewidth]{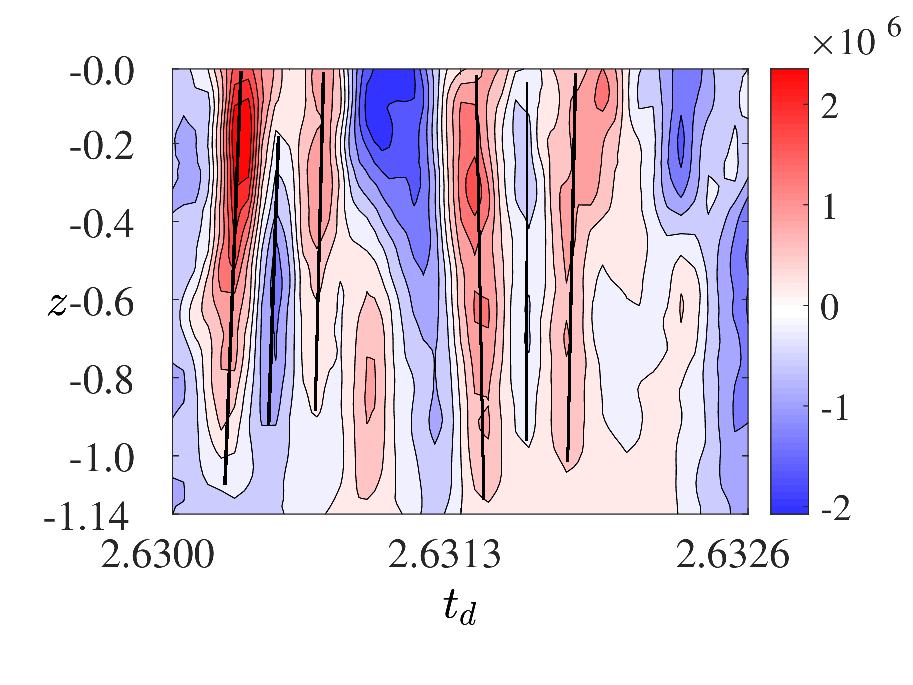}\\
\caption{(a-b) Contour plots of $\partial{u}_z/\partial t$ 
	at a cylindrical radius of $s=1$ are shown for 
	the scales $l \leq l_E$ 
	over short time intervals in the saturated state of 
	two dynamo simulations with 
	 (a) $q^*=17$ and 
	 (b) $q^*=18$
	for the equatorially 
		anti-symmetric $Y_2^1$ heat flux boundary condition. 
	The parallel black lines represent the primary wave 
	travel direction, with their slope giving the 
	group velocity $U_{g,z}$. Table \ref{tablecg} lists 
	the estimated group velocities for the fast and slow 
	MAC waves ($U_f$ and $U_s$, respectively) and $U_{g,z}$. 
	 The other dynamo parameters are 
	$Ra_V=2500, E = 1.2 \times 10^{-5}$, and $Pm=Pr=1$. }
\label{cg2500}
\end{figure}
	
\begingroup
\setlength{\tabcolsep}{2pt} 
\renewcommand{\arraystretch}{1.25} 
\begin{table}   
	\centering
	\begin{tabular}{cccccccccccccc}
		SL& $E$& $Ra_V$&$q^*$& Fig.&$\omega_n^2$&
		$\omega_C^2$&$\omega_M^2$&$-{\omega_A}^2$
		&$\omega_f$&$\omega_s$& $U_f$& $U_s$& $U_{g,z}$\\
		&&&& No.&$(\times10^{10})$& $(\times10^{8})$
		& $(\times10^{8})$& $(\times10^{8})$&$(\times10^{4})$
		&$(\times10^{4})$&&&\\
		1&$1.2 \times 10^{-5}$ &2500 &17 &\ref{cg2500}
		(a) &2.5&13.02&6.59&3.94&4.63&0.90& 8989 &525& 456\\	
		2&$1.2 \times 10^{-5}$ &2500 &18 &\ref{cg2500}
		(b) &2.7&11.51&2.22&4.32&3.46&0.0& 9651 &0& 9274\\
	\end{tabular}
	\caption{Summary of the data 
		for MAC wave measurement in the dynamo models. 
		The sampling frequency $\omega_n$ is selected to 
		ensure that the fast MAC waves are
		captured when measuring group velocity. The 
		values of $\omega_M^2$, $-\omega_{A}^2$, 
		and $\omega_C^2$ are computed 
		using \eqref{om2}, based on the averaged
		wavenumbers 
		$m$, $k_s$, and $k_z$ in the energy-containing 
		scales where $l \leq l_E$. The group velocity in 
		the $z$ direction ($U_{g,z}$) is then compared 
		with the estimated velocities of the fast 
		($U_f$) and slow ($U_s$) MAC waves.}
	\label{tablecg}
\end{table}
\endgroup
Figure \ref{cg2500} shows the measurement 
of wave velocities from their propagation
paths in the saturated state of dynamos at 
$q^* = 17$ and $q^* = 18$ for 
	the equatorially anti-symmetric $Y_2^1$ 
	heat flux pattern at the outer boundary. 
Contours of the 
fluctuating $z$-velocity, represented by 
$\partial u_z / \partial t$ at cylindrical 
radius $s = 1$, are plotted over short time 
intervals where the ambient magnetic field and 
wavenumbers in the energy-containing scales 
$l \le l_E$ remain approximately constant. 
By examining the slope of the black lines, 
we determine the axial group velocity $U_{g,z}$ 
and compare it with the estimated axial 
group velocities of the fast ($U_f$) and slow ($U_s$) waves. 
These velocities are derived from the respective 
frequencies in the diffusionless limit 
by taking their derivative with respect to $k_z$ \citep{aditya2022}. 
In the dipole-dominated run at $q^* = 17$, 
 the slow MAC waves are predominant although the fast MAC waves
also exist
(figure \ref{cg2500}(a)). In contrast, at $q^* = 18$, 
the slow MAC waves are nearly absent while
the fast waves are abundant (figure \ref{cg2500}(b)).  
\subsection{Complementarity of vertical and horizontal 
	buoyancies in the dipole--multipole transition}
\label{complement}
\begin{figure}
\centering
\hspace{-2 in} (a) \hspace{2 in} (b)\\
\includegraphics[width=0.45\linewidth]{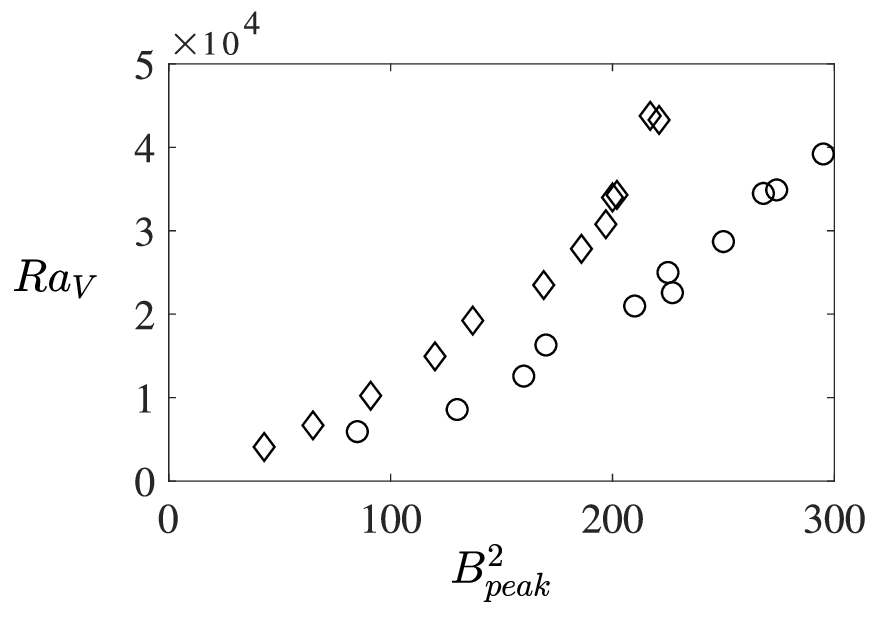}
\includegraphics[width=0.45\linewidth]{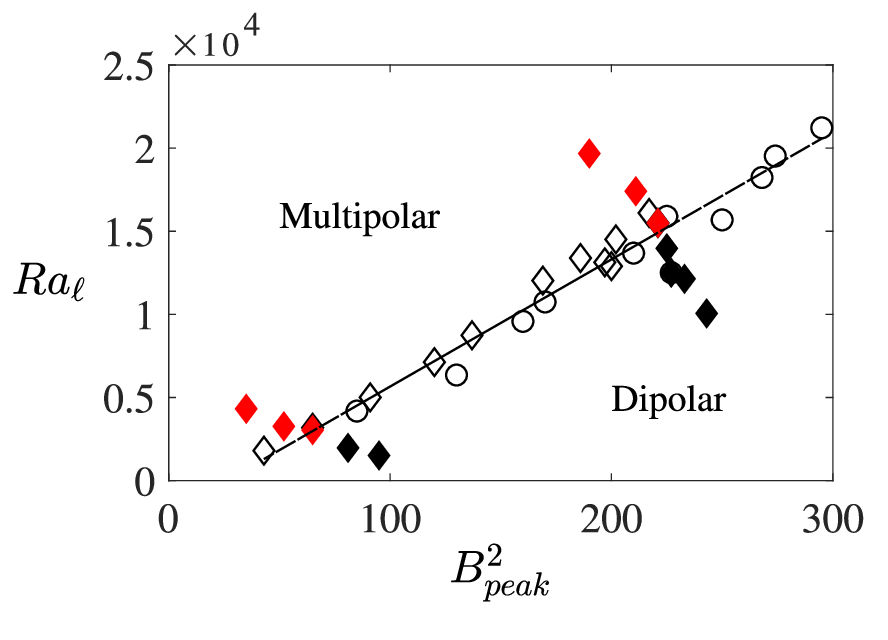}\\
\hspace{-2 in} (c) \hspace{2 in} (d)\\
\includegraphics[width=0.45\linewidth]{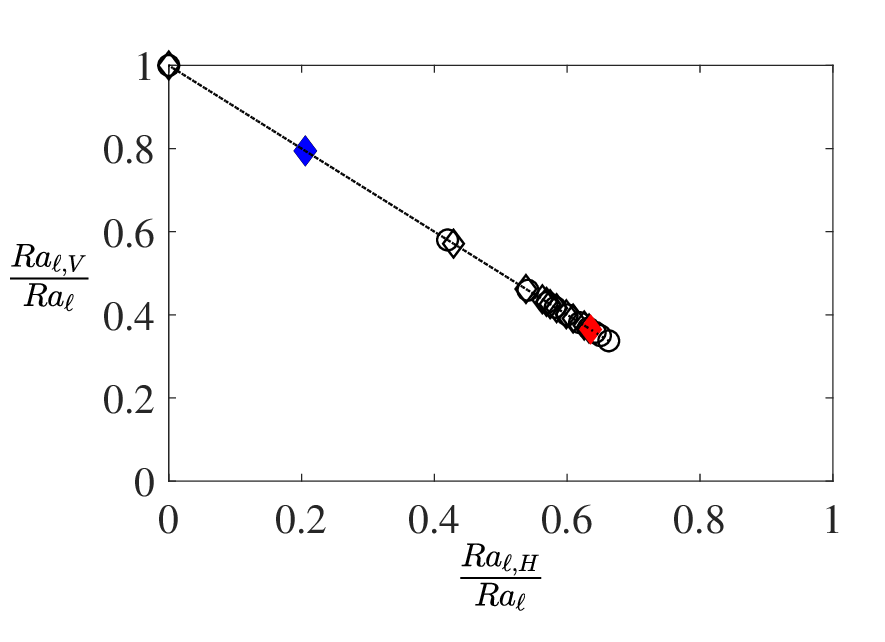}
\includegraphics[width=0.45\linewidth]{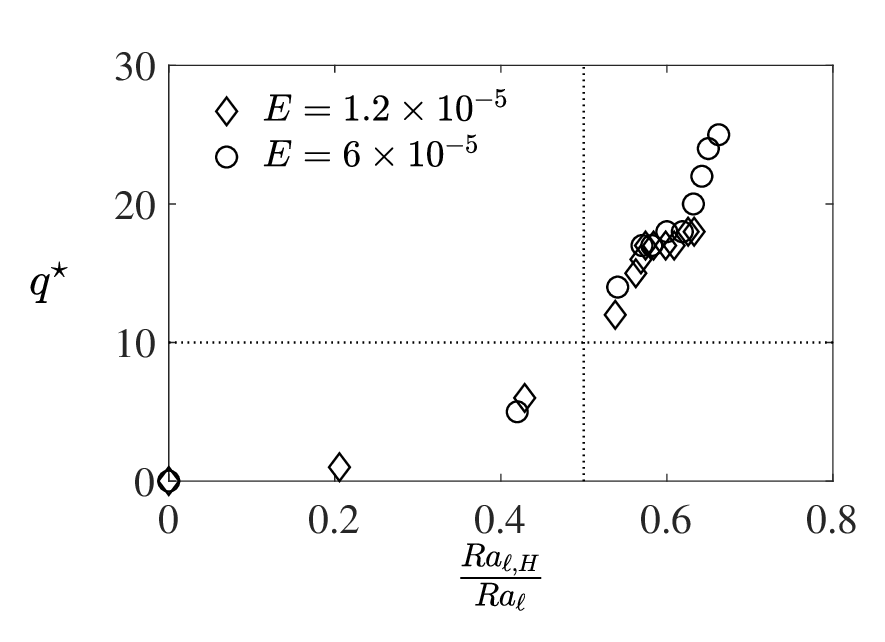}\\
\caption{(a) Variation of the modified vertical Rayleigh number $Ra_V$  
	with the square of the peak magnetic field, $B^2_{peak}$ 
	at the 
	suppression of slow MAC waves. (b) Variation 
	of the local Rayleigh number $Ra_\ell$, defined 
	in \eqref{ravrah2}, with $B^2_{peak}$. 
	The values of $Ra_V$, $Ra_\ell$ and $B^2_{peak}$ 
	in the plots are given in table \ref{tablereversal}. 
	The parameters of the two dynamo series and their 
	symbolic representations are as follows: 
	$E=6\times10^{-5},Pm=Pr=5$ (circles), 
	$E=1.2\times10^{-5},Pm=Pr=1$ (diamonds). 
	The filled symbols show the evolution path 
	of the dynamo with increasing heat flux heterogeneity. 
	Black symbols represent the dipolar state 
	and red symbols represent the reversing or 
	multipolar state. (c) The states of polarity
	reversals are shown in a plot of local relative vertical 
	versus relative horizontal Rayleigh numbers, normalized
	by the resultant local Rayleigh number $Ra_\ell$.
	For $E=1.2 \times 10^{-5}$,
reversals occur at
$Ra_V=27000$ and $Ra_{\ell,H}/Ra_\ell=0.2$ 
(blue diamond) as well as at $Ra_V=1500$ and
 $Ra_{\ell,H}/Ra_\ell=0.64$ (red diamond).
	(d) Variation of $q^*$ with respect to
	$Ra_{\ell,H}/Ra_{\ell}$. The horizontal line 
	at $q^* = 10$ and the vertical line at 
	$Ra_{\ell,H}/Ra_{\ell} = 0.5$ indicate 
	that $Ra_{\ell,H}/Ra_{\ell} > 0.5$
	corresponds to $q^* = O(10)$.
}
\label{ralsimilar}
\end{figure}
		

In the presence of rapid rotation and a magnetic field, 
the buoyant forcing generates
fast and slow MAC waves when 
the inequality 
$ |\omega_{C}| > |\omega_{M}| > |\omega_{A}| > |\omega_{\eta}| $ 
is satisfied. The magnetic field intensity increases 
with the vigour of convection in this dipole-dominated
regime; however, a sufficiently strong 
forcing suppresses the slow MAC waves when 
$ |\omega_{A}| \approx |\omega_{M}| $, leading to 
dipole collapse. 
The nature of buoyancy could be either vertical,
measured by $ Ra_{V} $, or horizontal,
measured by $ Ra_{H} $. 
In a dynamo with homogeneous boundary heat flux, 
the dipolar regime transitions in succession 
to polarity-reversing and multipolar states 
with increasing $Ra_{V}$ \citep{jfm24}.
In this study,  
$ Ra_H $ is progressively increased 
for a fixed $ Ra_{V} $ until the dynamo 
undergoes the polarity transition (table \ref{tableruns}). 
The square of the peak magnetic field
measured at the polarity transition, 
$B^2_{peak}$, is plotted against $Ra_V$ 
for two different Ekman numbers in figure 
\ref{ralsimilar}(a). The approximately linear
variation of $Ra_V$ 
with $B^2_{peak}$ is different in 
the two cases. On the other hand, the resultant local 
Rayleigh number $Ra_\ell$, defined by equation 
\eqref{raresultant}, exhibits a self-similar 
variation with $B^2_{peak}$ 
(figure \ref{ralsimilar}(b)).
%
This self-similar line separates
the dipolar and multipolar regimes.
The filled symbols represent the 
	evolution path of the dynamo
	with increasing order of heat flux heterogeneity for 
	two vertical Rayleigh numbers
	 ($Ra_V=2500$ and $Ra_V=20000$).
The filled black symbols here represent the states where
$ |\omega_{M}| > |\omega_{A}| $ while the filled 
red symbols represent 
$ |\omega_{M}| \lesssim |\omega_{A}| $.
In figure \ref{ralsimilar}(c), all the 
polarity-reversing states
 are plotted in the $ Ra_{\ell,H}$ versus 
 $Ra_{\ell, V}$ space, normalized by
the value of $ Ra_\ell $.
This plot also demarcates the 
dipolar and multipolar regimes and furthermore
demonstrates the complementarity between the 
horizontal and vertical buoyancies in producing
the polarity transition through the condition of
vanishing slow MAC waves, as suggested by figure 
\ref{freqlinear}(e). For $E=1.2 \times 10^{-5}$,
reversals occur in a strongly driven dynamo at
$Ra_V=27000$ and
$Ra_{\ell,H}/Ra_\ell=0.2$ (blue diamond) as well as
a relatively 
weakly driven dynamo at $Ra_V=1500$ and
 $Ra_{\ell,H}/Ra_\ell=0.64$ (red diamond);
 see also table \ref{tablereversal}.
Since the self-generated magnetic field depends 
on the vertical Rayleigh number $Ra_{\ell,V}$, 
only weak-field dynamos of $B^2_{rms} < O(1)$
 exist for $Ra_{\ell,V}/Ra_\ell < 0.25$. 
Therefore, solutions for 
$Ra_{\ell,V}/Ra_\ell < 0.25$ (i.e. $Ra_{\ell,H}/Ra_\ell > 0.75$)
are not considered.
Figure
\ref{ralsimilar}(d) relates 
the two measures of heterogeneity,
$q^*$ and $ Ra_{\ell, H}/Ra_\ell$ and indicates that
values of $ Ra_{\ell, H}/Ra_\ell >0.5$
 correspond to values
of $q^*$ of $O(10)$.

\subsection{Combination of symmetric and 
	anti-symmetric boundary heat flux}
\label{composite}
Since the heat flux heterogeneity in
the lower mantle is a combination of both
equatorially symmetric and anti-symmetric variations,
it is instructive to consider the effect of composite
heat flux boundary conditions on polarity transitions.
The analysis of the heat flux distribution based on
 the seismic shear wave velocity in the Earth's 
lower mantle \citep{masters1996} suggests a 
complex heat flux pattern at the CMB which 
features a well-defined $Y_2^2$ component. That said, the 
analysis of peak-to-peak heat flux variations 
reveals that the ratio of the symmetric to 
anti-symmetric heat flux variation is approximately 
1.65.
%
This ratio is obtained by calculating the 
peak-to-peak variation in heat flux for the 
symmetric and anti-symmetric contributions separately, 
within the seismic tomography pattern.
With a composite heat flux heterogeneity at the outer boundary,
a dipole-dominated solution is observed when the inequality 
$|\omega_{C}| > |\omega_{M}| > |\omega_{A}|$ is satisfied 
in the presence of the slow MAC waves. 
For a given vertical buoyant forcing ($Ra_V \sim 10^2$ times
its value for nonmagnetic onset of convection), the progressive
increase of horizontal buoyancy produces polarity transitions
for composite patterns where the equatorially
symmetric and anti-symmetric variations are comparable.
For example, a heat flux pattern with $Y_2^2:Y_2^1 = 2:1$ 
and the Earth-like 
heat flux pattern derived from seismic tomography both
produce reversals for $q^*$ of $O(10)$ 
(table \ref{tableruns}). For the same
$Ra_V$, the values of $Ra_{\ell,H}/Ra_\ell$ at which the transition
occurs in the two cases are nearly equal, which points to
the comparable ratios of the symmetric to 
anti-symmetric heat flux variation.
Notably, a heterogeneity consisting of
comparable magnitudes of symmetric and anti-symmetric
variations induces the polarity transition at a value
of $Ra_{\ell,H}/Ra_\ell$ (and $q^*$) which is of the same order
as that for the transition induced by a purely anti-symmetric variation.

The polarity transition with the boundary heat flux
derived from seismic tomography is analysed further
in figure \ref{basicstate1} of
Appendix \ref{tomogrev}. In the dipole-dominated
state at $q^*=10$, $|\omega_M|^2 >
|\omega_A|^2$ in several regions, where $\omega_A^2$
is based on the resultant basic state
temperature gradient $\beta$, defined in \eqref{betadyn}. 
In these regions, the Alfv\'en wave
velocity is greater than the velocity of the buoyancy
perturbations and the slow MAC waves
are generated. On the other hand,
the polarity-reversing state
at $q^*=13$ shows approximate parity between 
 $|\omega_M|^2$ and $|\omega_A|^2$ so that the slow
 waves are suppressed. This analysis also explains why
a composite boundary heterogeneity made up of
a dominant equatorially symmetric variation
does not induce polarity transitions.
For example, the heterogeneity with the ratio 
$Y_2^2:Y_2^1 = 5:1$ does not admit
polarity transitions
 even at high $q^*$ (table \ref{tableruns})
 since the condition 
$|\omega_{A}|^2 \gtrsim |\omega_{M}|^2$ is not met.
\section{Two-component magnetoconvection}
\label{twocomp}
Convection in the cores of planets like Earth are driven by
compositional and thermal buoyancy. The compositional part
in Earth,
arising from the progressive growth of the inner core, is dominant.
 With the knowledge
of the peak magnetic field intensity, a useful constraint on the
lateral variation in heat flux can be obtained from an analysis
of the evolution of an isolated disturbance under rapid rotation
and both compositional and thermal buoyancy.
%
\subsection{Evolution of a density disturbance in two-component 
magnetoconvection}
\label{twoclinear}
In addition to the variables $\bm{u}$, $\bm{b}$ and $\Theta$
considered in \eqref{vars}(a--c), the composition $C$ is also
decomposed into its mean and perturbation parts, $C= C_0+\gamma$.
Here, the lengths are scaled by the perturbation size $\delta$ 
and time is scaled by magnetic diffusion time 
$\delta^2/\eta$.
The velocity $\bm{u}$ and magnetic field 
$\bm{B}$ are scaled by $\eta/\delta$ and $(2\Omega\rho\mu\eta)^{1/2}$,
respectively. 
The temperature is scaled by $\beta_y^T \delta$ and composition is 
scaled by $\beta_y^C \delta$, where $\beta_y^T$ and $\beta_y^C$ are the 
vertical temperature and composition gradients, respectively.
The dimensionless equations for $\bm{u}$, $\bm{b}$, $\Theta$ and $\gamma$
are given by,
\begin{align}
		E_\eta \bigg(\frac{\partial \bm{u}}{\partial t}
		+u_c\, z \frac{\partial \bm{u}} {\partial x}
		+u_c u_z \hat{\bm{e}}_x\bigg)
		+\hat{\bm{e}}_z\times\bm{u}=
		-\nabla p^\star+ 
		(\nabla\times \bm{b}) \times \bm{B}_0 \nonumber\\
		+Ra_{\ell,V}^T \Theta \hat{\bm{e}}_y 
		+ Ra_{\ell,V}^C \gamma \hat{\bm{e}}_y+E \nabla^2 \bm{u},
		\label{ddmom}\\
 \frac{\partial \bm{b}}{\partial t} + 
u_c \,z \frac{\partial \bm{b}}{\partial x} = 
u_c b_z\hat{\bm{e}}_x
+(\bm{B}_0\cdot\nabla)\bm{u}+\eta \nabla^2 \bm{b},\label{ddind}\\
		 \frac{\partial \Theta}{\partial t}
		+u_c z \frac{\partial\Theta}{\partial x}
		+u_y+u_z\beta^\star=q^T\nabla^2 \Theta,
		\label{ddtemp}\\
		 \frac{\partial \gamma}{\partial t}+u_c z \frac{\partial\gamma}{\partial x}
		+u_y=q^C \nabla^2 \gamma, \label{ddcomp}\\
		 \nabla\cdot\bm{u}=\nabla\cdot\bm{b}=0.
		 \label{dddiv}
	\end{align}

The dimensionless parameters in \eqref{ddmom}, based on
the length scale of the density disturbance, are
the Ekman number $E=\nu/2\varOmega \delta^2$, 
magnetic Ekman number $E_\eta=\eta/2\varOmega \delta^2$, 
and local vertical compositional and thermal  Rayleigh numbers,
$Ra_{\ell,V}^C = g \alpha^C |\beta_{y}^C| \delta^2/2 \varOmega \eta$ and 
$ Ra_{\ell,V}^T = g \alpha^T |\beta_{y}^T| \delta^2/2 \varOmega \eta$, respectively.
Here, $\alpha^T$ and $\alpha^C$ are the thermal and 
compositional expansion coefficients, respectively.
Additionally, $q^T=\kappa^T/\eta$, $q^C=\kappa^C/\eta$ and 
$\beta^\star=\beta_{z}/\beta_{y}^C$, where $\kappa_T$ and
$\kappa_C$ are the diffusivities of temperature and
composition and $\beta_z$ is the horizontal
temperature gradient.

For the rapidly rotating system considered in \S \ref{linear},
the limit $\nu, \kappa_C, \kappa_T \ll \eta$ results in 
$E, q^C, q^T \to 0$ while $E_\eta \ll 1$ retains a 
small but finite value. In this limit, a plane
wave solution of the form 
$\hat{u}_z \sim \mbox{e}^{\mathrm{i} \lambda t}$ 
gives the  characteristic equation,
\begin{equation}
	\begin{aligned}
		\lambda^5&- 2 \mathrm{i}\omega_\eta \lambda^4
		- (\omega_{C}^2+2 \omega_{M}^2+{\omega^C_{A,V}}^2
		+{\omega^T_{A,V}}^2-2{\omega_{A,H}^2}+\omega_{\eta}^2) \lambda^3\\&
		+2 \mathrm{i} \omega_\eta(\omega_C^2+\omega_M^2 
		+{\omega^C_{A,V}}^2+{\omega^T_{A,V}}^2-2{\omega_{A,H}^2}) \lambda^2\\&
		+\big(\omega_C^2\omega_\eta^2+\omega_M^4+({\omega^C_{A,V}}^2
		+{\omega^T_{A,V}}^2-2{\omega_{A,H}^2})(\omega_{M}^2
		+\omega_{\eta}^2)\big) \lambda\\&
		-\mathrm{i} \omega_\eta\omega_M^2 ({\omega^C_{A,V}}^2
		+{\omega^T_{A,V}}^2-{\omega_{A,H}^2})=0,
		\label{ddcmbchar}
	\end{aligned}
\end{equation}
where the dimensionless fundamental frequencies are,
\begin{subequations}
\begin{gather} \label{freqddcmb}
	\begin{aligned}
		&\omega_{C}^2=\frac{1}{E_\eta^2}\frac{k_z^2}{k^2},\quad 
		\omega_{M}^2=\frac{(\bm{B}\cdot \bm{k})^2}{E_\eta}, \quad 
		{\omega_{A,V}^T}^2=\frac{Ra_{\ell,V}^T}{E_\eta}  
		\frac{k_z^2}{k^2},\quad  
		{\omega_{A,V}^C}^2=\frac{Ra_{\ell,V}^C}{E_\eta}  
		\frac{k_z^2}{k^2}, \quad \\ 
		&\omega_{A,H}^2=\frac{Ra_{\ell,H}}{2 E_\eta} 
		\frac{k_z^2}{k^2}, \quad 
		\omega_{\eta}^2=k^4.
	\end{aligned}
	\tag{\theequation a--f}
\end{gather}
\end{subequations}
The local horizontal Rayleigh number
$Ra_{\ell,H}$ has the same definition as in
\eqref{radelta}. 
For the inequality $|\omega_C| \gg |\omega_M| \gg |\omega_{A,V}^C|, \, 
|\omega_{A,V}^T|, \, |\omega_{A,H}| \gg |\omega_\eta|$, 
the roots of the characteristic equation \eqref{ddcmbchar}
are
approximated by,
\begin{eqnarray}
	\lambda_{1,2} &\approx& \pm \bigg(\omega_C
	+ \frac{\omega_M^2}{\omega_C} \bigg) 
	+\mathrm{i} \, \frac{\omega_M^2\omega_\eta}{\omega_C^2}, 
	\label{l12}\\ 
	\lambda_{3,4} &\approx& \pm \bigg(\frac{\omega_M^2}{\omega_C}+
	\frac{{\omega_{A,V}^C}^2+{\omega_{A,V}^T}^2
		-2\omega_{A,H}^2}{2\omega_C}\bigg) 
	+ \mathrm{i} \, \omega_\eta \,
	\bigg(1-\frac{{\omega_{A,V}^C}^2+{\omega_{A,V}^T}^2
		-\omega_{A,H}^2}{2\omega_M^2}\bigg),
	\label{l34} \\
	\lambda_{5} &\approx& \mathrm{i}\,\omega_{\eta} 
	\frac{{\omega_{A,V}^C}^2+{\omega_{A,V}^T}^2-\omega_{A,H}^2}{\omega_M^2},
	\label{l5}
\end{eqnarray}
Here, $\lambda_{1,2}$ and $\lambda_{3,4}$ give the 
approximate frequencies of the 
oppositely travelling
fast ($\omega_f$) and slow ($\omega_s$) MAC waves, respectively. 
From \eqref{l34}, the resultant buoyancy frequency is given by
\begin{equation} 
	\omega_{A}^2={\omega_{A,V}^C}^2+{\omega_{A,V}^T}^2-2\omega_{A,H}^2,
	\label{rfreqdd}
\end{equation}
which indicates the complementarity between the vertical buoyancies
of composition and temperature and the horizontal
buoyancy of temperature. The resultant local 
Rayleigh number is then given by
\begin{equation}
Ra_\ell =\frac{g \big|\alpha^C \beta_y^C+\alpha^T \beta_y^T-2 
	\alpha^T \beta_z k_y/k_z\big|\delta^2}{2 \varOmega \eta},
\label{raldd}
\end{equation}

%
%
The solutions for $\hat{u}_z$, $\hat{\Theta}$ and $\hat{\gamma}$ are,
\begin{equation}
	\big[\hat{u}_z,\hat{\Theta},\hat{\gamma}\big] = 
	\sum_{m=1}^{5}\big[D_{m},G_{m},Q_{m}\big] 
	\mbox{e}^{\mathrm{i} \lambda_{m}t},
	\label{waveddcmb}
\end{equation}
where the coefficients $D_{m}$, $G_{m}$ and $Q_{m}$ are evaluated 
from the initial conditions for
$\hat{u}_z$, $\hat{\Theta}$, $\hat{\gamma}$ and their 
time derivatives.
For instance, 
the initial conditions 
for $\hat{u}_z$ and its time derivatives are given by
\begin{equation}
	\begin{aligned}
		\mathrm{i}^n \sum_{m=1}^{5}D_m \lambda_{m}^n
		=\bigg(\frac{\partial^n\hat{u}_z}{\partial t^n}\bigg)_{t=0}
		=d_{n+1}, \ n=0,1,2,3,4,
	\end{aligned}
	\label{uzcoff1}
\end{equation}
where
\begin{equation}
	\begin{aligned}
		d_1&=\hat{u}_z\big|_{t=0}=0,\\
		d_2&=\frac{\partial{\hat{u}_z}}{\partial{t}}\big|_{t=0}=
		\bigg({\omega_{A,V}^C}^2 \hat{\Theta}_0+{\omega_{A,V}^T}^2 
		\hat{\gamma}_0\bigg)\frac{k_y}{k_z},\\
		d_3&=\frac{\partial^2 \hat{u}_z}{\partial{t^2}}\big|_{t=0}=0,\\
		d_4&=\frac{\partial^3{\hat{u}_z}}{\partial{t^3}}\big|_{t=0}
		=-d_2 (\omega_M^2+\omega_C^2+\omega_{A}^2),\\
		d_5&=\frac{\partial^4{\hat{u}_z}}{\partial{t^4}}\big|_{t=0}
		=d_2 \omega_{\eta}\omega_{M}^2,
		\label{}
	\end{aligned}
\end{equation}
where $\hat{\Theta}_0$ and $\hat{\gamma}_0$
are the initial perturbations of temperature and composition
respectively, both of which have the Gaussian distribution
as in \eqref{pert} in Cartesian coordinates.
The coefficients of the fast and slow wave parts of $\hat{u}_z$ 
are then obtained from the roots of \eqref{ddcmbchar},
as follows:
\begin{eqnarray}
	D_{1}&=\dfrac{d_{5}- \mathrm{i} \, d_{4}(\lambda_2
		+\lambda_3+\lambda_4+\lambda_5)
		+\mathrm{i} \, d_{2}(\lambda_2\lambda_4\lambda_5
		+\lambda_3\lambda_4\lambda_5
		+\lambda_2\lambda_3\lambda_4
		+\lambda_2\lambda_3\lambda_5)}{(\lambda_1-\lambda_2)
		(\lambda_1-\lambda_3)(\lambda_1
		-\lambda_4)(\lambda_1-\lambda_5)},
	\label{d1coeff2c}\\
	D_{3}&=\dfrac{d_{5}- \mathrm{i} \, d_{4}(\lambda_1
		+\lambda_2+\lambda_4+\lambda_5)
		+ \mathrm{i} \,d_{2}(\lambda_1\lambda_4\lambda_5
		+\lambda_2\lambda_4\lambda_5
		+\lambda_1\lambda_2\lambda_4+
		\lambda_1\lambda_2\lambda_5)}{(\lambda_3-\lambda_1)
		(\lambda_3-\lambda_2)(\lambda_3
		-\lambda_4)(\lambda_3-\lambda_5)},
	\label{d3coff2c}
\end{eqnarray}
Further, $\hat{u}_y$ is calculated from equation \eqref{uy}.
The spectral 
	coefficients of $\hat{\Theta}$
	and $\hat{\gamma}$ are 
	obtained in a similar 
	way (Appendix \ref{ddspcoeffs}).

In two-component convection, both thermal and compositional buoyancies 
drive the convection and are responsible for magnetic field generation. 
The fraction of thermal power relative to the total 
convective power is given by,
\begin{equation}
	f^T =\frac{P^T}{P^T+P^C}\times100 \%,
	\label{pratio}
\end{equation}
where $P^T$ and $P^C$ are the dimensionless thermal and 
compositional powers respectively, defined by

	\begin{equation}
		P^T=
		Ra_{\ell,V}^T \int_{-\infty}^{\infty}
		\int_{-\infty}^{\infty}  \bm{u} \cdot  
		\Theta \hat{\bm{e}}_y \, \mbox{d}y\mbox{d}z,\quad 
		P^C=
		Ra_{\ell,V}^C \int_{-\infty}^{\infty}
		\int_{-\infty}^{\infty}\bm{u}\cdot  
		\gamma \hat{\bm{e}}_y \, \mbox{d}y\mbox{d}z.
		\label{power}
	\end{equation}

The integrals in \eqref{power} are computed at $x=0$
for the limits $\pm20 $ in $(y,z)$
(The integrals
of $u_y \Delta T_0$ and $u_y \Delta C_0$ vanish).


\subsection{Complementarity of vertical and horizontal 
	buoyancies in two-component convection}
\label{complementarity2comp}

\begin{figure}
	\centering
	\includegraphics[width=0.6\linewidth]{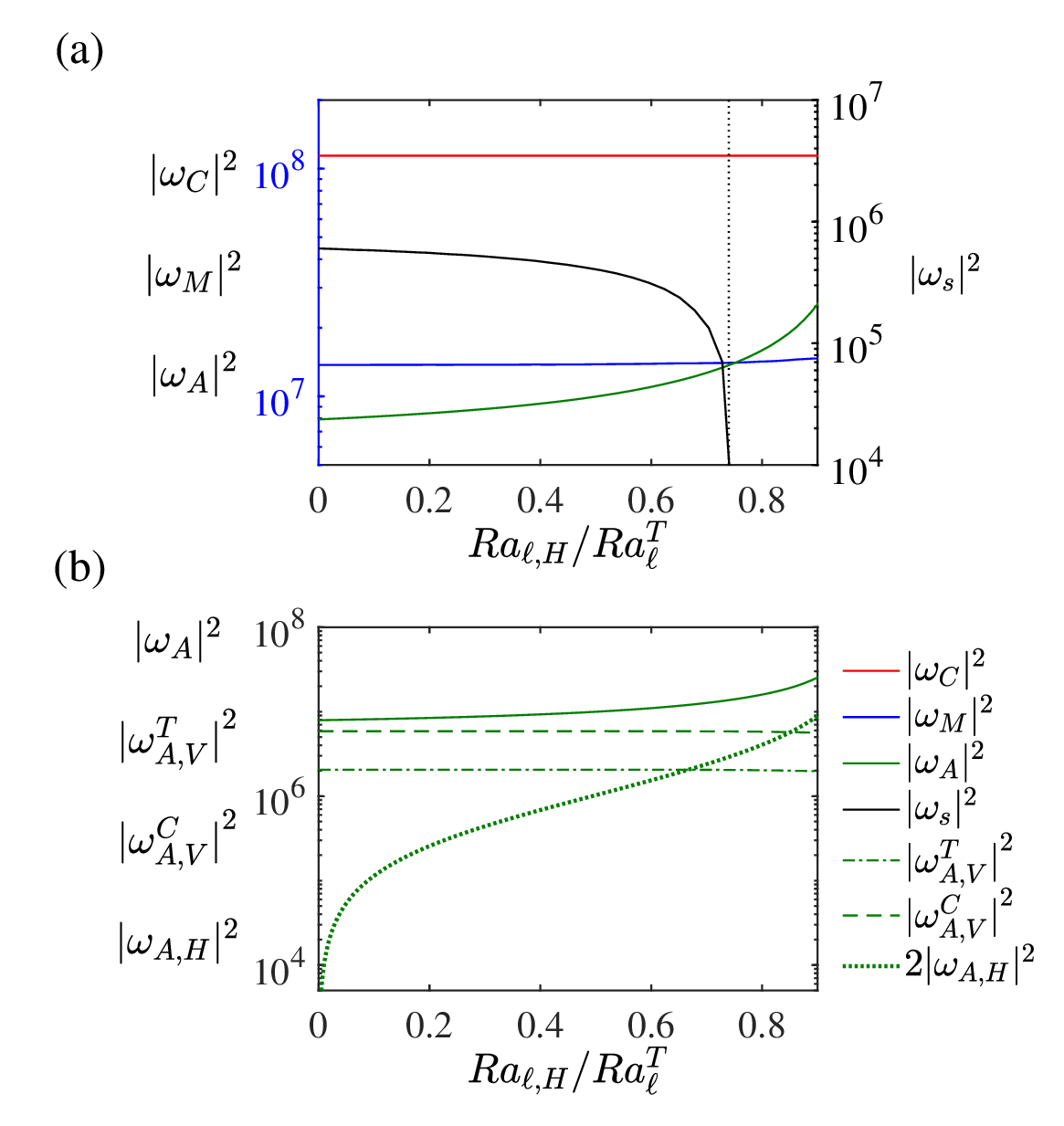}
	\caption{Variation of the squares of frequencies with
		relative horizontal buoyancy in two-component
		magnetoconvection. The dotted vertical
		line in (a) corresponds to $|\omega_A| \approx |\omega_M|$, when
		the slow wave frequency $\omega_s$ goes to zero.
		Panel (b) gives the
		decomposition of $\omega_A^2$ into its three parts consisting
		of the vertical buoyancies of composition and temperature
		(${\omega^C_{A,V}}^2$, ${\omega^T_{A,V}}^2$) and
		the horizontal buoyancy of temperature ($\omega_{A,H}^2$).
		The parameters used are $E_\eta = 2 \times 10^{-5}$,
		$B^2_{peak}=200$ and $f^T= 10$\% at time $t/t_\eta= 10^{-2}$.}
	\label{freqdd}
\end{figure}
In \S \ref{compl1}, the complementarity between 
vertical and horizontal 
buoyancies at the state of vanishing slow MAC
waves was noted. In two-component convection, 
a complementarity exists between
the vertical buoyancies of composition and 
temperature and the horizontal buoyancy of temperature in 
suppressing the slow MAC waves (figure \ref{freqdd}). 
Figure \ref{freqdd}(a) shows the variation of the squares 
of the fundamental frequencies and the slow MAC wave 
frequency with relative horizontal buoyancy, 
whereas figure \ref{freqdd}(b) presents the decomposition 
of $\omega_{A}^2$ into its three constituent parts. 
While the compositional Rayleigh number $Ra^C_{\ell,V}$ is held
constant (see below), the thermal Rayleigh number is
determined by the power ratio $f^T$, given by \eqref{pratio}.
In figure \ref{freqdd}(a), $f^T$ is set to 
 10\% and the horizontal buoyancy is 
progressively increased. When the resultant $\omega_{A}$
obtained from \eqref{rfreqdd} matches $\omega_{M}$, 
the slow MAC waves disappear. 

The analysis in figure \ref{freqdd} is then performed for a
range of $f^T$, and the results are summarized in table
\ref{ddlineartable}.

\subsubsection{A constraint on the lateral heat flux variation
in Earth's lower mantle}
\label{qstarbound}
 In one-component convective dynamos, the heterogeneity factor
$q^*$ that causes the suppression of slow MAC waves
may take values of $O(1)$--$O(10)$; see figure \ref{ralsimilar}(d). 
In two-component convection, we aim to 
place an upper bound on $q^*$ for Earth using plausible
values of the peak magnetic field intensity
 and the Rayleigh number in the core.
For the dipole-dominated regime given by $|\omega_C| > 
|\omega_M| > |\omega_A|$,
 the square of the 
peak value of the scaled magnetic field should take
 values of $O(10^2)$ \citep{jfm21,aditya2022} while the 
 mean square value of the field could be $O(1)$, as suggested
 	by observations \citep{gillet2010fast}.  Furthermore,
 if strong compositional buoyancy in the core
 by itself maintains the dynamo in a dipolar state below
 the threshold for polarity transitions, a value of $Ra^C_{\ell,V}$
 of $O(10^3)$ is reasonable \citep{jfm24}. 
 For Earth, this
 value of the local Rayleigh number corresponds to
 a vertical 
 Rayleigh number $O(10^3)$ times the critical
 Rayleigh number for onset of nonmagnetic convection, essential
 for reproducing the observed polar circulation of
 0.6--0.9$^\circ$ yr$^{-1}$ \citep{hulot2002} in the low-inertia
 geodynamo \citep[see][]{pepi23}. Now, using the
 complementarity of the buoyancy frequencies \eqref{rfreqdd}, the value
 of the relative horizontal buoyancy
 $Ra_{\ell,H}/Ra^T_{\ell}$ for polarity transitions
 based on the condition of 
 vanishing slow MAC waves, $|\omega_A| \approx |\omega_M|$,
 may be
 obtained for a range of thermal power fractions $f^T$.
Since thermal buoyancy 
is not expected to contribute more than 25\% of the
total buoyancy power for Earth's core \citep{lister1995strength}, 
the geodynamo can exist
below the threshold for polarity reversals for 
$Ra_{\ell,H}/Ra^T_{\ell}>0.5$ 
(table \ref{ddlineartable}), 
which must in turn correspond
to large lower-mantle heat flux variations  
$q^*$ of $O(10)$ as per figure \ref{ralsimilar}(d). 
The self-consistent
two-component dynamo model, subject to heterogeneous 
	outer boundary heat flux derived
	from the seismic shear wave 
	velocity in Earth's lower mantle, further substantiates this 
	point (see table \ref{tablehetero} in Appendix \ref{dynamodd}).
The	estimated value of $q^*$ 
	is regarded as an upper bound 
	because it corresponds to the suppression 
	of the slow MAC waves, at which state
	the dynamo exhibits reversals 
	consistent with the occasional reversals 
	observed in Earth. Beyond this limit, 
	the dynamo transitions to a multipolar state.	
For relatively small horizontal buoyancy,
the dynamo would be comfortably placed in the
dipole-dominated regime without polarity transitions.
In short, for nearly invariant vertical Rayleigh numbers
of composition and temperature, the magnitude of
the horizontal Rayleigh number 
would determine whether the dynamo operates in a non-reversing
dipolar state, a reversing state or a multipolar state.

\begingroup
\setlength{\tabcolsep}{6pt} 
\renewcommand{\arraystretch}{1} 
\begin{table}
	\centering
	\begin{tabular}{lcccccccc}
		SL	&  $B^2_{peak}$ 
		& $f^T\%$& $Ra^C_{\ell,V}$  & 
		$Ra^T_{\ell,V}$    
		& $Ra_{\ell,H}$    & $-\omega_{A, V}^2/\omega_{M}^2$ 
		&$\omega_{A, H}^2/\omega_{M}^2$ & $Ra_{\ell,H}/Ra^T_{\ell}$ \\
		1&200&10&2500&875&2461&0.57&0.215&0.74\\
		2&200&12&2500&966&2370&0.59&0.207&0.71\\
		3&200&15&2500&1045&2291&0.61&0.195&0.64\\
		4&200&20&2500&1230&2106&0.63&0.185&0.62\\
		5&200&25&2500&1637&1699&0.68&0.160&0.51\\
		6&200&30&2500&1791&1545&0.72&0.140&0.46\\
		7&200&40&2500&2098&1238&0.78&0.110&0.37\\
		8&200&50&2500&2500&836&0.83&0.085&0.25\\
		\hline
	\end{tabular}
	\caption{Calculation of the relative intensity of
		horizontal buoyancy (last column) in two-component linear magnetoconvection
		for states where the slow MAC waves disappear. 
		The thermal power ratio $f^T$
		 is defined in \eqref{pratio}; 
		$Ra^C_{\ell,V}$, $Ra^T_{\ell,V}$
		are the local 
		vertical compositional and thermal Rayleigh numbers
		 respectively. In addition, $Ra_{\ell,H}$ is the 
		local horizontal Rayleigh number. The
		resultant local thermal Rayleigh number  is
		 $Ra_\ell^T=Ra^T_{\ell,V}+Ra_{\ell,H}$.
		The parameters are $E_\eta=2\times10^{-5}$ and 
		$t=0.01$.}
	\label{ddlineartable}
\end{table}
\endgroup


\section{Concluding remarks}
\label{concl}

The present study investigates the dipole--multipole 
transition through the analysis of MHD wave motions in rapidly
rotating dynamos subject to inhomogeneous heat flux 
at the outer boundary. The regime
in focus is that of low inertia, where the nonlinear
inertial force is small relative to the Coriolis
force not only
on the length scale of the planetary core
but also on the characteristic length scale of
convection. 
The dipole--multipole transition corresponds
to the approximate 
parity between $|\omega_{M}|$ and $|\omega_{A}|$, 
where $|\omega_{M}|$ is based on the peak field intensity 
and $|\omega_{A}|$ is a resultant buoyancy frequency 
 derived from the vertical and horizontal
(lateral) buoyancy frequencies. 
The present study focuses on non-axisymmetric 
outer boundary heat flux, motivated by mantle 
convection models.
That said, as indicated by earlier studies 
\citep[e.g.][]{glatzmaier1999role}, a
	high equatorial axisymmetric boundary heat flux can 
	also induce the dipole--multipole transition in
	rotating dynamos. 
	An axisymmetric positive equatorial heat flux anomaly
	is analogous to the enhanced equatorial heat
	flux in a model with homogeneous boundary heat flux \citep{jfm24},
	where the parity 
	between $|\omega_{M}|$ and $|\omega_{A}|$
	in equation \eqref{wsapprox} causes the disappearance of
	the slow waves, and in turn, the polarity transition.
For the axisymmetric, equatorially anti-symmetric 
$Y_1^0$ heterogeneity, the resultant $\beta$ 
	in equation \eqref{betadyn} must be evaluated with $\beta_{z}$ 
	taken under a modulus sign in regions where 
	$\partial T/\partial r < 0$, such that an increase in 
	$q^*$ (or, equivalently, in the magnitude of $\beta_{z}$) 
	leads to an increase in the resultant buoyancy frequency 
	$|\omega_{A}|$. Under this procedure, preferential cooling 
	of either the Northern or the Southern hemisphere
	has the same outcome, likely inducing polarity transitions.
	However, the theoretical model in \S \ref{linear} does
	not predict the total suppression of convection under stable
	stratification, the analysis of which should be the subject of a
	future study. 

%
A dipole-dominated, strong-field dynamo with a mean 
square magnetic field intensity 
$B^2 / (2 \varOmega \rho \mu \eta) = O(1)$
undergoes a polarity 
transition through the progressive increase 
of horizontal buoyancy induced by an 
equatorially anti-symmetric heat flux 
heterogeneity at the outer boundary. 
The transition occurs not by the loss of
equatorial symmetry of the convection columns, 
but by the selective suppression of
 the slow MAC waves.
Polarity reversals lie in a range
of horizontal buoyancies between the dipolar and
multipolar regimes. 
A non-axisymmetric, equatorially symmetric
heat flux variation does not induce the transition
even at high values of the dimensionless heterogeneity $q^*$
since the mean temperature gradient
at the equator is unaffected for a fixed vertical
buoyancy.
A boundary heterogeneity consisting of
comparable magnitudes of symmetric and anti-symmetric
variations induces the polarity transition at a value of
$q^*$ which is of the same order
as that for the transition induced by a purely 
anti-symmetric variation.
Whether the rapid collapse of the axial 
	dipole during reversals may be correlated with
	the decay of the slow MAC waves when $|\omega_A| \approx
	|\omega_M|$ is a problem that requires 
	further investigation.

The present study proposes
a dynamical constraint on the upper bound of
the heat flux heterogeneity based on the 
complementarity between the 
vertical and horizontal buoyancies in suppressing
the slow waves, noted in both the linear magnetoconvection
and nonlinear dynamo models. In a single-component
(thermal) buoyancy-driven dynamo, a dipole solution
may be possible for any value in the range 0--0.75
of the relative horizontal buoyancy, $Ra_{\ell,H}/Ra_\ell$
(figure \ref{ralsimilar}(c)). However, the fact that
compositional buoyancy is dominant in the two-component
geodynamo, together with the known
order of magnitude of the peak field
intensity in the inertia-free limit,
suggest $Ra_{\ell,H}/Ra_\ell > 0.5$, corresponding
to $q^*$ of $O(10)$
at which an axial dipole can exist.
From the point of view of global
mantle convection \citep{olson2015} and 
the regional stratification of the outer
core \citep{mound2019}, the maximum variation
of heat flux at Earth's CMB could be
higher
than the mean superadiabatic heat flux.

For comparable thermal
power ratios $f^T$, the magnitudes of the
upper bound of $q^*$ predicted by
the two-component magnetoconvection model 
(\S \ref{twoclinear}) and obtained from
the 
self-consistent two-component dynamo simulations 
(Appendix \ref{dynamodd}) are comparable.
A lower bound
for the heterogeneity may also be obtained
from dynamo simulations, based on the minimum
variation needed to obtain the hemispherical
(east--west)
variability in the high-latitude magnetic
flux in Earth \citep{sahoo2020response} and the
mantle-induced
heat flux heterogeneity
 at the inner core boundary \citep{pepi11}. 

In \S \ref{twoclinear}, the chosen value of
the local compositional
Rayleigh number $Ra_{\ell,V}^C$ is well supported by
observations of the polar core flow \citep{pepi23}. 
That said, the value of $Ra_{\ell,V}^C$ in Earth's core may 
have varied in the past. Based on the
complementarity of buoyancy frequencies in \eqref{rfreqdd},
small vertical Rayleigh numbers
require large horizontal Rayleigh numbers, and in turn,
large $q^*$ to achieve reversals.
For the present-day Earth, with nearly invariant vertical 
Rayleigh numbers of composition and temperature,
values of the relative horizontal buoyancy much
lower than that needed to suppress the slow MAC waves
in the core
would place the dynamo in a non-reversing dipolar state.
This provides a way to explain the existence of
geomagnetic superchrons, the long
periods in Earth's history without polarity reversals. 
That said, even for
relative horizontal buoyancies $>0.5$,
the ratio of symmetric to
anti-symmetric heterogeneity in heat flux at Earth's CMB
can vary in geological time. For example,
\citet{courtillot2007mantle} suggest
a relation between superchrons
 and the formation 
of mantle plumes, which control
the heat flow across the CMB. If a plume 
forms near the equator, the heat flux at the CMB is dominated by 
an equatorially symmetric heat flux pattern, promoting a 
dipole-dominated non-reversing state. 
On the other hand, if a plume forms far from the equator, 
the resulting anti-symmetric heat flux pattern suppresses 
the slow MAC waves and produces a regime conducive to reversals. 

\section*{Acknowledgments}
 This study was supported in part by Research Grant
 MoE-STARS/STARS-1/504 under Scheme for
 Transformational and Advanced Research 
in Sciences awarded by the Ministry of Education (India)
and in part by Research grant CRG/2021/002486 awarded
by the Science and Engineering Research Board (India).
The computations were performed on Param Pravega, the supercomputer at the 
Indian Institute of Science, Bangalore.

\section*{Declaration of interests}
The authors report no conflict of interest.
\appendix

\section{Dynamo with outer boundary
heat flux derived from seismic tomography}
\label{tomogrev}

\begin{figure}
	\centering
	\includegraphics[width=0.5\linewidth]{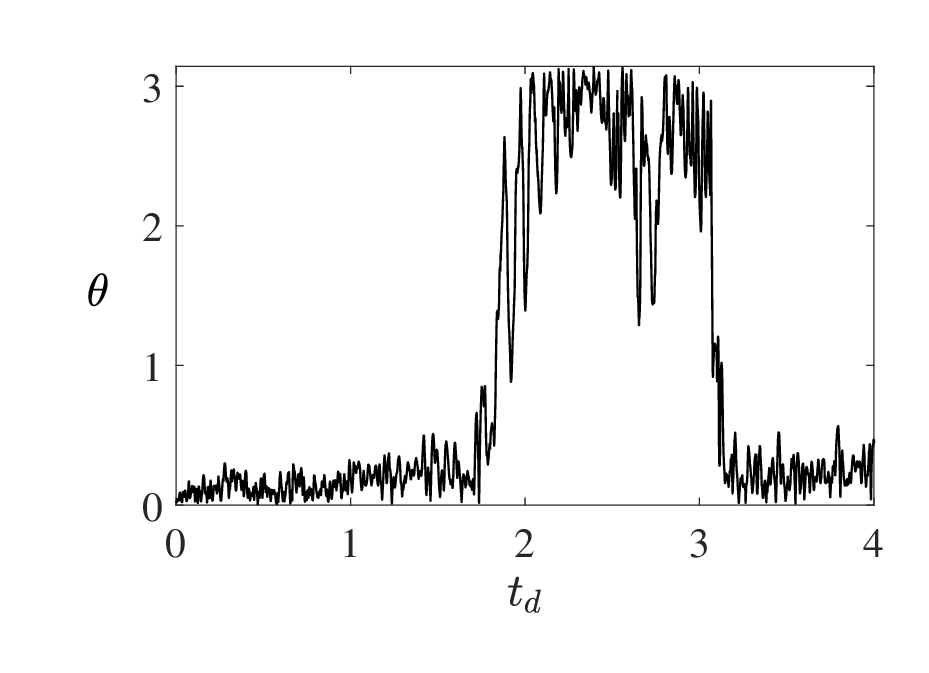}
	\caption{ Evolution of dipole colatitude with time 
	(measured in units of the magnetic 
		diffusion time) for the dynamo subject to
		heterogeneous outer boundary heat flux based on
		 the seismic shear wave velocity in
		Earth's lower mantle 
		at $q^*=13$. 
		The other parameters 
		are $ Ra_V=2500$, $E = 1.2 \times 10^{-5} $, 
		and $ Pm = Pr = 1 $.}
	\label{tomogtilt1}
\end{figure}

\begin{figure}
	\centering
	\hspace{-2 in}	(a)  \hspace{1.8 in} (b) \\
	\includegraphics[width=0.35\linewidth]{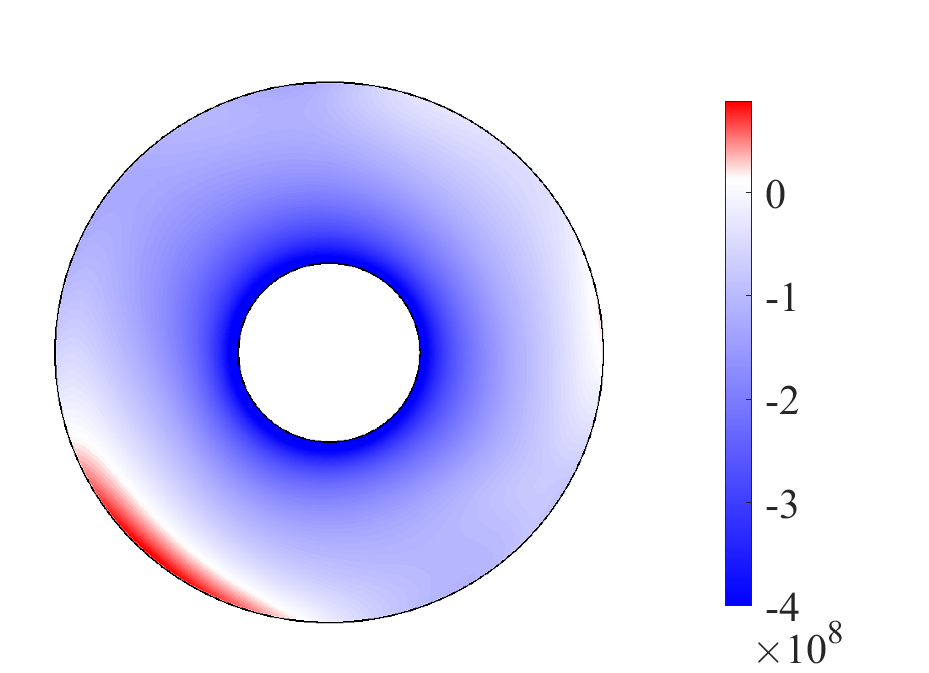}
	\includegraphics[width=0.35\linewidth]{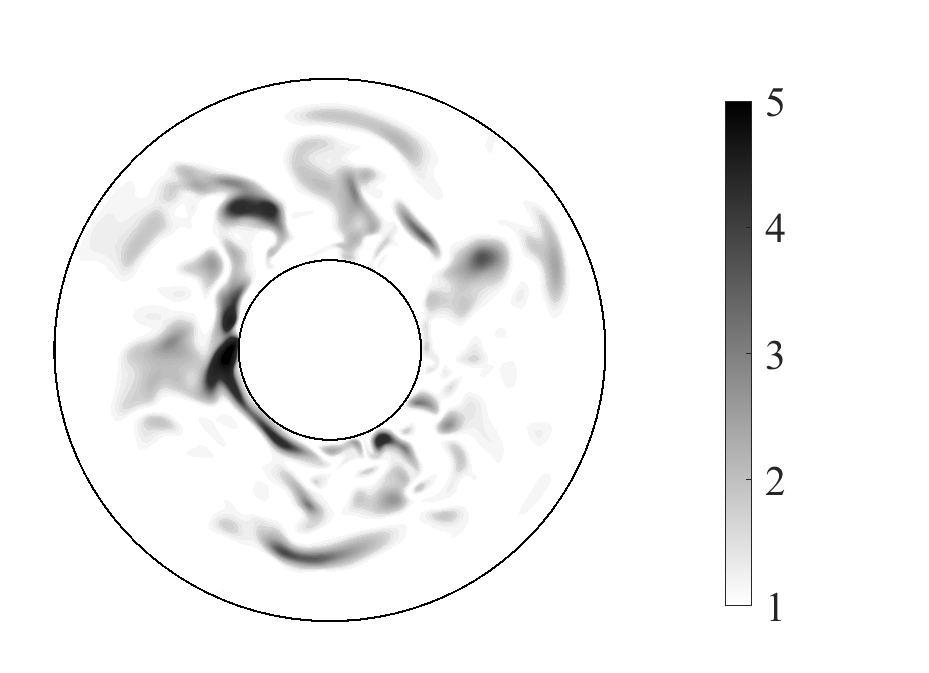}\\
	\hspace{-2 in}	(c)  \hspace{1.8 in} (d) \\
	\includegraphics[width=0.35\linewidth]{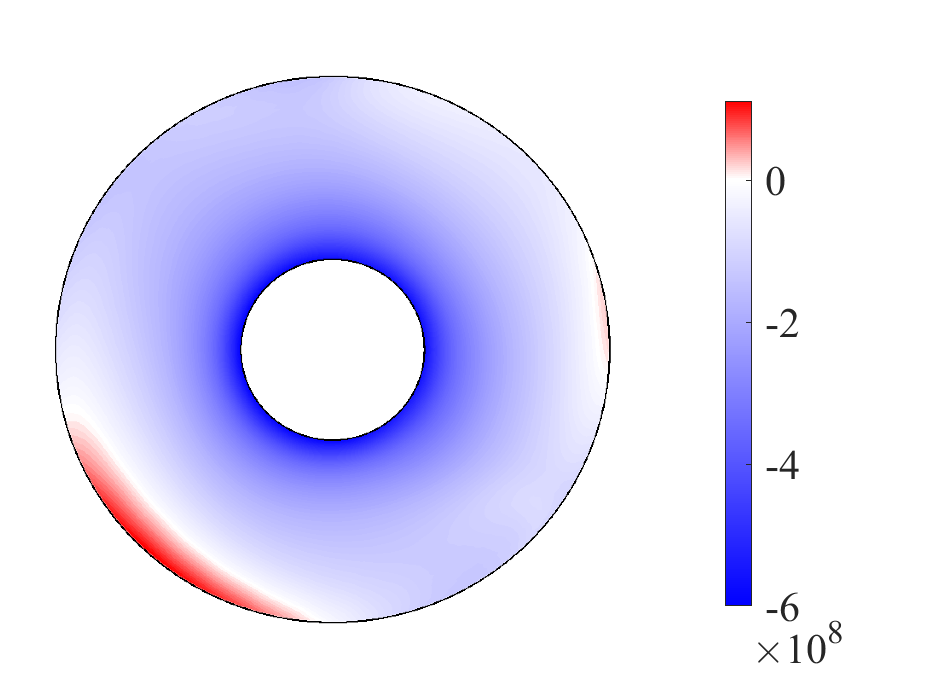}
	\includegraphics[width=0.35\linewidth]{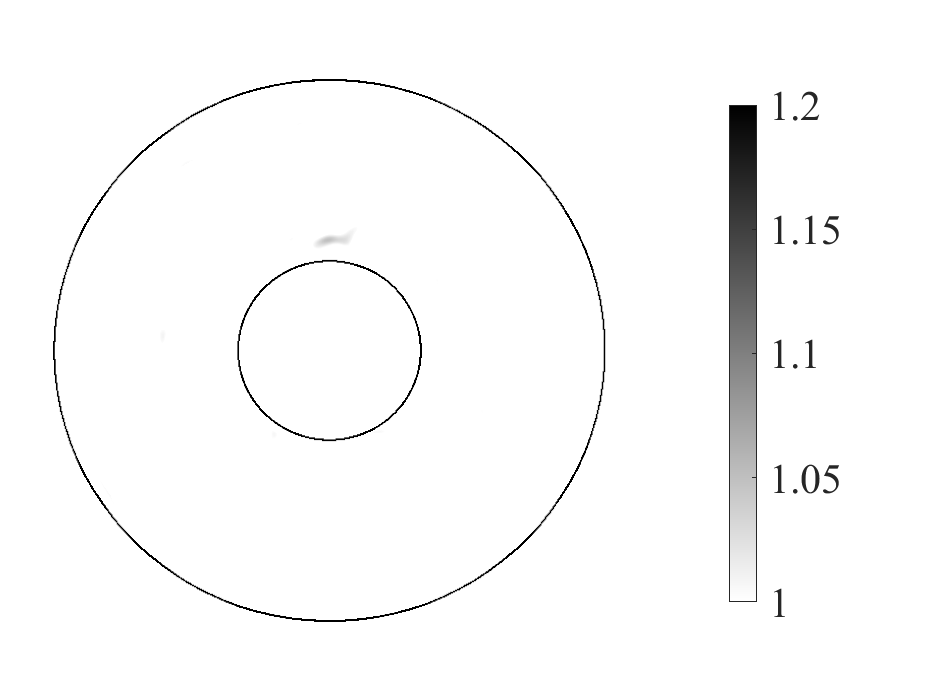}\\
	\caption{Section 
	plots at height $ z=0.2 $ below the equator showing  
		$ \omega_{A}^2 $ (a, c) 
		and $|\omega_{M}^2/\omega_{A}^2|$ (b, d)
		for the composite
		outer boundary heat flux heterogeneity
		based on the seismic shear wave velocity
		in Earth's lowermost mantle. Two values
		of the heterogeneity are considered, 
		$ q^* = 10 $ (a, b) 
		and $ q^* = 13 $ (c, d). The other parameters 
		are $ Ra_V=2500$, $E = 1.2 \times 10^{-5} $, 
		and $ Pm = Pr = 1 $.}
	\label{basicstate1}
\end{figure}

A dynamo subject to an outer boundary
heat flux varying linearly as the
seismic shear wave velocity in the lower mantle
\citep{masters1996} produces polarity
reversals at sufficiently large heat flux
heterogeneity, as shown by the evolution
of the axial dipole colatitude in figure \ref{tomogtilt1}.
This heat flux heterogeneity is a combination
of equatorially symmetric and anti-symmetric
variations of comparable magnitude. The resultant basic
state (mean) 
temperature gradient $\beta$ is given by  
\eqref{betadyn}, wherein $\beta_s$ measures
the basic state gradient with homogeneous
boundary heat flux and the non-zero 
$\beta_z$ at the equator
measures the anti-symmetric part of the variation. 
The square of the resultant buoyancy frequency $\omega_A$, shown 
in figures \ref{basicstate1} (a) \& (c) at a section
$z=0.2$ below the equatorial plane, is proportional to
the resultant gradient $\beta$. In the non-reversing
dipolar run at $q^*=10$, $|\omega_M|^2 >
|\omega_A|^2$ in several regions (figure \ref{basicstate1}c), 
suggesting the generation of slow MAC waves. However,
in the reversing run at $q^*=13$, $|\omega_M|^2 \approx
|\omega_A|^2$ in this section (figure \ref{basicstate1}d), 
suggesting the suppression
of the slow waves.

\section{Spectral coefficients for the perturbations of
temperature and composition}
\label{ddspcoeffs}
From \eqref{waveddcmb}, the initial conditions 
for $\hat{\Theta}$ and its time derivatives are given by
\begin{equation}
	\begin{aligned}
		\mathrm{i}^n \sum_{m=1}^{5}G_m \lambda_{m}^n
		=\bigg(\frac{\partial^n\hat{\Theta}}{\partial t^n}\bigg)_{t=0}
		=g_{n+1}, \ n=0,1,2,3,4,
	\end{aligned}
	\label{thetacoff1}
\end{equation}
where
\begin{equation}
	\begin{aligned}
		g_1&=\hat{\Theta}\big|_{t=0}=\hat{\Theta}_0,\\
		g_2&=\frac{\partial{\hat{\Theta}}}{\partial{t}}\big|_{t=0}=0,\\
		g_3&=\frac{\partial^2 \hat{\Theta}}{\partial{t^2}}\big|_{t=0}=
		a_2\bigg(\beta_{y}^T\frac{k_y}{k_z}-\beta_{z}\bigg),\\
		g_4&=\frac{\partial^3\hat{\Theta}}{\partial{t^3}}\big|_{t=0}=
		a_3\bigg(\beta_{y}^T\frac{k_y}{k_z}-\beta_{z}\bigg),\\
		g_5&=\frac{\partial^4\hat{\Theta}}{\partial{t^4}}\big|_{t=0}=
		a_4\bigg(\beta_{y}^T\frac{k_y}{k_z}-\beta_{z}\bigg).
		\label{thetacoff2}
	\end{aligned}
\end{equation}
Here, the coefficients $a_2$, $a_3$, $a_4$ are defined in \eqref{initial_an}.

The coefficients of the fast and slow wave components of $\hat{\Theta}$ 
are obtained using the roots of equation \eqref{ddcmbchar}. For example,

\begin{eqnarray}
	G_{1}&=\dfrac{g_{5}- \mathrm{i} \, g_{4}(\lambda_2
		+\lambda_3+\lambda_4+\lambda_5)
		+\mathrm{i} \, g_{2}(\lambda_2\lambda_4\lambda_5
		+\lambda_3\lambda_4\lambda_5
		+\lambda_2\lambda_3\lambda_4
		+\lambda_2\lambda_3\lambda_5)}{(\lambda_1-\lambda_2)
		(\lambda_1-\lambda_3)(\lambda_1
		-\lambda_4)(\lambda_1-\lambda_5)},
	\label{g1coeff}\\
	G_{3}&=\dfrac{g_{5}- \mathrm{i} \, g_{4}(\lambda_1
		+\lambda_2+\lambda_4+\lambda_5)
		+ \mathrm{i} \,g_{2}(\lambda_1\lambda_4\lambda_5
		+\lambda_2\lambda_4\lambda_5
		+\lambda_1\lambda_2\lambda_4+
		\lambda_1\lambda_2\lambda_5)}{(\lambda_3-\lambda_1)
		(\lambda_3-\lambda_2)(\lambda_3
		-\lambda_4)(\lambda_3-\lambda_5)},
	\label{g3coff}
\end{eqnarray}


From \eqref{waveddcmb}, the initial conditions 
for $\hat{\gamma}$ and its time derivatives are given by
\begin{equation}
	\begin{aligned}
		\mathrm{i}^n \sum_{m=1}^{5}Q_m \lambda_{m}^n
		=\bigg(\frac{\partial^n\hat{\gamma}}{\partial t^n}\bigg)_{t=0}
		=q_{n+1}, \ n=0,1,2,3,4,
	\end{aligned}
	\label{gammacoff1}
\end{equation}
where
\begin{equation}
	\begin{aligned}
		q_1&=\hat{\gamma}\big|_{t=0}=\hat{\gamma}_{0},\\
		q_2&=\frac{\partial{\hat{\gamma}}}{\partial{t}}\big|_{t=0}=0,\\
		q_3&=\frac{\partial^2 \hat{\gamma}}{\partial{t^2}}\big|_{t=0}
		=a_2 \beta_{y}^C \frac{k_y}{k_z},\\
		q_4&=\frac{\partial^3\hat{\gamma}}{\partial{t^3}}\big|_{t=0}
		=a_3\beta_{y}^C \frac{k_y}{k_z},\\
		q_5&=\frac{\partial^4\hat{\gamma}}{\partial{t^4}}\big|_{t=0}
		=a_4\beta_{y}^C \frac{k_y}{k_z}.
		\label{qcoff}
	\end{aligned}
\end{equation}
The coefficients of the fast and slow wave components of 
$\hat{\gamma}$ are obtained using the roots of 
equation \eqref{ddcmbchar}. For example,

\begin{eqnarray}
	Q_{1}&=\dfrac{q_{5}- \mathrm{i} \, q_{4}(\lambda_2
		+\lambda_3+\lambda_4+\lambda_5)
		+\mathrm{i} \, q_{2}(\lambda_2\lambda_4\lambda_5
		+\lambda_3\lambda_4\lambda_5
		+\lambda_2\lambda_3\lambda_4
		+\lambda_2\lambda_3\lambda_5)}{(\lambda_1-\lambda_2)
		(\lambda_1-\lambda_3)(\lambda_1
		-\lambda_4)(\lambda_1-\lambda_5)},
	\label{q1coeff}\\
	Q_{3}&=\dfrac{q_{5}- \mathrm{i} \, q_{4}(\lambda_1
		+\lambda_2+\lambda_4+\lambda_5)
		+ \mathrm{i} \,q_{2}(\lambda_1\lambda_4\lambda_5
		+\lambda_2\lambda_4\lambda_5
		+\lambda_1\lambda_2\lambda_4
		+\lambda_1\lambda_2\lambda_5)}{(\lambda_3-\lambda_1)
		(\lambda_3-\lambda_2)(\lambda_3-\lambda_4)(\lambda_3-\lambda_5)},
	\label{q2coff}
\end{eqnarray}

\section{Two-component nonlinear dynamo model}
\label{dynamodd}

\begin{figure}
	\centering
	\hspace{-2.1 in}	(a)  \hspace{2.1 in} (b) \\
	\includegraphics[width=0.45\linewidth]{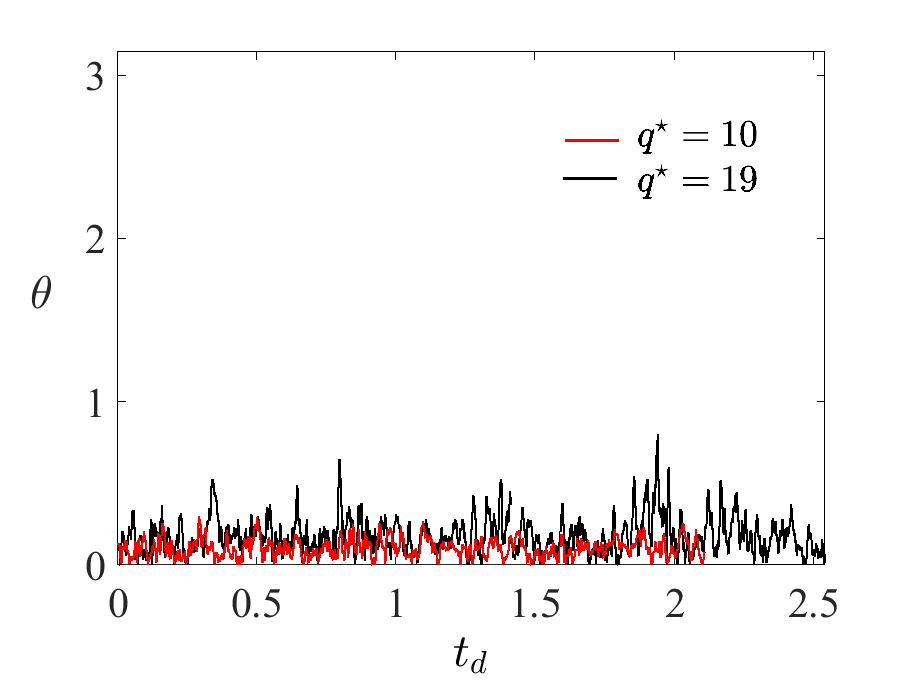}
	\includegraphics[width=0.45\linewidth]{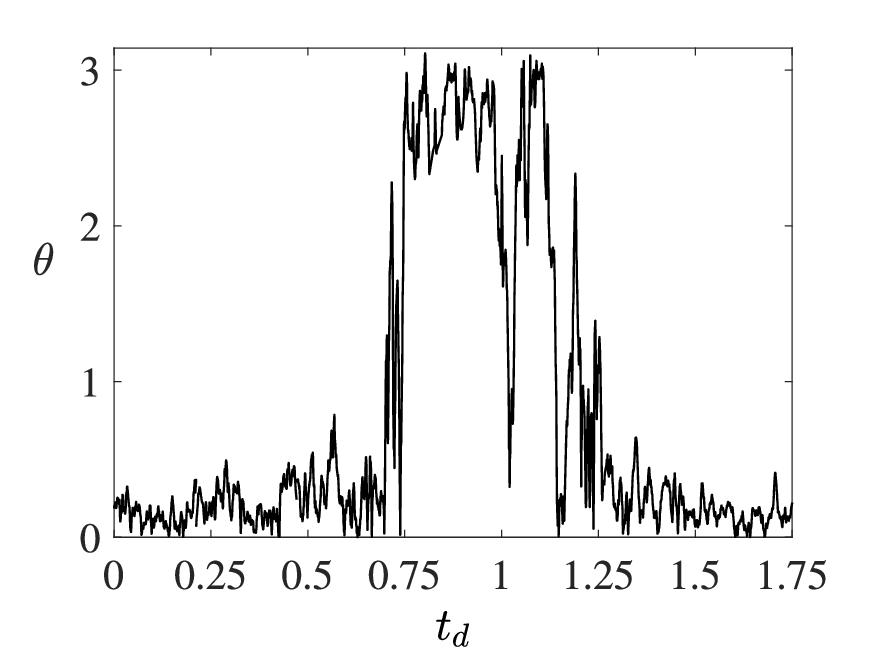}\\
	\hspace{-2.1 in}	(c)  \hspace{2.1 in} \\
	\hspace*{0.75 cm}\includegraphics[width=0.45\linewidth]{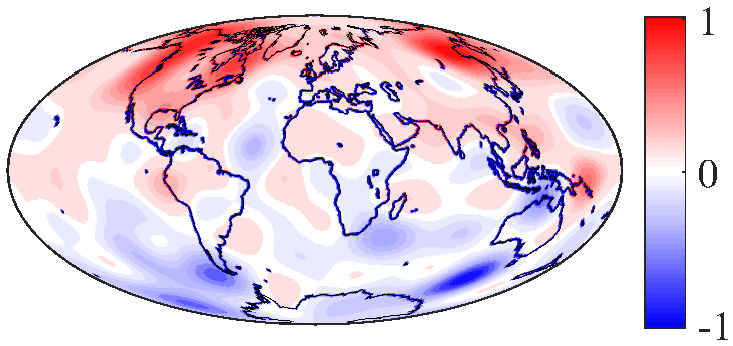}\\
	\caption{Evolution of dipole colatitude 
		with magnetic diffusion time for 
		(a) $q^*=10, 19$ (stable dipolar) and 
		(b) $q^*=20$ (reversing) for the dynamo subject to
		heterogeneous outer boundary heat flux based on
		the seismic shear wave velocity in
		Earth's lower mantle. (c) Snapshot of
		the radial magnetic
		field at the outer boundary for $q^*=10$.
		The dynamo parameters 
		are  $E = 6 \times 10^{-5},Pm=5,Sc=5, Pr=0.5, Ra^T=130, 
		Ra^C=18000$.}
	\label{ddtilt}
\end{figure}

\begingroup
\setlength{\tabcolsep}{2.5pt} 
\renewcommand{\arraystretch}{1} 
\begin{table}
	\centering
	\begin{tabular}{ccccccccccccccccc } 
		\multicolumn{17}{c}{ }\\
		\centering
		$q^*$&$N_r$&$l_{max}$&$Rm$
		&$Ro_l$&$l_E$&$l_C$&$\bar{m}$&$\bar{k}_s$&$\bar{k}_z$&$E_k$&$E_m$&$B^2_{peak}$
		& $B^2_{rms}$&$f_{dip}$&$\dfrac{Ra_{\ell,H}}{Ra_\ell^T}$&Type\\
		&&&&&&&&&&$\times10^5$&$\times10^5$&&&&&D/R/M\\
		\multicolumn{17}{c}{ }\\
		0 &144&144&450&0.025&24&15&7&4.23&4.47&14.68&25.03&223&4.36&0.63&0.00&D\\
		1 &144&144&455&0.026&25&15&7&4.39&4.45&17.97&22.52&216&4.15&0.62&0.07&D\\ 
		3 &144&144&461&0.027&25&16&7&4.46&4.44&15.08&18.32& 211&2.79&0.62&0.19&D\\
		5 &144&144&464&0.027&25&16&7&4.32&4.37&15.12&18.41& 205&2.85&0.61&0.28&D\\ 
		7 &144&144&469&0.027&26&16&7&4.38&4.43&15.25&18.53& 202&2.83&0.61&0.35&D\\
		10&144&144&470&0.028&28&16&7&4.48&4.33&15.55&18.84& 190&2.98&0.59&0.43&D\\
		12&144&144&472&0.031&29&17&7&4.45&4.43&16.31&16.95& 186&2.54&0.57&0.47&D\\ 
		15&144&144&473&0.031&30&17&7&4.49&4.30&16.43&17.86& 183&2.57&0.53&0.51&D\\
		18&144&144&482&0.031&31&17&7&4.43&4.35&16.49&17.02& 179&2.63&0.47&0.57&D\\ 
		19&144&144&496&0.032&31&17&7&4.30&4.31&16.29&16.87& 175&3.42&0.46&0.58&D\\ 
		20&144&144&505&0.032&32&17&7&4.46&4.33&18.46& 8.87& 136&1.09&0.31&0.60&R\\ 
		25&144&144&516&0.032&33&17&7&4.48&4.46&18.74& 8.42& 107&1.08&0.30&0.62&R\\
		\hline
	\end{tabular}
	\caption{ Summary of the main input and output parameters 
		of the two-component
		dynamo simulations considered in this study. 
		Here, $q^*$
		is the dimensionless measure of the boundary heterogeneity, 
		defined in \eqref{qstar}, $N_r$ is the number 
		of radial grid points, $l_{max}$ is the maximum spherical 
		harmonic degree, $Rm$ is the magnetic Reynolds number,
		$Ro_\ell$ is the local Rossby number, $l_C$ and $l_E$ are 
		the mean spherical harmonic degrees of convection and energy 
		injection, $\bar{m}$ is the mean spherical harmonic order in 
		the range $l \le l_E$ , $\bar{k}_s$ and $\bar{k}_z$ are the 
		mean $s$ and $z$ wavenumbers in the range $l \le l_E$, 
		$E_k$ and $E_m$ are the time-averaged total
		kinetic and magnetic energies, $f_{dip}$ is the relative 
		axial dipole field strength, 
		$B^2_{peak}$ is the square of the 
		peak field in the saturated dynamo, and
		$B^2_{rms}$ is the measured mean square value of the field 
		in the spherical shell. Type `D' and `R' denotes dipolar 
		and reversing dynamos, respectively.
		The 
		local horizontal Rayleigh number is given 
		by $Ra_{\ell,H}$ and the local
		thermal Rayleigh number is	$Ra_\ell^T$.
		The dynamo parameters 
		are 	$E = 6 \times 10^{-5},Pm=5,Sc=5, Pr=0.5, Ra^T=130, 
		Ra^C=18000$ and the 
		thermal power ratio $f^T= 11\%$, where
		$Ra^T$ and $Ra^C$ are the modified thermal and 
		compositional Rayleigh numbers.
		Here, $Ra^T$ is set to its critical value for the onset of
nonmagnetic convection with homogeneous boundary heat flux while 
$Ra^C$ is $\approx 900 \times$ its critical value for the onset of convection.}
	\label{tablehetero}
\end{table}
\endgroup

A thermochemically driven dynamo is considered within an 
electrically conducting fluid, which is confined between two 
concentric, co-rotating spherical surfaces corresponding to 
the inner core boundary (ICB) and the core--mantle boundary (CMB). 
The ratio of the inner radius $r_i$ to the outer radius 
$r_o$ is set to 0.35. Lengths are scaled by $L$ and  time 
is scaled by $L^2/\eta$, where $\eta$ is the magnetic diffusivity. 
The velocity is scaled by $\eta/L$ and the magnetic 
field is scaled by $(2\varOmega \mu\eta\rho)^{1/2}$, 
where $\varOmega$ is the angular velocity of rotation, 
$\mu$ is the magnetic permeability and $\rho$ is the fluid density. 
The temperature is scaled by $\beta^T L$ and the composition is 
scaled by $\beta^C L$, where $\beta^T$ and $\beta^C$ are
the mean thermal and compositional gradients 
within the shell, respectively. 
In the Boussinesq approximation, the governing 
non-dimensional magnetohydrodynamic (MHD) equations for velocity, 
magnetic field, temperature, and composition are as follows:

\begin{align}
	E Pm^{-1}  \Bigl(\frac{\partial {\bm u}}{\partial t} + 
	(\nabla \times {\bm u}) \times {\bm u}
	\Bigr)+  {\hat{\bm{z}}} \times {\bm u} = - \nabla p^\star  
	+ Ra^T \, T \, {\bm r} \,  \nonumber\\  
	+ Ra^C \, C \, {\bm r} +(\nabla \times {\bm B})
	\times {\bm B} +E\nabla^2 {\bm u}, \label{momentumdd} \\
	\frac{\partial {\bm B}}{\partial t} = 
	\nabla \times ({\bm u} \times {\bm B}) 
	+ \nabla^2 {\bm B},  \label{inductiondd}\\
	\frac{\partial T}{\partial t} +({\bm u} \cdot \nabla) T =  
	Pm Pr^{-1} \,
	\nabla^2 T+S_o,  \label{heat1dd}\\
	\frac{\partial C}{\partial t} +({\bm u} \cdot \nabla) C =  
	Pm Sc^{-1} \,
	\nabla^2 C+S_i,  \label{comp1dd}\\
	\nabla \cdot {\bm u}  =  \nabla \cdot {\bm B} = 0.  
	\label{divdd}
\end{align}

The modified pressure, $p^*$ is expressed as 
$p + \frac{1}{2} E \, Pm^{-1} \, |\bm{u}|^2$. 
The dimensionless parameters governing the system are 
the Ekman number $E = \nu/2\varOmega L^2$, the Prandtl number 
$Pr = \nu/\kappa^T$, the Schmidt number $Sc = \nu/\kappa^C$ 
and the magnetic Prandtl number, $Pm = \nu/\eta$. 
The modified thermal and compositional Rayleigh numbers are given 
by $Ra^T = g \alpha^T \beta^T L^2/2\Omega\eta$ and 
$Ra^C = g \alpha^C \beta^C L^2/2\Omega\eta$, respectively. 
Here, $g$ is the gravitational acceleration, $\nu$ 
is the kinematic viscosity, $\kappa^T$ and $\kappa^C$ 
are the thermal and compositional diffusivities, 
and $\alpha^T$ and $\alpha^C$ are the coefficients of thermal 
and compositional expansion, respectively.

No-slip and electrically insulating conditions are imposed at the
two boundaries. 
The basic state temperature profile consists
of a combination of basal and internal heating,
\begin{equation}
	\frac{\partial T_0}{\partial r} = - \frac{A}{r^2} - B \, r,
	\label{tprofile}
\end{equation}
where the first and second terms on the right-hand side
of \eqref{tprofile} represent basal heating 
and internal heating respectively, and $A/B=2.23$.
The inner boundary is isothermal while the 
mean heat flux
at the outer boundary is $-r_o$.
The heterogeneous heat flux imposed at the outer boundary is
derived from
		the seismic shear wave velocity variation
		in the Earth's lower mantle \citep{masters1996}.
For composition, a uniform volumetric sink $S_i=-1$ is considered, 
with a constant flux at the ICB and zero flux at the CMB. 
The basic state compositional gradient is given by
\begin{equation} 
	\frac{\partial C_0}{\partial r} = 
	\frac{Sc}{Pm} \frac{S_i}{3} \bigg(\frac{r_o^3}{r^2} - r\bigg).
	\label{cprofile}
\end{equation}
By progressively increasing the value of $q^*$ at
a fixed Rayleigh numbers $Ra^T$ and $Ra^C$, the relative
horizontal buoyancy $Ra_{\ell,H}/Ra_\ell^T$ at which a
dipolar (D) dynamo undergoes the transition to a reversing (R)
dynamo is noted  (table \ref{tablehetero}). For a thermal
power ratio $f^T = 11\%$, the polarity transition occurs
at relative horizontal buoyancy $Ra_{\ell,H}/Ra_\ell^T=0.6$ in the dynamo, which
compares favourably with the values $>0.5$ predicted for vanishing slow
MAC waves in the linear magnetoconvection model (table 
\ref{ddlineartable}). The two-component dynamo
model indicates that the value of $Ra_{\ell,H}/Ra_\ell^T$
at the polarity transition corresponds to $q^* =O(10)$. 

The evolution of the dipole colatitude $\theta$ in figures
\ref{ddtilt} (a) \& (b) shows the onset of polarity
reversals at $q^*=20$. In the dipole-dominated
 regime at $q^*=10$,
the radial magnetic field at the outer boundary shows
well-defined 
high-latitude magnetic flux lobes  symmetrically placed
about the equator in the Eastern and
Western hemispheres, as in present-day Earth
 (figure \ref{ddtilt} c).

\clearpage
\newpage
\bibliography{cmbbib}

@article{loper2003buoyancy,
  title={Buoyancy-driven perturbations in a rapidly rotating,
   electrically conducting fluid: part {I}--flow and magnetic field},
  author={Loper, D. E. and Chulliat, A. and Shimizu, H.},
  journal={Geophys. Astrophys. Fluid Dyn.},
  volume={97},
  number={6},
  pages={429--469},
  year={2003},
  publisher={Taylor \& Francis}
}

@article{hathaway1979convective,
  title={Convective instability when the temperature gradient
   and rotation vector are oblique to gravity. {I}. {F}luids without diffusion},
  author={Hathaway, D. H. and Gilman, P. A. and Toomre, J.},
  journal={Geophys. Astrophys. Fluid Dyn.},
  volume={13},
  number={1},
  pages={289--316},
  year={1979},
  publisher={Taylor \& Francis}
}

@article{courtillot2007mantle,
  title={Mantle plumes link magnetic superchrons 
  to Phanerozoic mass depletion events},
  author={Courtillot, V. and Olson, P.},
  journal={Earth Planet. Sci. Lett.},
  volume={260},
  number={3-4},
  pages={495--504},
  year={2007},
  publisher={Elsevier}
}

@article{takahashi2008effects,
  title={Effects of thermally heterogeneous structure 
  in the lowermost mantle on the geomagnetic field strength},
  author={Takahashi, F. and Tsunakawa, H. and 
  Matsushima, M. and Mochizuki, N. and Honkura, Y.},
  journal={Earth Planet. Sci. Lett.},
  volume={272},
  number={3-4},
  pages={738--746},
  year={2008},
  publisher={Elsevier}
}

@article{stanley2008mars,
  title={Mars' paleomagnetic field as the 
  result of a single-hemisphere dynamo},
  author={Stanley, S. and Elkins, T. L. and Zuber, 
  M. T. and Parmentier, E. M.},
  journal={Science},
  volume={321},
  number={5897},
  pages={1822--1825},
  year={2008},
  publisher={American Association for the Advancement of Science}
}

@article{jfm24,
  title={Self-similarity of the dipole--multipole 
  transition in rapidly rotating dynamos},
  author={Majumder, D. and Sreenivasan, B. and Maurya, G.},
  journal={J. Fluid Mech.},
  volume={980},
  pages={A30},
  year={2024},
  publisher={Cambridge University Press}
  }

@article{jfm21,
  title={Evolution of forced magnetohydrodynamic 
  waves in a stratified fluid},
  author={Sreenivasan, B. and Maurya, G.},
  journal={J. Fluid Mech.},
  volume={922},
  year={2021},
  publisher={Cambridge University Press}
}

@article{pepi23,
  title={The role of magnetic waves in tangent cylinder convection},
  author={Majumder, D. and Sreenivasan, B.},
  journal={Phys. Earth Planet. Inter.},
  volume={344},
  pages={107105},
  year={2023},
  publisher={Elsevier}
}

@ARTICLE{07willis,
        author = {A. P. Willis and B. Sreenivasan and
D. Gubbins},
        title = {Thermal core-mantle interaction: {E}xploring regimes
for \lq locked' dynamo action},
        journal = {Phys. Earth Planet. Inter.},
        volume = {165},
        pages = {83--92},
        year = {2007},
}

@article{brag1967,
  title={Magnetic waves in the {E}arth's core},
  author={Braginsky, S. I.},
  journal={Geomagn. Aeron.},
  volume={7},
  pages={851--859},
  year={1967}
}

@article{chraub2006,
  title={Scaling properties of convection-driven dynamos 
in rotating spherical shells and application to planetary magnetic fields},
  author={Christensen, U. R. and Aubert, J.},
  journal={Geophys. J. Int.},
  volume={166},
  number={1},
  pages={97--114},
  year={2006},
  publisher={Blackwell Publishing Ltd Oxford, UK}
}

@article{teed2010rapidly,
  title={Rapidly rotating plane layer convection with zonal flow},
  author={Teed, R. J and Jones, C. A. and Hollerbach, R.},
  journal={Geophys. Astrophys. Fluid Dyn.},
  volume={104},
  number={5-6},
  pages={457--480},
  year={2010},
  publisher={Taylor \& Francis}
}

@article{masters1996,
  title={A shear-velocity model of the mantle},
  author={Masters, G. and Johnson, S.and Laske, G. and Bolton, H.},
  journal={Phil. Trans. R. Soc. Lond., A},
  volume={354},
  number={1711},
  pages={1385--1411},
  year={1996},
  publisher={The Royal Society London}
}

@article{zhong2007supercontinent,
  title={Supercontinent cycles, true polar wander, and 
  very long-wavelength mantle convection},
  author={Zhong, S. and Zhang, N. and Li, Z. X. and Roberts, J. H.},
  journal={Earth Planet. Sci. Lett.},
  volume={261},
  number={3-4},
  pages={551--564},
  year={2007},
  publisher={Elsevier}
}

@article{yoshida2008mantle,
  title={Mantle convection with longest-wavelength thermal 
  heterogeneity in a 3-{D} spherical model: {D}egree one or two?},
  author={Yoshida, M.},
  journal={Geophys. Res. Lett.},
  volume={35},
  number={23},
  year={2008},
  publisher={Wiley Online Library}
}

@article{glatzmaier1999role,
  title={The role of the Earth's mantle in controlling the 
  frequency of geomagnetic reversals},
  author={Glatzmaier, G. A. and Coe, R. S. and 
  Hongre, L. and Roberts, P. H.},
  journal={Nature},
  volume={401},
  number={6756},
  pages={885--890},
  year={1999},
  publisher={Nature Publishing Group UK London}
}

@article{kutzner2004simulated,
  title={Simulated geomagnetic reversals and 
  preferred virtual geomagnetic pole paths},
  author={Kutzner, C. and Christensen, U. R.},
  journal={Geophys. J. Int.},
  volume={157},
  number={3},
  pages={1105--1118},
  year={2004},
  publisher={Blackwell Science Ltd Oxford, UK}
}

@book{glatz2013,
  title={Introduction to Modeling Convection in Planets
  and Stars: Magnetic Field, Density Stratification, Rotation},
  author={Glatzmaier, G. A.},
  year={2013},
  publisher={Princeton University Press},
}

@article{buffett2002energetics,
  title={Energetics of numerical geodynamo models},
  author={Buffett, B. A. and Bloxham, J},
  journal={Geophys. J. Int.},
  volume={149},
  number={1},
  pages={211--224},
  year={2002},
  publisher={Blackwell Publishing Ltd Oxford, UK}
}

@article{bullard1954,
  title={Homogeneous dynamos and terrestrial magnetism},
  author={Bullard, E. C. and Gellman, H.},
  journal={Philos. Trans. R. Soc. London, Ser. A},
  volume={247},
  number={928},
  pages={213--278},
  year={1954},
  publisher={The Royal Society London}
}

@article{olson2002time,
  title={The time-averaged magnetic field in 
  numerical dynamos with non-uniform boundary heat flow},
  author={Olson, P. and Christensen, U. R.},
  journal={Geophys. J. Int.},
  volume={151},
  number={3},
  pages={809--823},
  year={2002},
  publisher={Blackwell Publishing Ltd Oxford, UK}
}

@Incollection{07bussechapter,
  title={Dynamics of rotating fluids},
  author={Busse, F. and Dormy, E. and Simitev, R. and Soward, A.},
  booktitle={Mathematical Aspects of Natural Dynamos},
  volume={13},
  pages={165--168},
  year={2007},
editor = {Dormy, E. and Soward, A. M.},
series = {The Fluid Mechanics of Astrophysics and Geophysics},
  publisher={CRC Press}
}

@article{lister1995strength,
  title={The strength and efficiency of thermal 
  and compositional convection in the geodynamo},
  author={Lister, J. R. and Buffett, B. A.},
  journal={Phys. Earth Planet. Inter.},
  volume={91},
  number={1-3},
  pages={17--30},
  year={1995},
  publisher={Elsevier}
}

@article{sahoo2016dynamos,
  title={Dynamos driven by weak thermal convection 
  and heterogeneous outer boundary heat flux},
  author={Sahoo, S. and Sreenivasan, B. and Amit, H.},
  journal={Phys. Earth Planet. Inter.},
  volume={250},
  pages={35--45},
  year={2016},
  publisher={Elsevier}
}

@article{sreeni2014,
    author = "Sreenivasan, B. and Sahoo, S. and Dhama, G.",
    title = "The role of buoyancy in polarity reversals of the geodynamo",
    journal = "Geophys. J. Int.",
    volume = "199",
    number = "3",
    pages = "1698--1708",
    year = "2014",
    publisher = "Oxford University Press"
}

@article{sahoo2017,
  title={On the effect of laterally varying boundary heat flux on rapidly rotating spherical shell convection},
  author={Sahoo, S. and Sreenivasan, B.},
  journal={Phys. Fluids},
  volume={29},
  pages={086602},
  year={2017},
}

@article{mcfadden1984lower,
  title={Lower mantle convection and geomagnetism},
  author={McFadden, P. L. and Merrill, R. T.},
  journal={J. Geophys. Res. Solid Earth},
  volume={89},
  number={B5},
  pages={3354--3362},
  year={1984},
  publisher={Wiley Online Library}
}

@article{gubbins1993persistent,
  title={Persistent patterns in the geomagnetic 
  field over the past 2.5 Myr},
  author={Gubbins, D. and Kelly, P.},
  journal={Nature},
  volume={365},
  number={6449},
  pages={829--832},
  year={1993},
  publisher={Nature Publishing Group UK London}
}

@article{carlut1998complex,
  title={How complex is the time-averaged geomagnetic 
  field over the past 5 Myr?},
  author={Carlut, J. and Courtillot, V.},
  journal={Geophys. J. Int.},
  volume={134},
  number={2},
  pages={527--544},
  year={1998},
  publisher={Blackwell Publishing Ltd Oxford, UK}
}

@article{johnson2003mapping,
  title={Mapping long-term changes in earth's magnetic field},
  author={Johnson, C. L. and Constable, C. G. and Tauxe, L.},
  journal={Science},
  volume={300},
  number={5628},
  pages={2044--2045},
  year={2003},
  publisher={American Association for the 
  Advancement of Science}
}

@article{jones1977thermal,
  title={Thermal interaction of the core and the mantle 
  and long-term behavior of the geomagnetic field},
  author={Jones, G. M.},
  journal={J. Geophys. Res.},
  volume={82},
  number={11},
  pages={1703--1709},
  year={1977},
  publisher={Wiley Online Library}
}

@article{larson1991mantle,
  title={Mantle plumes control magnetic reversal frequency},
  author={Larson, R. L. and Olson, P.},
  journal={Earth and Planetary Science Letters},
  volume={107},
  number={3-4},
  pages={437--447},
  year={1991},
  publisher={Elsevier}
}

@article{lhuillier2013statistical,
  title={Statistical properties of reversals and chrons 
  in numerical dynamos and implications for the geodynamo},
  author={Lhuillier, F. and Hulot, G. and Gallet, Y.},
  journal={Earth Planet. Sci. Lett.},
  volume={220},
  pages={19--36},
  year={2013},
  publisher={Elsevier}
}

@article{zhang1992convection,
  title={On convection in the {E}arth's core driven by 
  lateral temperature variations in the lower mantle},
  author={Zhang, K. and Gubbins, D.},
  journal={Geophys. J. Int.},
  volume={108},
  number={1},
  pages={247--255},
  year={1992},
  publisher={Blackwell Publishing Ltd Oxford, UK}
}

@article{aubert2007detecting,
  title={Detecting thermal boundary control in surface 
  flows from numerical dynamos},
  author={Aubert, J. and Amit, H. and Hulot, G.},
  journal={Phys. Earth Planet. Inter.},
  volume={160},
  number={2},
  pages={143--156},
  year={2007},
  publisher={Elsevier}
}

@article{olson2010geodynamo,
  title={Geodynamo reversal frequency and heterogeneous 
  core--mantle boundary heat flow},
  author={Olson, P. and Coe, R. S. and Driscoll, P. E. 
  and Glatzmaier, G. A. and Roberts, P. H.},
  journal={Phys. Earth Planet. Inter.},
  volume={180},
  number={1-2},
  pages={66--79},
  year={2010},
  publisher={Elsevier}
}

@article{sahoo2020response,
  title={Response of {E}arth's magnetic 
  field to large lower mantle heterogeneity},
  author={Sahoo, S. and Sreenivasan, B.},
  journal={Earth Planet. Sci. Lett.},
  volume={549},
  pages={116507},
  year={2020},
  publisher={Elsevier}
}

@article{laj1991geomagnetic,
  title={Geomagnetic reversal paths},
  author={Laj, C. and Mazaud, A. and Weeks, R. and 
  Fuller, M. and Herrero-Bervera, E.},
  journal={Nature},
  volume={351},
  number={6326},
  pages={447--447},
  year={1991}
}

@article{love2000statistical,
  title={Statistical assessment of preferred transitional 
  {VGP} longitudes based on palaeomagnetic lava data},
  author={Love, J. J.},
  journal={Geophys. J. Int.},
  volume={140},
  number={1},
  pages={211--221},
  year={2000},
  publisher={Blackwell Publishing Ltd Oxford, UK}
}

@article{fisk1931isopors,
  title={Isopors and isoporic movement},
  author={Fisk, H. W.},
  journal={Bull. Int. Geod. Geophys. Union},
  volume={8},
  pages={280--292},
  year={1931}
}

@article{holme2015large,
  title={Large-scale flow in the core},
  author={Holme, R. and Olson, P. and Schubert, G.},
  journal={Treatise Geophys.},
  volume={8},
  pages={107--130},
  year={2015}
}

@article{christensen2003secular,
  title={Secular variation in numerical
   geodynamo models with lateral variations of boundary heat flow},
  author={Christensen, U. R. and Olson, P.},
  journal={Phys. Earth Planet. Inter.},
  volume={138},
  number={1},
  pages={39--54},
  year={2003},
  publisher={Elsevier}
}

@article{mound2023longitudinal,
  title={Longitudinal structure of Earth’s magnetic field controlled 
  by lower mantle heat flow},
  author={Mound, Jonathan E and Davies, Christopher J},
  journal={Nat. Geosci.},
  volume={16},
  number={4},
  pages={380--385},
  year={2023},
  publisher={Nature Publishing Group UK London}
}

@article{su1994degree,
  title={Degree 12 model of shear velocity heterogeneity in the mantle},
  author={Su, W. j. and Woodward, R. L. and Dziewonski, A. M.},
  journal={J. Geophys. Res. Solid Earth},
  volume={99},
  number={B4},
  pages={6945--6980},
  year={1994},
  publisher={Wiley Online Library}
}

@article{olson2014magnetic,
  title={Magnetic reversal frequency scaling in 
  dynamos with thermochemical convection},
  author={Olson, P. and Amit, H.},
  journal={Phys. Earth Planet. Inter.},
  volume={229},
  pages={122--133},
  year={2014},
  publisher={Elsevier}
}

@article{pepi11,
  title={On mantle-induced heat flow 
  variations at the inner core boundary},
  author={Sreenivasan, B. and Gubbins, D.},
  journal={Phys. Earth Planet. Inter.},
  volume={187},
  number={3-4},
  pages={336--341},
  year={2011},
  publisher={Elsevier}
}

@article{coe2006symmetry,
  title={Symmetry and stability of the geomagnetic field},
  author={Coe, R. S. and Glatzmaier, G. A.},
  journal={Geophys. Res. Lett.},
  volume={33},
  number={21},
  year={2006},
  publisher={Wiley Online Library}
}

@article{amit2011influence,
  title={The influence of degree-1 mantle 
  heterogeneity on the past dynamo of Mars},
  author={Amit, H. and Christensen, U. R. and Langlais, B.},
  journal={Phys. Earth Planet. Inter.},
  volume={189},
  number={1-2},
  pages={63--79},
  year={2011},
  publisher={Elsevier}
}

@article{tassin2021,
  title={Geomagnetic semblance and dipolar--multipolar 
  transition in top-heavy double-diffusive geodynamo models},
  author={Tassin, T. and Gastine, T. and Fournier, A.},
  journal={Geophys. J. Int.},
  volume={226},
  number={3},
  pages={1897--1919},
  year={2021},
  publisher={Oxford University Press}
}

@article{prf18,
  title={Scale dependence of kinetic helicity and selection 
  of the axial dipole in rapidly rotating dynamos},
  author={Sreenivasan, B. and Kar, S.},
  journal={Phys. Rev. Fluids},
  volume={3},
  number={9},
  pages={093801},
  year={2018},
  publisher={APS}
}

@article{aditya2022,
  title = {The role of slow magnetostrophic waves 
in the formation of the axial dipole
in planetary dynamos},
  author = {A. Varma and B. Sreenivasan},
  journal = {Phys. Earth Planet. Inter.},
  volume = {333},
  pages = {106944},
  year = {2022},
}

@article{jackson2000,
	title={Four centuries of geomagnetic secular variation from historical records},
	author={Jackson, A. and Jonkers, A.R.T. and Walker, M.R.},
	journal={Philos. Trans. Royal Soc. A},
	volume={358},
	number={1768},
	pages={957--990},
	year={2000},
	publisher={The Royal Society}
}

@article{doell1971,
  title={Pacific geomagnetic secular variation},
  author={Doell, R. R. and Cox, A.},
  journal={Science},
  volume={171},
  number={3968},
  pages={248--254},
  year={1971},
  }

@article{mound2019,
	title={Regional stratification at the top of {E}arth's core due to core--mantle boundary heat flux variations},
	author={Mound, J. and Davies, C. and Rost, S. and Aurnou, J.},
	journal={Nat. Geosci.},
	volume={12},
	number={7},
	pages={575--580},
	year={2019},
	publisher={Nature Publishing Group UK London}
}

@ARTICLE{olson2015,
  author = {Olson, P. and Deguen, R. and Rudolph, M.L. and Zhong, S.},
  title = {Core evolution driven by mantle global circulation},
  journal = {Phys. Earth Planet. Inter.},
  year = {2015},
  volume = {243},
  pages = {44--55},
  publisher = {Elsevier}
}

@article{kutzner2002,
  title={From stable dipolar towards reversing numerical dynamos},
  author={Kutzner, C. and Christensen, U. R.},
  journal={Phys. Earth Planet. Inter.},
  volume={131},
  number={1},
  pages={29--45},
  year={2002},
  publisher={Elsevier}
}

@article{hulot2002,
  title={Small-scale structure of the geodynamo inferred from
 {O}ersted and {M}agsat satellite data},
  author={Hulot, G. and Eymin, C. and Langlais, B. and Mandea, M. and Olsen, N.},
  journal={Nature},
  volume={416},
  number={6881},
  pages={620--623},
  year={2002},
  publisher={Nature Publishing Group UK London}
}

@article{frasson2025geomagnetic,
  title={Geomagnetic dipole stability and zonal 
  flows controlled by mantle heat flux heterogeneities},
  author={Frasson, T. and Schaeffer, N. and 
  Nataf, H.-C. and Labrosse, S.},
  journal={Geophys. J. Int.},
  volume={240},
  number={3},
  pages={1481--1504},
  year={2025},
  publisher={Oxford University Press}
}

@article{gillet2010fast,
  title={Fast torsional waves and strong magnetic field within the Earth’s core},
  author={Gillet, N. and Jault, D. and Canet, E. and Fournier, A.},
  journal={Nature},
  volume={465},
  number={7294},
  pages={74--77},
  year={2010},
  publisher={Nature Publishing Group UK London}
}

@article{koelemeijer2012normal,
  title={Normal mode sensitivity to {E}arth's {D}$''$ layer and topography on the core-mantle boundary: what we can and cannot see},
  author={Koelemeijer, P. J. and Deuss, A. and Trampert, J.},
  journal={Geophys. J. Int.},
  volume={190},
  number={1},
  pages={553--568},
  year={2012},
  publisher={Blackwell Publishing Ltd Oxford, UK}
}

@article{gulcher2021coupled,
  title={Coupled dynamics and evolution of primordial and recycled heterogeneity in {E}arth's lower mantle},
  author={G{\"u}lcher, A. J. P. and Ballmer, M. D. and Tackley, P. J.},
  journal={Solid Earth},
  volume={12},
  number={9},
  pages={2087--2107},
  year={2021},
  publisher={Copernicus GmbH}
}

@Incollection{olson2015treatise,
  title={Core {D}ynamics: {A}n {I}ntroduction and {O}verview},
  author={Olson, P.},
  booktitle={Treatise on Geophysics},
  volume={8},
edition = {Second},
  year={2015},
editor = {Schubert, G.},
  publisher={Elsevier}
}
\bibliographystyle{jfm}


\clearpage
\newpage

\begin{center}
\Large Polarity transitions induced by
symmetry-breaking outer boundary heat 
flux in rapidly rotating dynamos\\
\vspace*{0.25cm}
\large
Debarshi Majumder and Binod Sreenivasan$^\dagger$\\

\vspace*{0.25cm}

SUPPLEMENTARY MATERIAL
\end{center}

\setcounter{figure}{0}
\renewcommand{\figurename}{Figure}
\renewcommand{\thefigure}{S\arabic{figure}}
\pagestyle{plain} 
\renewcommand{\thesection}{S\arabic{section}}

\section*{S1~~~~ The evolution of a dynamo from a seed magnetic field:
	The role of slow MAC waves in dipole formation}
\label{seed}

\begin{figure}
	\centering
	\hspace{-4 in}	(a) \\
	\includegraphics[width=0.8\linewidth]{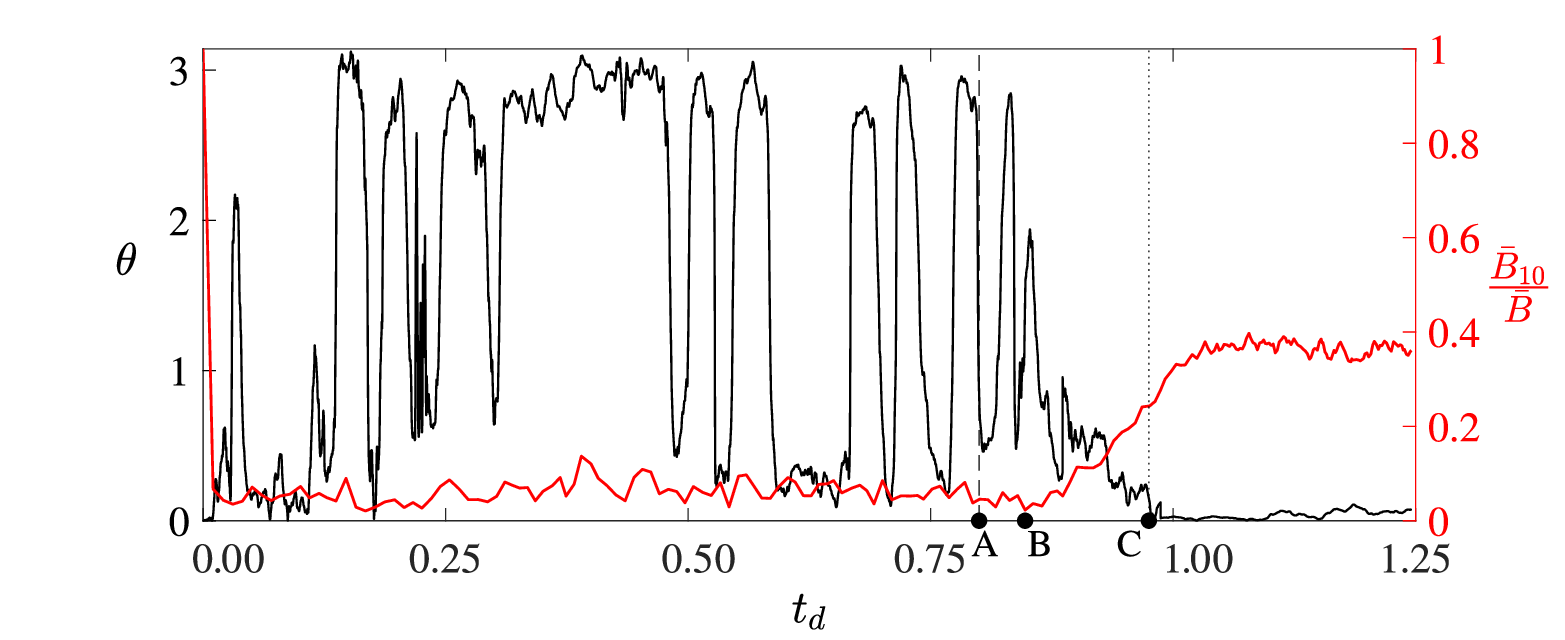}	\\
	\hspace{-4 in}	(b) \\
	\includegraphics[width=0.8\linewidth]{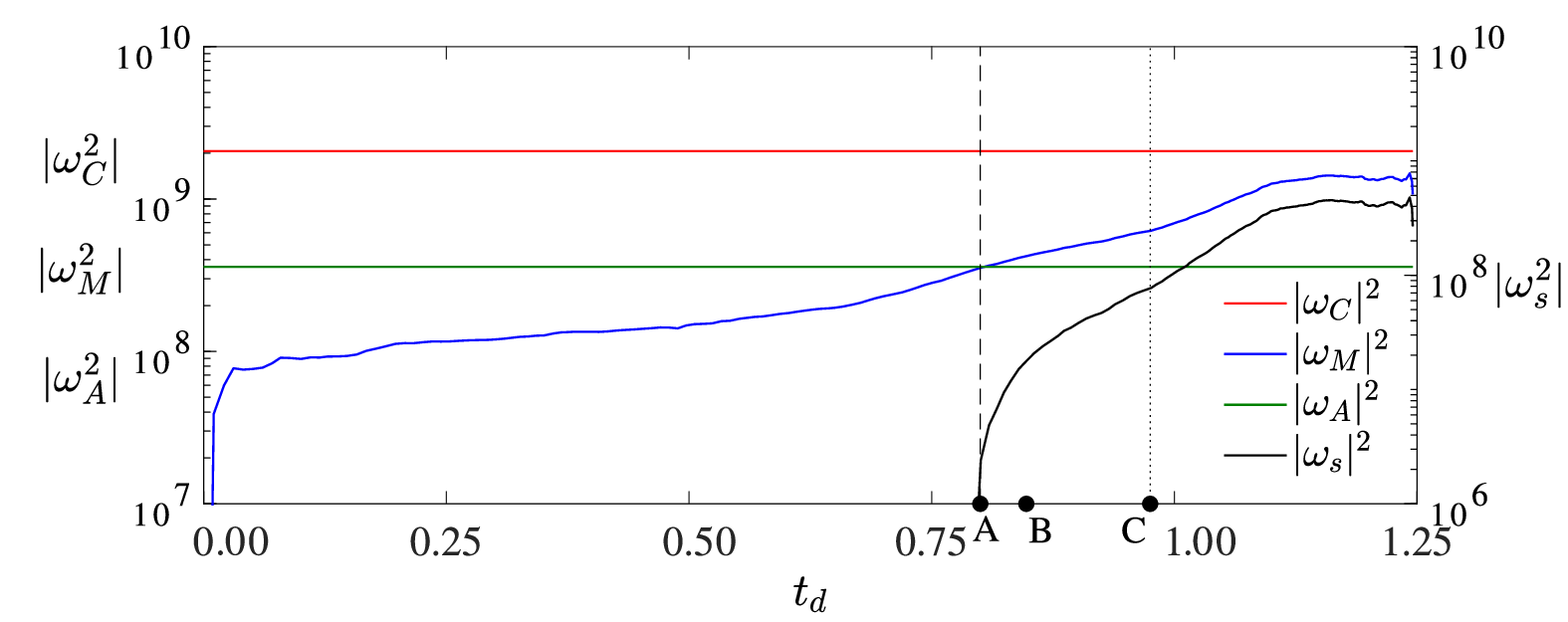}	\\
	\hspace{-2.75 cm}	(c) \hspace{1.75 cm} B \hspace{5 cm} C \\
	\hspace*{1cm}	\includegraphics[width=0.4\linewidth]{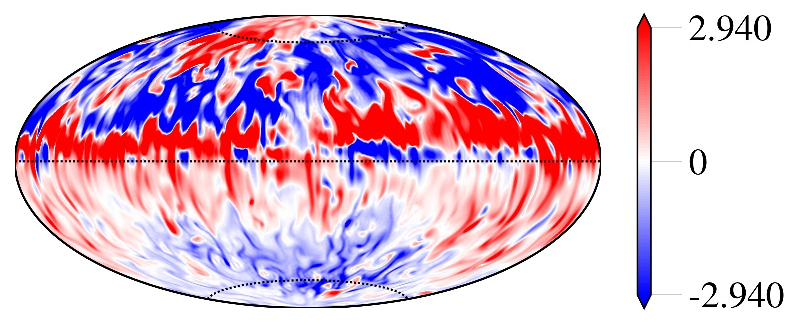}	
	\includegraphics[width=0.4\linewidth]{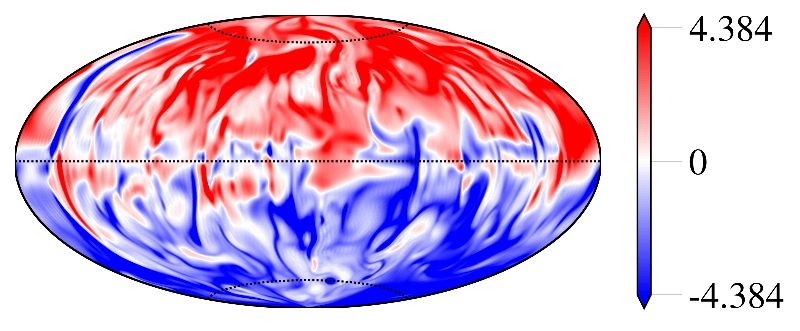}               \\
	\caption{(a) Evolution of the dipole colatitude $\theta$
		and the 
		ratio of the root mean square values of the axial 
		magnetic field ($\bar{B}_{10}$) to the 
		the total magnetic field ($\bar{B}$) 
		with time in units of the magnetic diffusion time. 
		The simulation begins from a seed magnetic field. 
		(b) The evolution in time of the fundamental frequencies. 
		The vertical dashed line at point ‘A’ indicates the 
		emergence of slow MAC waves, which occurs when 
		$|\omega_{M}| \approx |\omega_{A}|$, and the 
		dotted line at point ‘C’ indicates the time of 
		dipole formation. Point ‘B’ represents a multipolar 
		state where slow waves are present. (c) The contours of 
		radial magnetic field at the outer boundary
		is shown for points 
		`B' and 'C'. The dynamo 
		parameters are as follows: 
		Ekman number $E = 1.2 \times 10^{-5}$, 
		modified Rayleigh number $Ra = 5000$,
		Prandtl number $Pr=1$ and magnetic
		Prandtl number $Pm =1$. See also \cite{aditya2022}.}
	\label{seedsupp}
\end{figure}

\begin{figure}
	\centering
	\hspace{-1.5 in}	(a)  \hspace{1.5 in} (b)\hspace{1.5 in} (c) \\
	\includegraphics[width=0.32\linewidth]{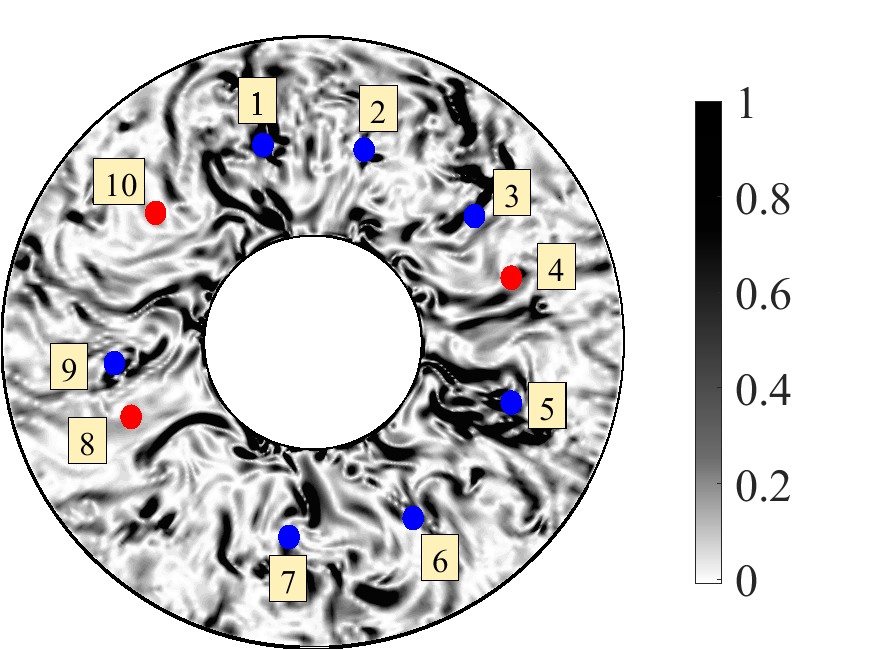}
	\includegraphics[width=0.32\linewidth]{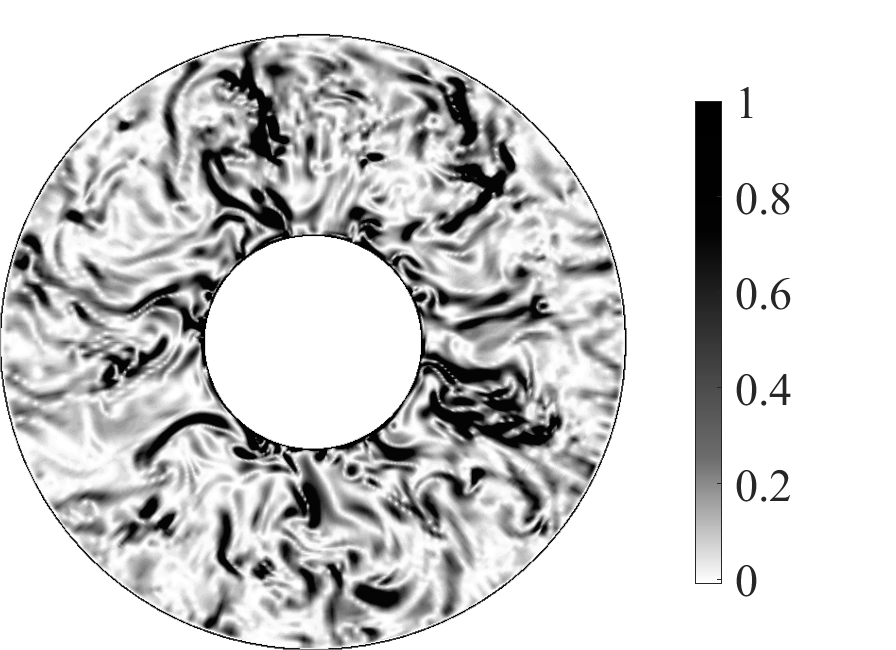}
	\includegraphics[width=0.32\linewidth]{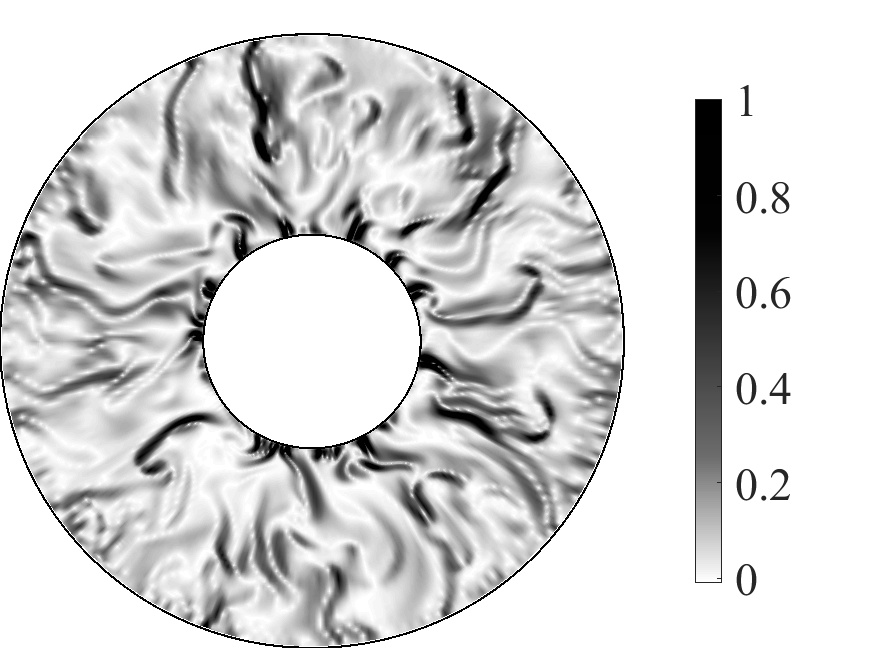}	\\
	\hspace{-1.5 in}	(d)  \hspace{1.5 in} (e)\hspace{1.5 in} (f) \\
	\hspace*{-0.6cm}	\includegraphics[width=0.32\linewidth]{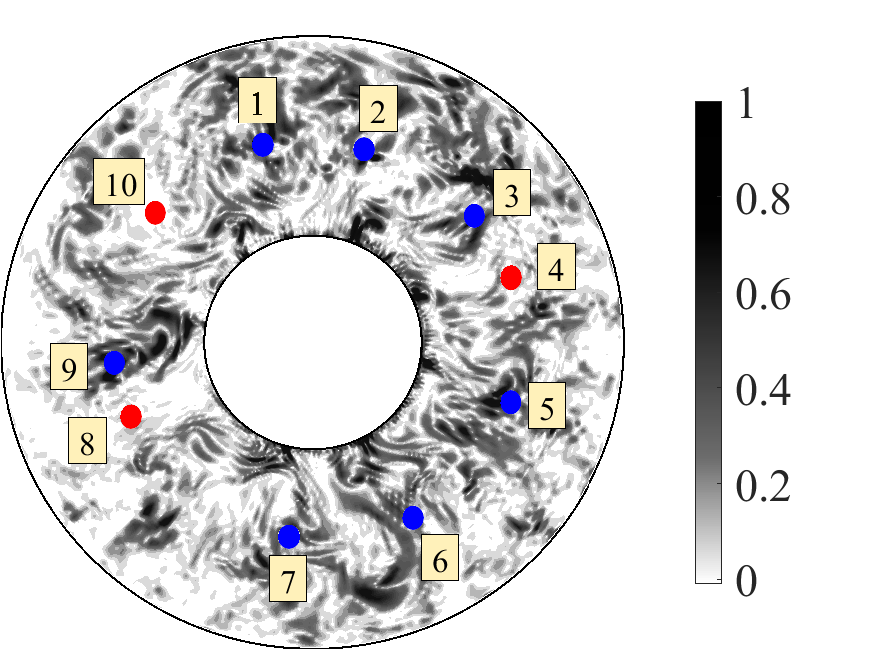}
	\includegraphics[width=0.32\linewidth]{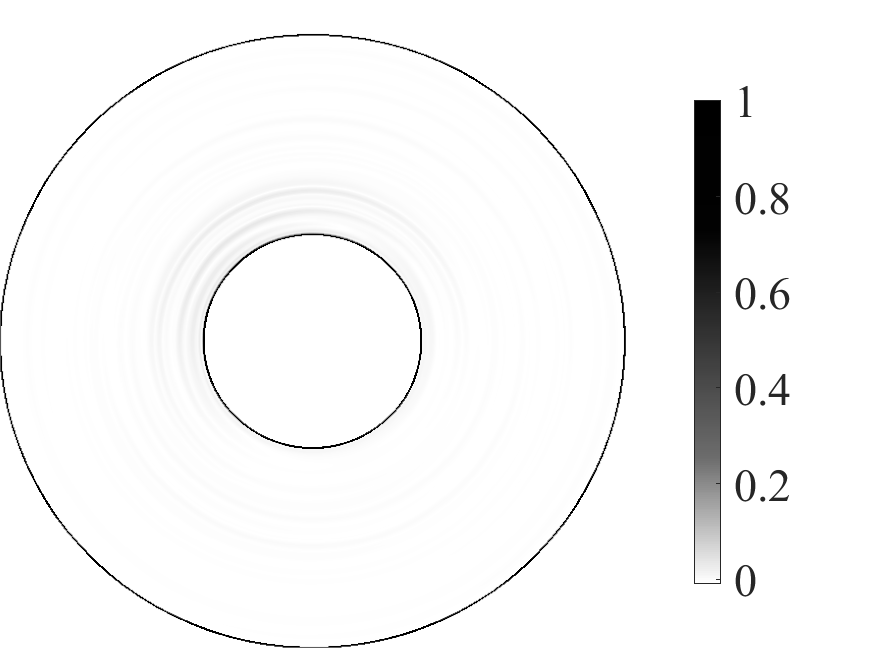}
	\hspace*{-0.6cm}	\includegraphics[width=0.32\linewidth]{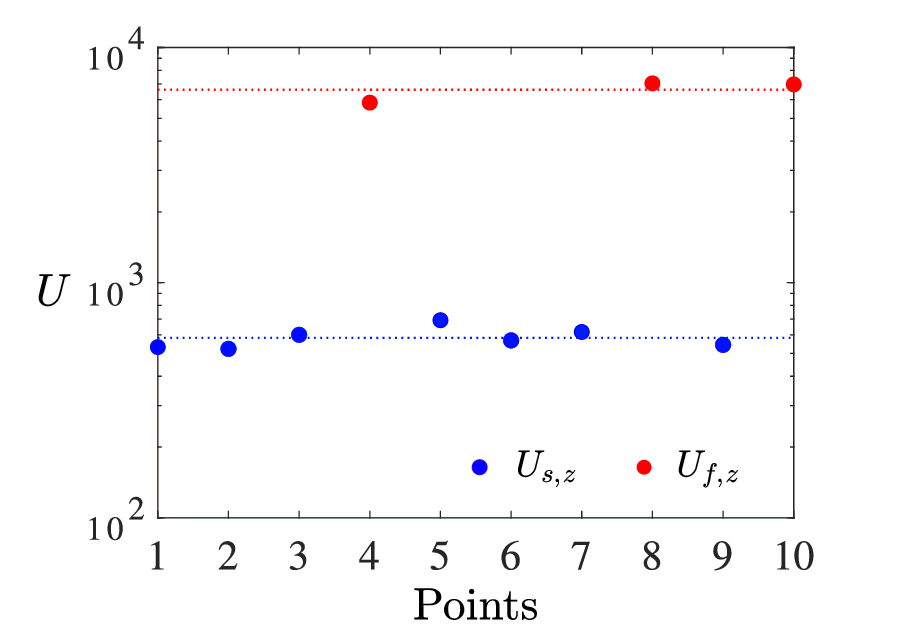}\\
	\caption{Greyscale plots on a horizontal section at $z = 0.2$ 
		below the equatorial plane at the
		instant `B' in figure \ref{seedsupp}, 
		showing the ratio of the magnitudes of (a) Coriolis, 
		(b) Lorentz, (c) buoyancy, (d) non-dipolar Lorentz, and (e) 
		dipolar Lorentz force terms in the $z$-vorticity equation to 
		the magnitude of the largest force among them. 
		(f) The estimated (lines) and measured group velocities
		(points) for 
		the blue and red points shown for figure (d). 
		Here, $U_{f,z}$ and $U_{s,z}$ denote the 
		measured axial group velocities of the fast and 
		slow MAC waves, respectively.
		The dynamo parameters are $E = 1.2 \times 10^{-5}$, 
		$Ra = 5000$, and $Pm = Pr = 1$. }
	\label{forcessupp}
\end{figure}

\begin{figure}
	\centering
	\hspace{-2 in}	(a)  \hspace{2 in} (b) \\
	\includegraphics[width=0.45\linewidth]{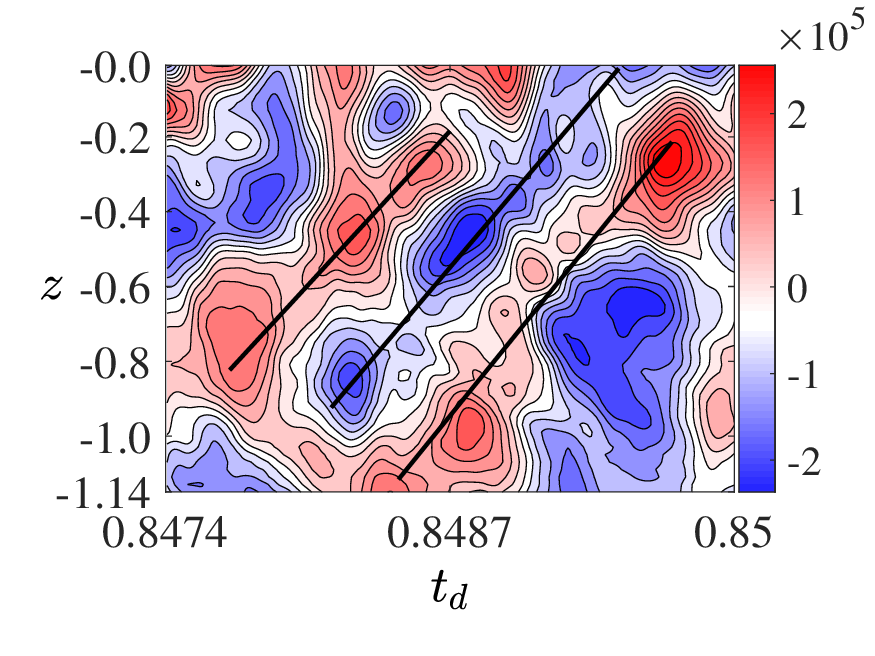}
	\includegraphics[width=0.45\linewidth]{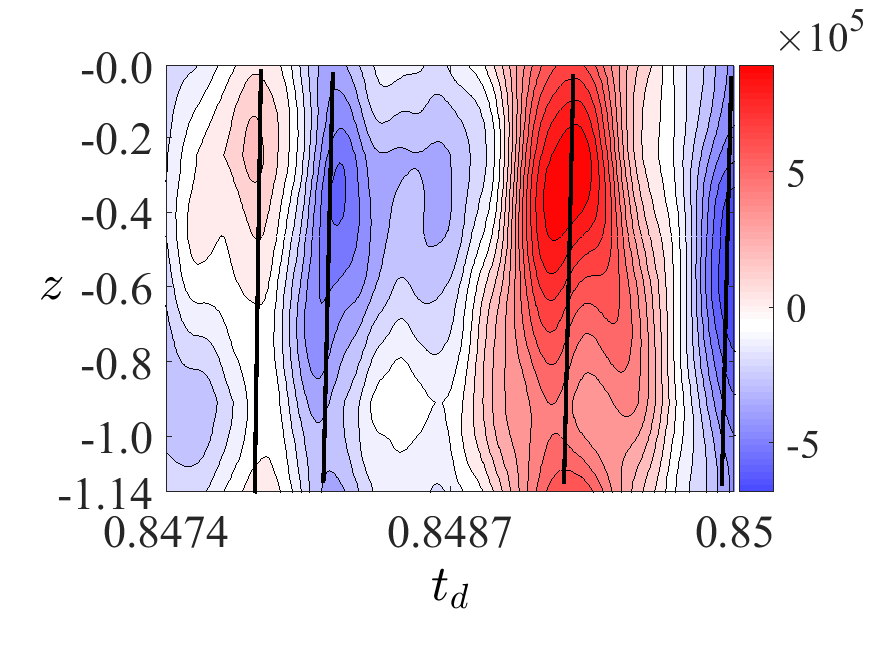}\\
	\caption{(a) \& (b) Contour plots of $\partial u_z / \partial t$ at a cylindrical 
		radius $s = 1$ for $l\le l_E$ presented 
		for the instant `B' at points `1' and `4' respectively in 
		figure~\ref{forcessupp}(d).		
		The parallel black lines indicate the primary wave travel direction, 
		with their slopes denoting the group velocity. 
		Panel (a) shows the propagation of slow MAC waves, 
		whereas panels (b) show the propagation of fast MAC waves. 
		The dynamo parameters are $E = 1.2 \times 10^{-5}$, $Ra = 5000$, $Pm = Pr = 1$.}
	\label{cgfsupp}
\end{figure}

We consider a convection-driven dynamo operating 
in a spherical shell, the boundaries
of which correspond to the inner core boundary 
and the core–mantle boundary. The
ratio of inner to outer radius is 0.35. Lengths are scaled by the thickness
of the spherical shell $L$ and time is 
scaled by magnetic diffusion time $L^2/\eta$. The velocity ${\bm u}$
and magnetic field ${\bm B}$
are scaled by $\eta/L$ and $(2\varOmega \rho \mu \eta)^{1/2}$, 
respectively. The temperature
is scaled by $\beta L$, where $\beta$
is the radial temperature gradient at the outer boundary.
The non-dimensional MHD equations for the velocity,
magnetic field and temperature are given in 
equations (3.1)--(3.4), \S 3. While $\eta$, $\rho$, $\mu$
and $\varOmega$ above are defined in \S 2, the dimensionless
parameters and the basic state
buoyancy profile used in the dynamo model
are described in \S 3. The velocity and
magnetic fields satisfy the no-slip and 
electrically insulating conditions respectively at the
two boundaries. 
The inner boundary is isothermal while the outer boundary has constant
heat flux.

Figure \ref{seedsupp} presents the evolution of the dynamo 
initiated from a seed magnetic field. The 
slow MAC waves onset at the point ‘A’, when the Alfv\'en
wave and buoyancy wave frequencies are approximately
equal, i.e. 
$|\omega_M| \approx |\omega_{A}|$. Point ‘B’ denotes a 
multipolar state in which slow waves are present, while 
the dotted line at point ‘C’ indicates the formation of the 
axial dipole. 
Figure \ref{forcessupp} shows the
relative magnitudes of the axial components of the 
curled Coriolis, buoyancy, and Lorentz
forces at point `B', where the slow MAC wave 
frequency $\omega_s$ is nonzero and yet the dynamo
is in a multipolar state. 
Here, the Coriolis force is balanced
by both the buoyancy and Lorentz forces at several
points within the volume. 
The Lorentz force essentially consists of
the non-dipolar part of the field since 
the dipole field has not formed and
therefore has a negligible contribution
(figures \ref{forcessupp} d \& e).
Figure \ref{cgfsupp} shows the measurement of wave motion 
in the dynamo at 
time `B'. Contours of $\partial u_z / \partial t$ at a cylindrical 
radius $s = 1$ for the energy-containing range of spherical harmonic
degrees $l\le l_E$ are plotted over short time windows during 
which the ambient magnetic field and wavenumbers remain 
approximately constant.
At the instant `B', both slow and fast MAC waves coexist.
The slow MAC waves are generated at several points where
the magnetic flux is concentrated so that localized balances
between the Lorentz, Coriolis and buoyancy forces exist.
The fast waves are present in regions where the magnetic
flux is relatively weak.
The measured and estimated axial group velocities of the 
waves at different points within the volume are shown in 
figure~\ref{forcessupp} (f). The method of their calculation
is given in \cite{aditya2022}.

The helicity of the slow MAC waves generated from the
non-dipolar magnetic flux is essential for formation of the axial
dipole \citep[see][]{aditya2022,jfm24}. In the hydrodynamic dynamo
at the same parameters where the Lorentz force is zero,
the field stays multipolar \citep{prf18}, which indicates the essential
role of the slow MAC waves in dipole formation.

\begin{figure}[h]
	\section*{S2~~~~ Mean resultant temperature gradient, $\beta$}
	\vspace*{0.5cm}
	\centering
	\hspace{-2 in}	(a) \hspace{2 in} (b) \\
	\includegraphics[width=0.4\linewidth]{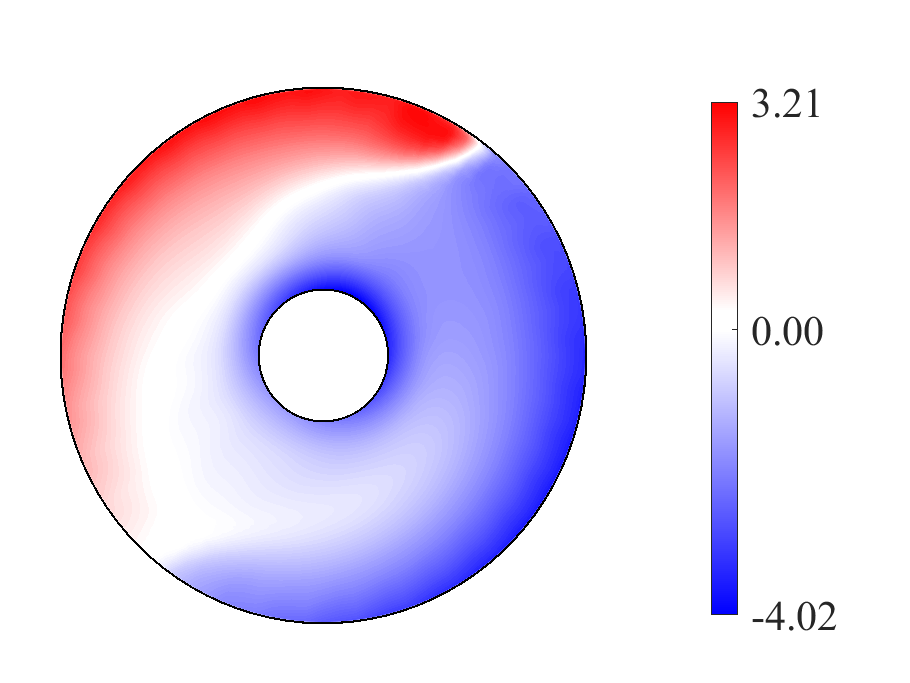}
	\includegraphics[width=0.4\linewidth]{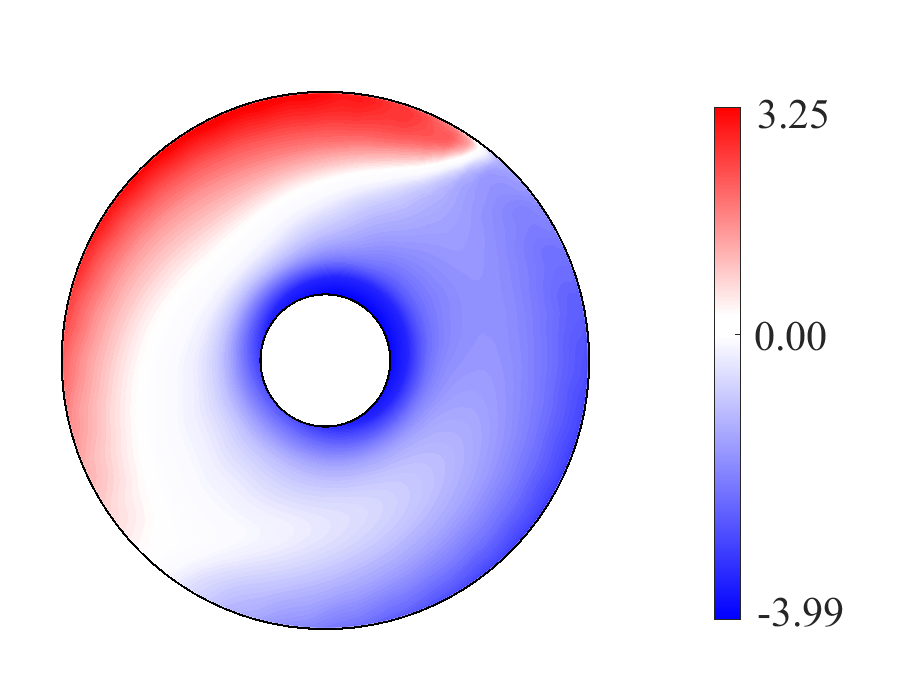}\\
	\caption{Horizontal section plots of the resultant 
		temperature gradient $\beta$ at 
		$z=0.4$ below the equator for (a) $Ra_V=30$ (slightly below onset) and 
		(b) $Ra_V=310$ ($Ra_V/Ra_{V,c} \approx 10$) at $q^\star=18$. 
		A time average of $\approx$ 1 magnetic diffusion time is taken 
		for the supercritical state to eliminate the effect of the perturbations in the convective state. 
		The $Y_2^1$ heat flux heterogeneity is applied
		at the outer boundary. 
		The other dynamo 
		parameters are $E = 1.2 \times 10^{-5}$ and $Pm = Pr = 1$.}
	\label{betaR2}
\end{figure}

\end{document}